\documentclass[
preprint,
 amsmath,amssymb,
 aps,
]{revtex4-1}

\usepackage{graphicx}
\usepackage{dcolumn}% Align table columns on decimal point
\usepackage{bm}

  \renewcommand{\arraystretch}{0.75}

\begin{document}

\preprint{}

\title{Prospects for multi-messenger extended emission from core-collapse supernovae in the Local Universe}% Force line breaks with \\

\author{Maurice H.P.M. van Putten}
\email{mvp@sejong.ac.kr}
\affiliation{Sejong University, 98 Gunja-Dong Gwangin-gu, Seoul 143-747, Korea, and OzGrav-UWA, Department of Physics, The University of Western Australia, Crawley, WA 6009, Australia}
\author{Amir Levinson}\affiliation{School of Physics and Astronomy, Tel Aviv University, 69978 Tel Aviv, Israel} 
\author{Filippo Frontera}\affiliation{Department of Physics and Earth Sciences, University of Ferrara, Via Saragat 1, I-44122 Ferrara, Italy, and INAF, IASF, Via Gobetti, 101, I-40129 Bologna, Italy}
\author{Cristiano Guidorzi}\affiliation{Department of Physics and Earth Sciences, University of Ferrara, Via Saragat 1, I-44122 Ferrara, Italy}
\author{Lorenzo~Amati}\affiliation{INAF, IASF, Via Gobetti, 101, I-40129 Bologna, Italy}
\author{Massimo~Della Valle}\affiliation{Istituto Nazionale di Astrofisica, Osservatorio Astronomico di Capodimonte, Salita Moiariello 16, I-80131 Napoli, Italy}

\date{\today}% It is always \today, today,

\begin{abstract}
Multi-messenger emissions from SN1987A and GW170817/GRB170817A suggest a Universe rife with multi-messenger transients 
associated with black holes and neutron stars. For LIGO-Virgo, soon to be joined by KAGRA, these observations promise 
unprecedented opportunities to probe the central engines of core-collapse supernovae (CC-SNe) and gamma-ray bursts. Compared to
neutron stars, central engines powered by black hole-disk or torus systems may be of particular interest to multi-messenger observations
by the relatively large energy reservoir $E_J$ of angular momentum, up to 29\% of total mass in the Kerr metric. These central engines
are expected from relatively massive stellar progenitors and compact binary coalescence involving a neutron star.
We review prospects of multi-messenger emission by catalytic conversion of $E_J$ by a non-axisymmetric disk or torus. 
Observational support for this radiation process is found in a recent identification of ${\cal E}\simeq (3.5\pm1)\%M_\odot c^2$ in Extended 
Emission to GW170817 at a significance of 4.2\,$\sigma$ concurrent with GRB170817A. A prospect on similar emissions 
from nearby CC-SNe justifies the need for all-sky blind searches of long duration bursts by heterogeneous computing.
\end{abstract}

\maketitle

%\newpage
\tableofcontents

%\mbox{}\\
\mbox{}\\
\begin{table}[h]
{\sc LIST OF SYMBOLS AND ACRONYMS} \\
\begin{tabular}{lll}
\mbox{}\\
$c$ & velocity of light ($3\times 10^{10}$ cm s$^{-1}$)\\
$c_s$ & sound speed\\
%$D$ & source distance\\ 
$E_B$ & energy in poloidal magnetic field\\
$E_{iso}$, $E_\gamma$ & isotropic equivalent and true energy in gamma-rays\\
%$E_{iso}$ & isotropic equivalent energy \\
$E_c$ & maximal spin energy neutron star $(3\times 10^{52}$ erg)\\
%$E_{res}$ & energy in reservoir \\
%$E_{rot}$ & energy in rotation\\
$\eta$ & efficiency\\
$h$ & dimensionless gravitational strain\\
$\xi$ & dimensionless mass-inhomogeneity\\
$L_j$ & luminosity in baryon poor jet (BPJ)\\
$L_0$ & $c^5/G=3.6\times 10^{59}$ erg s$^{-1}$\\
$\sigma$ & torus-to-black hole mass ratio $M_T/M$\\
$\dot{m}$ & accretion rate \\
$M_\odot$ & solar mass ($2\times 10^{33}$ g)\\
$\dot{N}$ & event rate \\
$\nu$ & kinematic viscosity\\
$\nu_H$ & black hole rotation frequency in Hz\\ 
$\Omega_H$, $\Omega_T$ & black hole and torus angular velocity\\ 
$\Omega_{ISCO}$ & angular velocity of matter at the ISCO\\
$\omega$ & frame dragging angular velocity\\
$q$  & index of rotation in accretion disk\\
%$R$ & branching ratio\\
$R_g$ & gravitational radius $(GM/c^2)$\\
$R_S$ & Schwarzschild radius ($2R_g$)\\
$r_{ISCO}$ & radius of ISCO\\
$r_c$ & transition radius to fragmentation by cooling\\
$r_d$ & Roche radius in accretion flows\\
$r_b$ & viscosity-to-radiation driven transition radius\\
$T_{spin}$ & lifetime of black hole spin\\
$\tau$ & coherence time scale\\
$t_{ff}$ & free fall time scale\\
$z$ 	& $r_{ISCO}/R_g$\\
$\theta_H$ & half-opening angle on horizon \\
(L,S)GRB(EE) & (long, short) gamma-ray burst (with Extended Emission)\\
QNM & quasi-normal mode
\end{tabular}
\end{table}

\newpage

\section{Introduction}

The breakthrough LIGO detection GW150914 opens a new window to the Universe \citep{LIG16} beyond electromagnetic radiation, 
neutrinos and (ultra-)high-energy cosmic rays. As broadband detectors covering 30 - 2000 kHz, LIGO-Virgo and KAGRA \citep{aku19} 
promise to probe an exceptionally broad class of astrophysical sources \citep{sat09, cut02}. While the black hole merger event GW150914 left no conclusive signature in the electromagnetic spectrum \citep{kal17}, it offers new results of direct astronomical interest with estimates of mass and spin in the progenitor binary, 
\begin{eqnarray}
M_1=35.7_{-3.8}^{5.4}M_\odot,~~a_1/M_1=0.31^{+0.48}_{-0.28},\\
M_2=29.1_{-4.4}^{3.8}M_\odot,~~a_2/M_2=0.46^{+0.48}_{-0.42},
\label{EQN_GW15}
\end{eqnarray}		
that are surprising given the more familiar observations on stellar mass black holes in X-ray binaries in the Milky Way. 
The inferred high mass $M_i$ of the black holes may originate from core-collapse of Population III stars \citep{kin14,ina17} and their modest 
dimensionless spin $a_i/M_i$ may result from classical isolated binary evolution \citep{qin19} except for those that are remnants of powering supernovae \citep[][]{van17a}.

Gravitational radiation has long since been observed in long time observations of binary evolution in the electromagnetic spectrum \citep{ver97}, notably
in the Hulse-Taylor pulsar PSR B1913+16 \citep{tay89,tay94,wei10}, the double pulsar PSR J0737-3039 \citep{lyn04}, and ultra-short period cataclysmic 
variables with He mass transfer from a degenerate dwarf directly onto a companion (low mass) white dwarf \citep{sma67,pac67,fau71,fau72,nel05,pos06,bil06}. 
The AM CVn source ES Cet ($d\simeq 350$ pc), for instance, has an orbital period of about 10 min, a mass ratio $q\simeq 0.094$ of the binary with a white dwarf of mass $M\simeq0.7M_\odot$ and a luminosity $L_{EM}\simeq 10^{34}$ erg s$^{-1}$ \citep{wou03,esp05}, whose expected gravitational wave-to-electromagnetic luminosity ratio 
satisfies $L_{gw}/L_{EM}\simeq0.3$. At an orbital period of about 5 min, current data on the AM CVn source RX J0806 \citep{bil06} suggest
\begin{eqnarray}
L_{gw} \simeq {L_{EM}},
\label{EQN_i1}
\end{eqnarray}
indicative of relativistic evolution that ultimately terminates in a binary merger. 

GW150914 is the most extreme example with an output $E_{GW}\simeq 3 M_\odot c^2$ at a peak luminosity $L_{gw}\simeq 200 Mc^2$\,s$^{-1}$, 
far surpassing anything featured in any electromagnetic transient event with a peak luminosity $L_{gw}\simeq 0.1\%\,L_0$,
where 
\begin{eqnarray}
L_0 = \frac{c^5}{G} = 3.6\times 10^{59} \,\mbox{erg s}^{-1} = 2\times 10^5 M_\odot c^2\,\mbox{s}^{-1}
\label{EQN_g5}
\end{eqnarray}
is the unit of luminosity defined by Newton's constant $G$ and the velocity of light $c$. In this light, PSR B1913+16 is a gentle whisper at 
$L_{gw}\simeq 10^{-28}L_0$. 
 
SN1987A \citep{bur87,gar87,her87,kun87} and the double neutron star merger GW170817 \citep{abb17} are the first multi-messenger sources of 
MeV-neutrinos and, respectively, gravitational radiation alongside electromagnetic radiation in optical and, respectively, a short gamma-ray burst GRB170817A. 
They give direct evidence for the formation of high-density matter and, respectively, the formation of a compact object in the immediate aftermath of the merger. 
These two events, therefore, point to radiative processes arising from high-density matter in strong gravitational interactions around neutron stars or stellar mass black holes believed to power the most relativistic transients in the sky: cosmological gamma-ray bursts (GRBs), discovered serendipitously by nuclear treaty monitoring satellites \citep{kle73}. 

For gamma-ray bursts, the potential of strong gravitational interactions may be inferred from compactness of their putative central engines, e.g., \citep{van00}
\begin{eqnarray}
\alpha_E = \frac{G E}{c^5\delta t}=2.75\times10^{-5} \,\left( \frac{E_{52}}{\delta t_{-3}}\right)
\label{EQN_alpha1}
\end{eqnarray}
for burst energies $E= E_{52} 10^{52}$ erg and variability times $\delta t=\delta t_{-3}$ ms. 
The observed isotropic equivalent energies $E_{iso}=10^{48}-10^{54}$ erg and variability times $\delta t$ down to 0.1 ms show $\alpha_E$ up to $10^{-4}$. 
Such values are extremely large compared to those of other transients, including GRB 980425 associated with SN 1998bw \citep{gal98} and galactic sources such as 
GRS 1915+105 \citep{mir94}. It implies engines in the form of neutron stars or stellar mass black holes, more likely so than aforementioned white dwarfs in CVs. Neutron star masses tend to cluster around $1.4M_\odot$ \citep{tho99}, from 1.25 $M_\odot$ of PSR J0737-303B \citep{lyn04} to 2.1 $M_\odot$ in the NS-WD binary PSR J075+1807 \citep{nic04}; masses of black hole candidates in X-ray novae are broadly distributed between about 5-20 $M_\odot$ \citep{bai98}. 

\begin{figure}
\centerline{\includegraphics[scale=0.5]{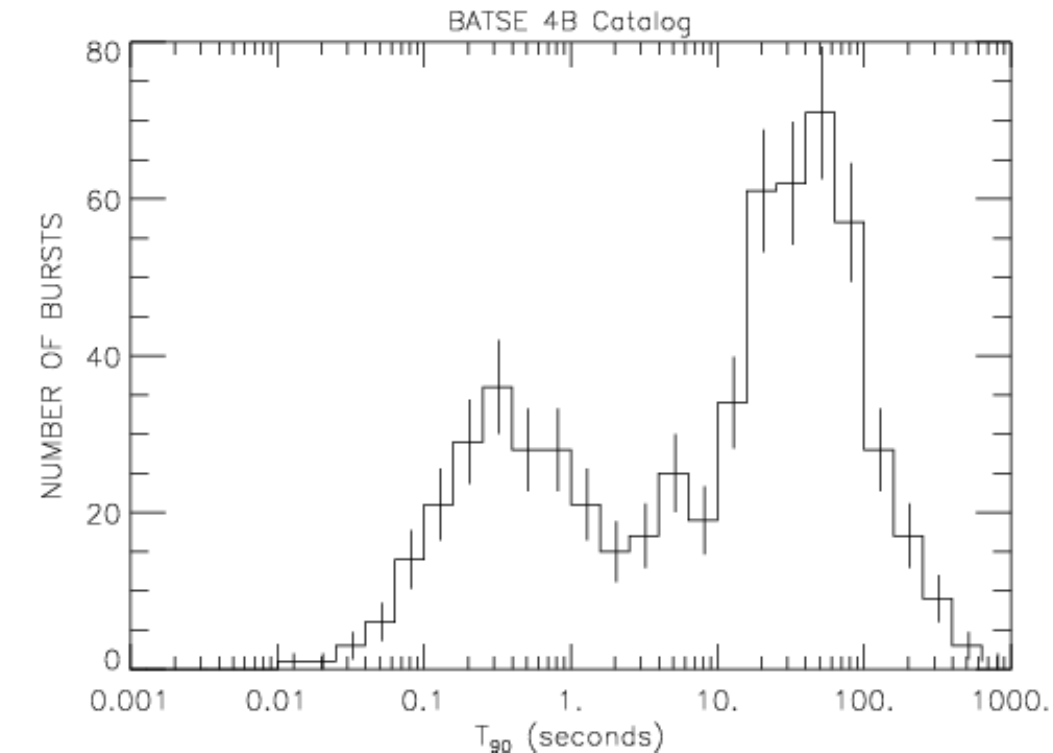}\includegraphics[scale=0.440]{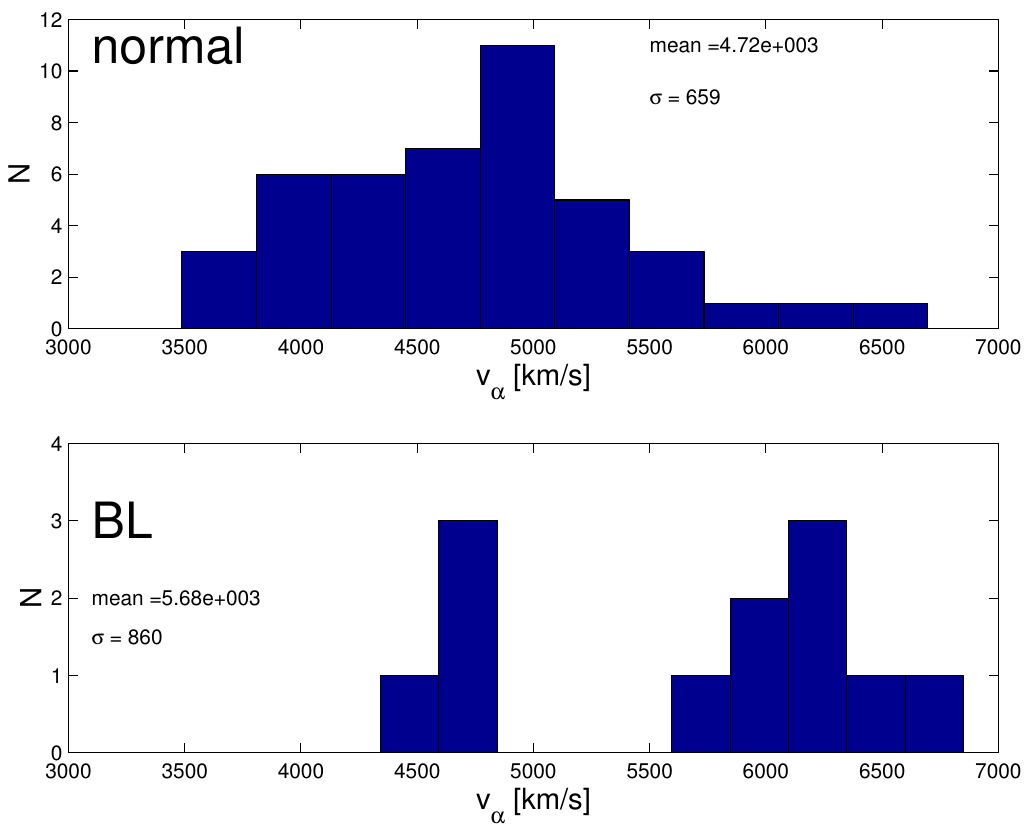}} 
\caption{(Left.) The bimodal distribution of durations in the BATSE 4B Catalog, showing a population of short GRBs (less than 2 s) 
and long GRBs (over 2 s) (Reprinted from \citep{bat01}.) (Right.) Core-collapse supernovae form a heterogeneous class of events, 
broadly partitioned in normal (narrow line) and relatively more energetic (broad line, BL) events. Reprinted from \citet{van11b}, data from \citet{mau10}.}
\label{figSN1987A3}\label{fig4B1}
\end{figure}

GRBs show anomalous Eddington luminosities of $L_\gamma \simeq 10^8-10^{14}L_{Edd}$, given limited durations of typically less than one minute (Fig. \ref{fig4B1}). 
These super-Eddington luminosities defy an origin in electromagnetic interactions in a baryonic energy source. 
The only physical processes known that might circumvent these limitations are neutrino emissions and gravitational interactions 
allowed by the theory of general relativity. In anisotropic emission, the true energy in gamma-rays 
$E_\gamma << E_{iso}$, e.g., when GRBs are produced in jet-like outflows at finite opening angles. Even thus, some events have 
$E_\gamma\simeq 10^{52}$ erg. Typical values of events that reveal collimation show a relatively narrow distribution around \citep{fra01,ghi06,ghi13}
\begin{eqnarray}
E_\gamma\simeq 9\times 10^{50}\,\mbox{ erg}.
\label{EQN_frail}
\end{eqnarray}

Normal long GRBs have an accompanying supernova explosion with kinetic energies $E_k$, typically greater than $E_\gamma$ but with some exceptions.
Particularly, $E_k$ may point to an energy reservoir $E_{res}$ that exceeds the maximal spin energy $E_c$ of a rapidly rotating neutron star \citep{van11b},
marking the birth of a stellar mass black hole or collapse of a neutron star thereto. The Burst and Transient Source Experiment (BATSE) discovered the populations 
of short (SGRB) and long GRBs (LGRB) with durations $T_{90}<2$ s and, respectively, $T_{90}>2$ s (Fig. \ref{fig4B1}), extended by {\em Swift} to short GRBs with 
Extended Emission (SGRBEE) lasting tens of seconds to well over a minute. Their soft EE is very similar to long GRBs with accompanying supernova. In attributing 
SGRBs to mergers, SGRBEEs defy the dynamical time scale $T_{merger}\simeq 10$ ms of NS-NS or NS-BH mergers by a large factor. 

Electromagnetic observations thus constrain the physical nature of GRBs by extreme values of (\ref{EQN_alpha1}) and 
\begin{eqnarray}
%\begin{array}{ll}
L_\gamma >> L_{Edd},~ ~
E_{res}      >E_c ~\mbox{(in some cases)},~ ~
T_{90}^{SGRBEE} >>T_{merger}.
%\end{array}
\label{EQN_i2}
\end{eqnarray}
GRB engines hereby should be ultra-relativistic, conceivably operating by {strong gravitational interactions with high density matter on the scale of 
the Schwarzschild radius} $R_S=2R_g$ of their total mass $M$,
\begin{eqnarray}
R_g = \frac{GM}{c^2}.
\label{EQN_g1}
\end{eqnarray}
If so, their engines might be luminous in gravitational waves over the lifetime of the inner engine, i.e., up to tens of seconds indicated by long 
GRBs \citep{van01b} - ``{\em If gravitational waves are detected from one or more gamma-burst triggers, the waves will almost certainly reveal the physical nature of the trigger}''  \citep{cut02}.
  
GRB170817A is a nearby SGRB that led to the first successful detection GW170817 of a binary neutron star merger \citep{abb17}.
At a distance of 40 Mpc, GW170817 was fortuitous, since the event rate of GRBs within the sensitivity distance of LIGO-Virgo and KAGRA is quite sparse. 
Even corrected for beaming, the true GRB event rate is about one per year within a distance of 100 Mpc, similar to the event rate of double neutron star coalescence,
or about one per century within the distance to Virgo $(D\simeq 20$ Mpc). Beaming is less severe in afterglow emission that follows the prompt phase as the blast wave slows down and, at late times, the emission is ultimately roughly isotropic. In such cases, an observer might  detect an orphan afterglow emission at radio 
wavelengths \citep{lev02}, a few months after the explosion. Identifying GRBs by afterglow emissions leaves uncertain the true time-of-onset of the trigger, 
however, hampering efficient search for an accompany gravitational wave burst.  
  
 A search for gravitational radiation from SN of type Ib/c \citep{mae02,mae08,fru06} appears opportune, given that they are
 far more numerous than LGRBs by about two orders of magnitude, representing about 20\% of all core-collapse supernovae (CC-SNe).
CC-SNe are remarkably heterogenous (Fig. \ref{figSN1987A3}) with a ratio of about two-to-one of narrow- to broad-line events with 
relatively high velocities of mass ejecta. The most energetic events of type Ib/c stand out as the parent population of normal LGRBs.
  
   {The small branching ratio of CC-SNe into successful GRBs is commonly attributed to the challenge of creating an energetic inner engine sufficiently 
    long lived, powering a supernova by an (ultra-)relativistic outflow. Jet powered supernovae for a broad class of events \citep{sto16,sto17,pir17}, that appears to
    include SN1987A (Fig. \ref{figSN1987A3}). Jet outflows may or may not successfully breakout of the progenitor remnant envelope, the first producing normal
    LGRB by collimated ultra-relativistic baryon-poor jet (BPJ). %\citep[e.g.][]{gra17}. 
    This diversity implies a broad class of CC-SNe with and without GRBs. 
    Unsuccessful jet breakout from the stellar envelope in a CC-SN event \citep{maz08,cou11,bro12} leads to so called ``choked GRBs,"  
    appearing as low-luminosity long GRB or, more broadly, as a class of X-ray transients \citep{sod08,mar14}.  Certain types of supernova explosions 
    may lead to a relativistic shock breakout that may explain (nearby) low-luminosity GRBs (LLGRBs) but not the prompt GRB emission of normal long GRBs 
    \citep{nak12,nak15,gra17}. Diversity may, in part, be attributed to engines harboring neutron stars or black holes, that is gradually becoming amenable 
    also to numerical simulations \citep{obe17}. The detailed properties of these engines leading to a successful breakout are not known, but a contributing factor 
    may be intermittency at the source increasing time-average luminosity of magnetically launched jets \citep{van15a}.
    Conceivably, therefore, the formation of energetic inner engines is more frequent than successful GRB-SNe. 
    Similar considerations might apply to their emission in neutrinos \citep{mes01,nak15}.      

\begin{figure}
\centerline{\includegraphics[scale=0.075]{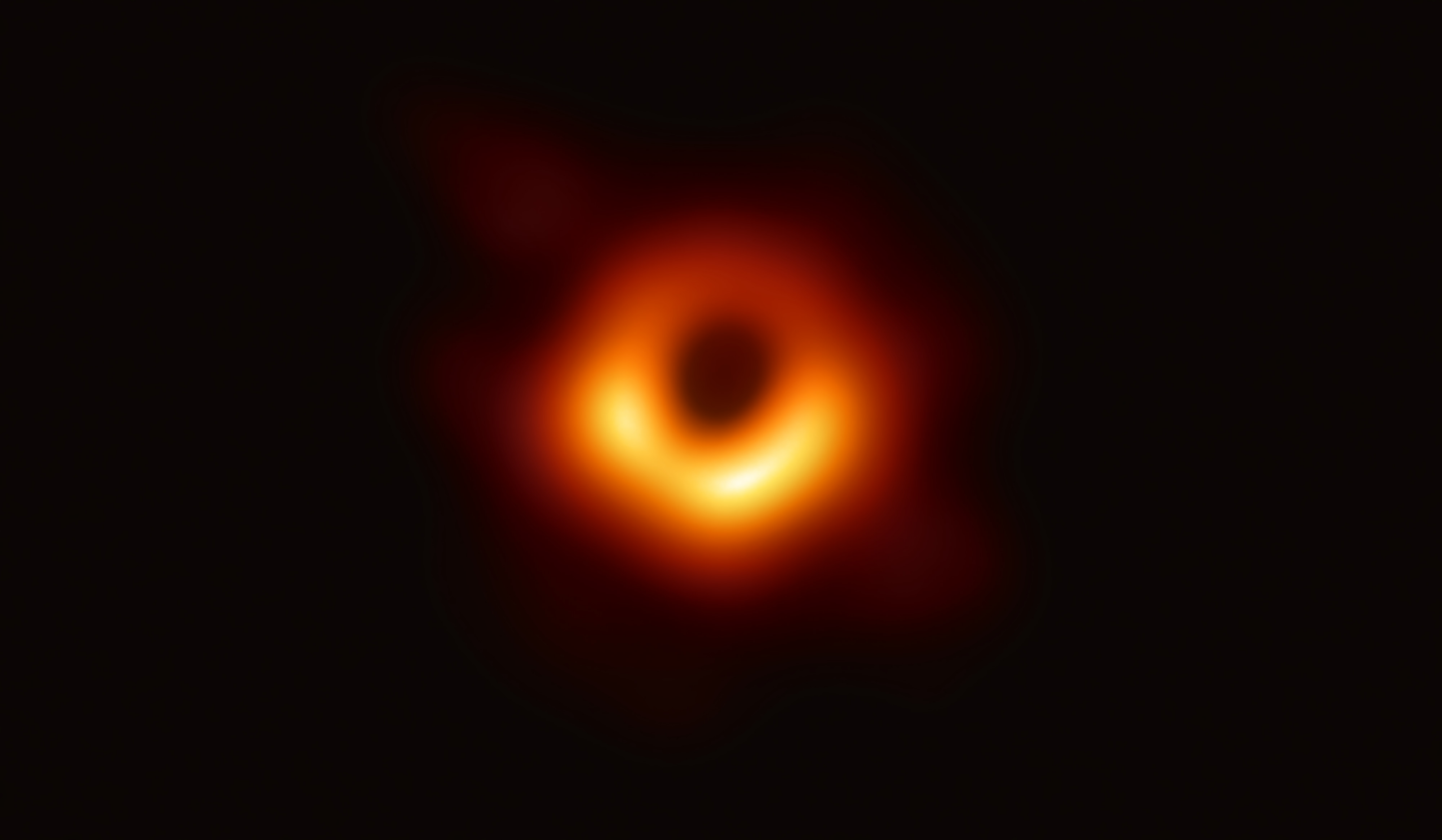}} 
\caption{Radio image of the disk of ionized gas around the black hole in the active galactic nucleus of M87. The high angular 
resolution reveals some clumpiness beyond what is expected from gravitational lensing by the gravitational field of the black hole
\citep{EHT19}, pointing to non-axisymmetric orbiting mass permitting radiation of gravitational waves of potential interest to LISA \citep{van12b}.
(Credit: Event Horizon Telescope Collaboration.)}
\label{EHT}
\end{figure}

    We shall review some specific prospects for gravitational radiation from CC-SNe as the parent population of normal long GRBs and LLGRBs, 
    from central engines harboring rotating black holes along with un-modeled searches for {broadband extended gravitational radiation} 
    using modern heterogeneous computing. {Our focus will be on %rapidly rotating black hole-disk or torus systems  
    potentially multi-messenger radiation from black hole-disk or torus systems and their naturally formation in core-collapse 
    of relatively more massive progenitor stars. For black holes, we recall their rotational energy $E_J$ in angular momentum $J$ 
    of up to $29\% M_\odot c^2$ according to the Kerr metric \citep{ker63}, far exceeding the limit of a few $\%M_\odot c^2$ of rapidly rotating neutron stars.
    For multi-messenger astronomy, this motivates all-sky blind searches for un-modeled broadband extended gravitational-wave emission (BEGE) searches from 
    CC-SNe and GRBs by novel pipelines exploiting modern high-performance computing alongside detailed energy constraints derived from 
    electromagnetic observations. Observational support for this approach can be found in the multi-messenger results obtained on GW170817/GRB170817A.}
  
  {We shall highlight prospects for gravitational-wave bursts from non-axisymmetric accretion flows onto rotating black holes (Fig. \ref{EHT}),
  here around stellar mass black holes in core-collapse of massive stars or (delayed) collapse of a hyper-massive neutron star formed in the
  immediate aftermath of a compact binary merger. A broad outlook is given to potentially long duration emission in gravitational radiation at relatively 
  low frequencies. This outlook is different from what may be expected in the first one or two seconds from core-collapse and bounce in complex short 
  duration bursts producing neutron stars \citep{ott09} with a LIGO sensitivity distance limited to a few Mpc \citep{rov09}. 
  Their connection to GRB-supernovae and extremely energetic supernovae, however, is not obvious \citep{bur07,des08}.
  In contrast, Extended Emission in LGRBs and SGREEs indicates central engine lifetimes of tens of seconds that conceivably indicates activity over
  similar durations in the central engines of the more frequent group of energetic type Ib/c supernovae. 
  It therefore appears opportune to search for un-modeled BEGE from, broadly speaking, nearby energetic CC-SNe by all-sky blind searches 
  by advanced computing.}
   
 The relatively energetic type Ib/c SNe have an event rate of about 100 per year within a distance of 100 Mpc. Found in optical surveys 
 using moderately sized telescopes, they and the larger class SNII provide readily available targets of opportunity of interest to LIGO-Virgo and KAGRA 
 provided that the (unknown) fraction successfully producing gravitational wave bursts exceeds their branching ratio into LGRBs \citep{heo15}. 
 Additionally, nearby galaxies such as M51 ($D\simeq 8$ Mpc) and M82 $(D\simeq 4$ Mpc) each with an event rate of over one core-collapse 
 supernova per decade. By their proximity, these events appear of interest as well, independently of any association with SNIb/c
 or GRBs \citep{and13,aas14}. 

\subsection{Energy in angular momentum $E_J$}

  Energetic output in BEGE from black hole-disk or torus systems may be large  by the ample energy reservoir $E_J$ of rotating black holes, 
   {exceeding that of rotating neutron stars by some two orders of magnitude.} 
   We recall the maximal rotational energy $E_c$ of a neutron star is attained at its break up frequency whereas the same of a black 
   hole of mass $M$ is attained when 
   its angular momentum reaches $GM^2/c$. Canonical bounds on the energy reservoir $E_J$ in angular momentum $J$ are 
   \begin{eqnarray}
   E_{J}\lesssim E_c & = 1.5\%M_\odot c^2 
   \label{EQN_EcEHa}
   \end{eqnarray}
   for neutron stars \citep{hae09}, and
   \begin{eqnarray} 
   E_J & \lesssim 29\%M c^2 M
   \label{EQN_EcEHb}
   \end{eqnarray}
   for rotating black holes of mass $M$ according to the Kerr metric \citep{ker63} (\S3 below).
      
   Of particular interest is {\em enhancement} of $E_J$ by up to two orders of magnitude in (angular momentum conserving) 
   core-collapse of a hyper-massive neutron star to a Kerr black hole, in the immediate aftermath of a merger or CC-SN.
   Such radically changes the outlook on potentially observable multi-messenger emission in catalytic conversion of $E_J$ by 
   a surrounding accretion disk or torus into gravitational radiation, MeV-neutrinos, mass-ejecta and electromagnetic radiation \citep{van03}.
    {If detected, LIGO-Virgo and KAGRA probes may reveal rotating black holes by calorimetry on their output in gravitational waves} \citep{van02b}. In general terms, 
   gravitational radiation from non-axisymmetries associated with core-collapse of high angular momentum progenitors and non-axisymmetric 
   collapse has been well appreciated \citep{bek73,thu74,nov75,eps76,det81}, see further \cite[e.g.][]{det81,kot06,ott09,fry11}. 
   Potentially extreme energy ${\cal E}$ in gravitational radiation may derive from non-axisymmetric accretion flows down to ISCO 
   powered by $E_J$ of the central black hole \citep{van01b}. Generated by matter in possibly turbulent motion, this calls
   for {\em un-modeled} searches for ascending chirps and descending chirps. % \citep{lev15}.
   
    Detailed spectral and temporal analysis of GRBs from BATSE, {\em BeppoSAX} and {\em Swift} give considerable support for 
    rotating black holes as common endpoints to energetic core-collapse events and mergers of neutron stars with neutron stars or stellar mass black holes alike.
    They hereby define a leading candidate as a universal inner engine to LGRBs and the {\em Swift} class of SGRBEE and LGRB's with no supernovae (LGRBN) 
    \citep{van14b}. In interaction with high density accretion flows, potentially powerful gravitational wave emission may ensue, powered by accretion or the 
    energy $E_J$ in the angular momentum $J$ of the black hole mediated by relativistic frame-dragging. 
    
   The existence of frame dragging is not in doubt: recent measurements of non-relativistic frame dragging around 
   Earth are in excellent agreement with general relativity \citep{ciu04,ciu07,ciu09,eve11}. (Gravity Probe-B measurement is 
   equivalent to that at 5.3 million Schwarzschild radii of an extremal Kerr black hole endowed with Earth's angular momentum 
   \citep{van13}.) Specifically, accretion flows onto rotating black holes offer a window to broadband extended gravitational-wave 
   emission, from non-axisymmetric accretion flows and high density matter accumulated at the Innermost Stable Circular Orbit (ISCO), 
   contemporaneously with two-component relativistic outflows that may drive an accompanying supernova explosion and GRB. 
   Some of these model considerations can be confronted with data from GRB catalogues of BATSE, {\em BeppoSAX} and {\em Swift.} 
   The resulting outlook on long duration {\em ascending} and {\em descending} chirps from accretion flows onto rotating black holes 
   \citep{lev15} suggests searches for BEGE using modern heterogeneous computing \citep{van17b}.

\subsection{Quadrupole gravitational radiation}

In central engines harboring neutron stars or black holes, gravitational radiation is expected from non-axisymmetric mass-motion by canonical instabilities \citep{van02,kob03,van03,pir07}, due to cooling in self-gravitating disks (e.g. \citep{gam01,ric05,mej05,lov14,had14}) or magnetic stresses \citep{tag90,tag99,tag01,lov14}, some of which may be pressure-induced by dissipation or magnetic fields in feedback from a rotating black hole \citep{van03,bro06}. 
Some of these mechanisms may be already seen at work in high frequency QPOs in micro-quasars \citep{tag06} or flaring in SgrA* \citep{tag06b},  

In what follows, we shall change to geometrical units and denote the gravitational radius (\ref{EQN_g1}) by $M$. Equivalently, we put $G=c=1$ in (\ref{EQN_g1}). Thus, $M$ parametrizes perturbations in space-time at a distance $D$ in terms of a {\em dimensionless strain}
\begin{eqnarray}
h = \frac{M}{D}+h_{GW},
\label{EQN_g2}
\end{eqnarray}
where $h_{GW}$ is the strain amplitude in gravitational radiation. At large distances, $h_{GW}$ satisfies the linearized Einstein equations in vacuo,
given by a second order wave equation for small amplitude perturbations that satisfies the same dispersion relation as electromagnetic waves (Appendix A).

At the lowest frequency, efficient gravitational radiation is described by the quadrupole gravitational-wave formula, a special case of higher frequency emissions from
rotating tidal fields from multipole mass moments $I_{lm}$, where $l$ and $m$ refer to the poloidal and azimuthal quantum numbers of spherical harmonics. 
Thorne \citep{tho80} gives a comprehensive overview of gravitational wave luminosity in $h_{GW}$ above from multipole mass moments defined by projections 
on the spherical harmonics $Y_{lm}$ $(l\ge m\ge2)$,
\begin{eqnarray}
L_{lm} = \frac{1}{32\pi} \frac{G}{c^{2l+1}} \left( \frac{d^{l+1}}{dt^{l+1}} I_{lm}\right)^2
\label{EQN_Tho1a}
\end{eqnarray}
by
\begin{eqnarray}
I_{lm} = \frac{16\pi}{(2l+1)!!} \left[ \frac{(l+1)(l+2)}{2(l-1)l} \right]^\frac{1}{2} \int_V Y_{lm}^* r^l dm,
\label{EQN_Tho1b}
\end{eqnarray}
where $dm=\rho\, d^3x$ over the source region $V$ expressed in spherical coordinates $(r,\theta,\varphi)$ as before. 

Relatively inefficient radiation derives from radial mass-motion $(m=0)$, introducing time-dependence with $m=0$ which, by (\ref{EQN_Tho1a}-\ref{EQN_Tho1b}), 
that does not radiate any angular momentum. Illustrative is the gravitational wave output of about 0.2\% from head-on collisions of two Schwarzschild black holes \citep{ann93} (cf. \citep{gib72}), { which reflects} the effective regularization by black hole event horizons of the singular behavior of Newton's law between point particles \citep{van12c}. In contrast, the $m\ne0$ tidal fields in binary mergers shows appreciable efficiency up to about 2\% (e.g. \cite{kyu13,szi15}) and slightly more in double neutron star coalescence \citep{ber15}. 

In binary of two masses $M$ with separation $a$, much greater than the Schwarzschild radius $R_g=2M$ of the system,
expressed in geometrical units $G=c=1$, wherein aforementioned $L_0=1$. 
The Newtonian potential energy $U_N=M^2/a$ between the two introduces a tidal field, whose quadrupole moment 
rotates at an angular velocity $\omega=2\Omega_b$, where $\Omega_b=\sqrt{2M/a}$ is the Keplerian orbital angular velocity of the binary. 
Combined, the dimensionless amplitude $\epsilon = U_N/M$ and angular velocity $M\omega$, define a luminosity 
\begin{eqnarray}
L \sim \epsilon^2 \omega^2 \sim \left( M\Omega_b\right)^\frac{10}{3}.
\label{EQN_LS}
\end{eqnarray}
For a binary of two masses $M_i$ $(i=1,2)$ in circular motion, a detailed derivation obtains the celebrated quadrupole gravitational radiation formula (Appendix A)
\begin{eqnarray}
L_{gw} = \frac{32}{5} \left({\cal M}\Omega_b\right)^\frac{10}{3},
\label{EQN_g5b}
\end{eqnarray}
further replacing $M$ with the {\em chirp mass} ${\cal M} = {M_1^\frac{3}{5}M_2^\frac{3}{5}}{(M_1+M_2)^{-\frac{1}{5}}}$ \citep[e.g.][]{sha83,tho02}.
An extension to non-circular orbits by incorporating enhanced emission at higher frequency harmonics obtains by including a factor $F(e)$ (Eqn. (32), Appendix A) as a function of ellipticity \citep{pet63,pos06}. 
It has been verified experimentally in long-term radio observations of the orbital decay of the Hulse-Taylor binary PSR 1913+16 to better than 0.1\% with $F(e)= 11.8568$ for the observed ellipticity $e=0.6171334$ \citep{tay94}. At the distance of 6.4 kpc, its $L_{gw}\simeq 8\times 10^{31}$ erg s$^{-1}$ produces an instantaneous dimensionless strain at the Earth that, as such, may be evaluated directly in geometrical units to give 
an equivalent isotropic equivalent strain
\begin{eqnarray}
h\simeq \frac{L_{gw}^{1/2}}{\Omega D} \simeq 1.38 \times 10^{-22},
\label{EQN_h1}
\end{eqnarray}
based on previous arguments with $k=1/16\pi$ (Appendix A). As a relatively compact binary, the Hulse-Taylor binary coalesces in about 310 Myr \citep{pos06}.

Coincidentally, (\ref{EQN_h1}) is very similar to the scale of maximal strain in the final merger of a circular binary of two neutron stars of total mass $M=M_1+M_2$
in the Local Universe. For an equal mass binary, $L_{gw}={2}/{5} \left({M}/{a}\right)^5$, 
averaging over the orientation of the source (e.g., \cite{pos06} for a more general discussion) gives 
\begin{eqnarray}
h = \sqrt{\frac{2}{5}}\frac{M^2}{aD}  = \sqrt{\frac{2}{5}} \frac{M}{D} \left(\pi Mf_{gw}\right)^\frac{2}{3},
\end{eqnarray}
where $f_{gw}$ is the quadrupole gravitational wave-frequency equal to twice the orbital frequency $f_{orb}$.
It shows that $h$ is the product of the Newtonian specific binding energy $U=M/a$ and distance scale factor $M/D$ (see also \cite{sat09}).
That is (e.g. \cite{tho92,ju00,pos06}),
\begin{eqnarray}
h= 6.3 \times 10^{-23} \left(\frac{M}{3M_\odot}\right)^\frac{5}{3} \left(\frac{D}{100\,\mbox{Mpc}}\right)^{-1} \left(\frac{f_{gw}}{1000\,\mbox{Hz}}\right)^\frac{2}{3}.
\label{EQN_h2}
\end{eqnarray}

In double neutron star coalescence, (\ref{EQN_h2}) holds true up to the instant when $h$ peaks at $f\simeq 800$ Hz \citep{bai08}. 
This ascending chirp has now been observed up to about 300 Hz in the NS-NS coalescence event GW170817 \citep{abb17}. 
Numerical simulations show that the neutron stars subsequently break up, merge into a hyper massive object (e.g. \citep{ber15a}) followed 
by prompt or delayed collapse to a stellar mass black hole accompanied by a burst of quasi-normal mode (QNM) ringing. The result is a rapidly 
rotating low-mass black hole of close to mass $M$ with an accretion disk of mass up to a few tenths of $M_\odot$ \citep{bai08}. It may give rise 
to a short GRB, but perhaps also a SGRB with Extended Emission \citep{van14b}.

\subsection{Multi-messenger emission from SN1987A}

SN1987A in the Large Magellanic Cloud (LMC, $D\simeq 50$ kpc (Fig. \ref{figSN1987A1}) stands out as a multi-messenger event by a luminous output in 
MeV-neutrinos and accompanying electromagnetic radiation. Characteristic for a core-collapse supernova, SN 1987A was radio-loud \citep{tur87} and aspherical \citep{pap89}. It featured relativistic radio jets \citep{nis99} with a possible black hole remnant, based on a lack of detection of a neutron star and on evidence for a black hole in the rather similar type IIL event SN1979C \citep{mat79,pat11}. Collectively, core-collapse supernova form a rather heterogeneous group \citep{fil97}, broadly in narrow line and broad line events (Fig. \ref{figSN1987A3}, right panel) with more energetic and relativistic ejection velocities. SN1987A has a relatively massive progenitor \citep{gil87,kir87}, possibly powered by an angular momentum rich wind or jet giving rise to its aspherical supernova remnant (Fig. 3).

The output $E_\nu\simeq 10^{53}$ erg in $>10$MeV neutrinos offered our most direct view yet on the innermost workings of a CC-SN.
Spectroscopic observations of SNe-Ibc \citep{maz05,tau09,mod14} reveal that geometry of ejecta of the stripped envelope supernovae is, in about 50\% of the 
observed events, strongly asymmetric. Non-axisymmetric angular momentum-rich explosion mechanisms inevitably produce gravitational waves, of interest as
candidate events in the Local Universe.
 
$E_\nu$ evidences the formation of high density matter. Combined with ample angular momentum in the progenitor, these are just the kind of conditions needed 
for an output in gravitational radiation. Such will be especially luminous, whenever the resulting non-axisymmetric high-density mass-motion takes place on the 
Schwarzschild scale of the system, e.g., the Innermost Stable Circular Orbit (ISCO) around a newly formed black hole. By virtue of the large value of $L_0$ in 
(\ref{EQN_g5}), the quadrupole formula (\ref{EQN_g5b}) predicts an appreciable luminosity for canonical values of non-axisymmetric mass inhomogeneities $\delta m$ about a central mass $M$,
\begin{eqnarray}
L_{gw}\simeq 10^{50}-10^{53}~\mbox{ erg~ s}^{-1}~~(\delta m/M = 0.1-1\%).
\label{EQN_LGW0}
\end{eqnarray}

More specifically, $\delta m = M_2<<M_1=M$ introduces a chirp mass $\mu \simeq \left({\delta m}/{M}\right)^\frac{3}{5} M$.
By virtue of (\ref{EQN_g5}), $L_{gw}$ reaches luminosities on par with $E_\nu$ of SN1987A for $\delta m/M\simeq 0.1\%$ at an orbital radius of a few times the 
Schwarzschild radius $R_S=2R_g$. As a mass perturbation in a torus or inner disk of mass $M_T\simeq 0.01 \,M$, the gravitational wave luminosity (\ref{EQN_g5b}) of $\delta m$ satisfies
\begin{eqnarray}
L_{gw} =\frac{32}{5} \left(\frac{\delta m}{M}\right)^2\left(\frac{M}{a}\right)^5 L_0
\end{eqnarray}
in the limit of a small chirp mass. Expressed in terms of $\xi = {\delta}{m}/{M_T}$ and $\sigma = {M_T}/{M}$, we have
\begin{eqnarray}
L_{gw} = 2\times 10^{51} \left(\frac{\xi}{0.1}\right)^2 \left(\frac{\sigma}{0.01}\right)^2\left(\frac{4M}{a}\right)^{5}~\mbox{erg~s}^{-1},
\label{EQN_g4b}
\end{eqnarray} 
where $a$ denotes the orbital separation. The observed dimensionless strain at a source distance $D$ satisfies
\begin{eqnarray}
h = \frac{L_{gw}^\frac{1}{2}}{\Omega D} = 4\sqrt{\frac{2}{5}} \xi \sigma \left(\frac{M}{D} \right)\left(\pi M f_{gw}\right)^\frac{2}{3}.
\label{EQN_h1987AA}
\end{eqnarray}
Generalized to a similar event in the Local Universe, e.g., SN1979C \citep{pat11}, scaled to a distance of 20 Mpc, we have
\begin{eqnarray}
h=3.4\times 10^{-23}\,M_1 \left(\frac{\xi}{0.1}\right)\,\left(\frac{\sigma}{0.01}\right)\left(\frac{D}{20\,\mbox{Mpc}}\right)^{-1} \left(\frac{f_{gw}}{600\,\mbox{Hz}}\right)^{\frac{2}{3}},
\label{EQN_h1987A}
\end{eqnarray}
where $f_{gw}=2f_{orb}$ in the Newtonian approximation $2 \pi f_{orb} = M^{-1} \left({M}/{a}\right)^{3/2}$ with $M=M_1\,10M_\odot$.

\begin{figure}
\centerline{\includegraphics[scale=0.33]{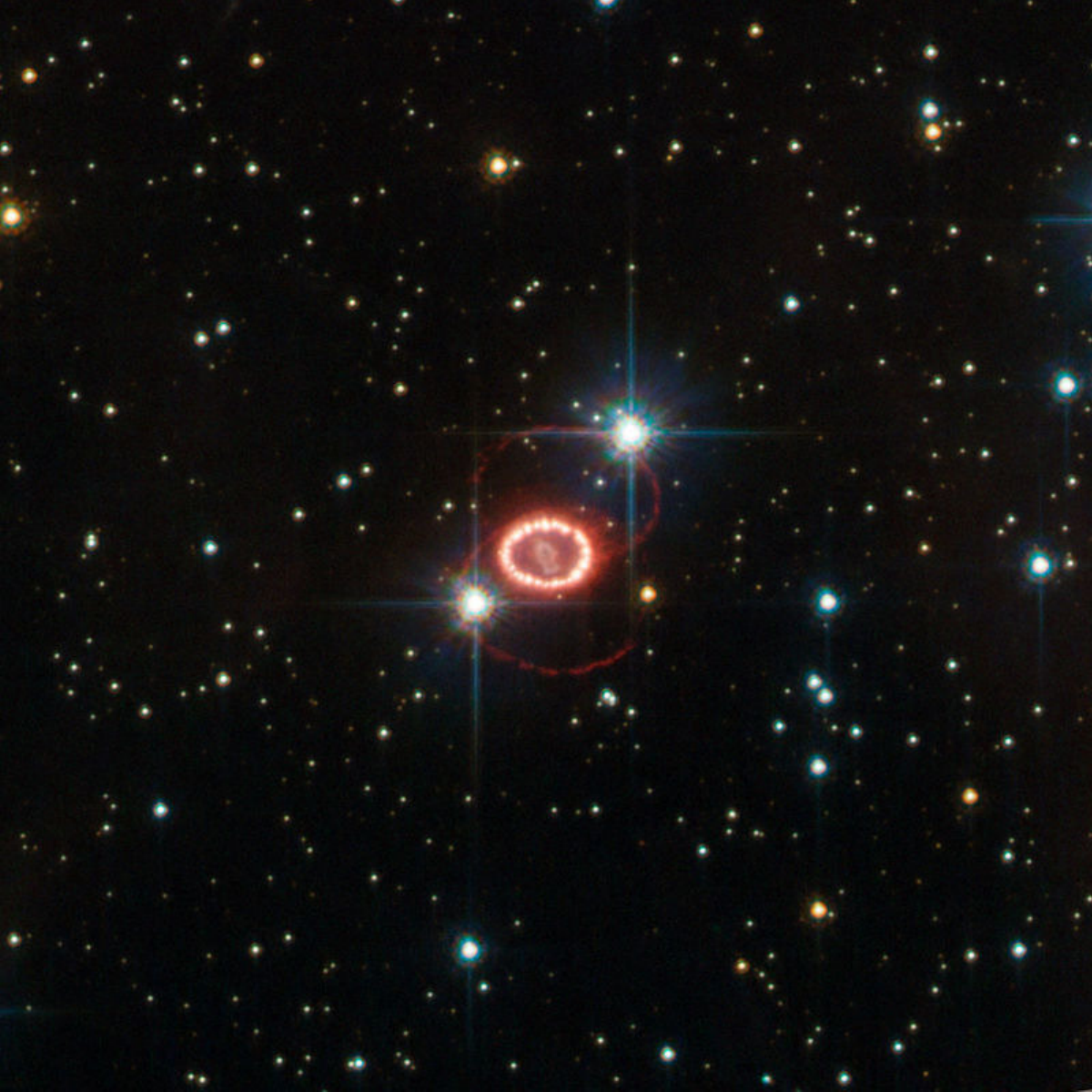}\hfill\includegraphics[scale=0.49]{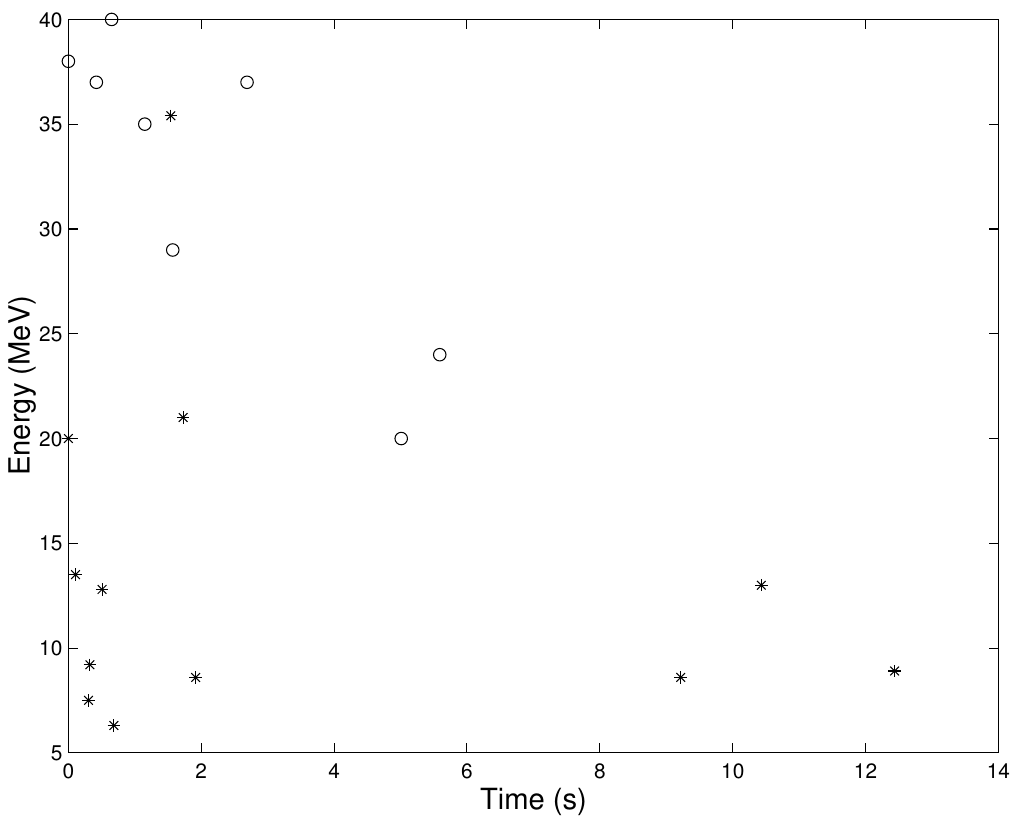}}
\caption{(Left.) SN 1987A is a Type II supernovae produced by core-collapse of the supergiant Sanduleak -69$^o$ 202 in the
Large Magellanic Cloud at a distance of about 50 kpc \citep{gil87,kir87}. (Credit: ESA/Hubble \& NASA.)
(Right.) Shown is the neutrino light curve compiled from Kamiokande (stars) and IMB (circles) listed in \citep{bur87} associated with the optical identification of SN 1987A \citep{gar87,her87,kun87}. The SN1987A neutrino light curve, showing an initial energy of $>$10 MeV representative for the formation of high density matter, possibly through continuing collapse of a protoneutron star. The final remnant is conceivably a stellar mass black hole, though undetected at present. (Reprinted from \cite{van05b}).}
\label{figSN1987A1}
\end{figure}

\subsection{Multi-messenger Extended Emission to GW170817/GRB170817A}

GW170817/GRB170817A \citep{abb17} is the first multi-messenger detection by LIGO-Virgo with accompanying electromagnetic radiation
in GRB170817A detected by {\em Fermi} and INTEGRAL \citep{con17,sav17,gol17,poz18,kas17b}. The merger is observed in 
gravitational radiation as a $\sim20$ long ascending chirp up to a few hundred Hz, satisfying 
\begin{eqnarray}
f_{gw}(t) = A \left(T_m - t\right)^{-\frac{3}{8}}
\label{EQN_f17}
\end{eqnarray}
with $A=138$\,s$^{-\frac{5}{8}}$ with $T_m=1842.43$\,s in the LIGO O2 data. From the multipole gravitational radiation
luminosity formula of a binary with ellipticity $e$ \citep{pet63}, the chirp mass satisfies \citep{abb17}
\begin{eqnarray}
{\cal M} = c\left(\frac{15}{768 F(e)}\right)^\frac{3}{5}\left(\pi A\right)^{-\frac{8}{5}} = 1.1188F(e)^{-3/5} M_\odot,
\label{EQN_Mc}
\end{eqnarray}
where $F(e)\simeq1$ at $e\simeq 0$ (\ref{EQN_Aforb1}) (Appendix A). By chirp mass, therefore, GW170817 is most
likely a double neutron star merger, though the merger of a neutron star with a low mass stellar black hole cannot
be excluded \citep{cou19}. 

\begin{figure}
\centerline{\includegraphics[height=81mm,width=140mm]{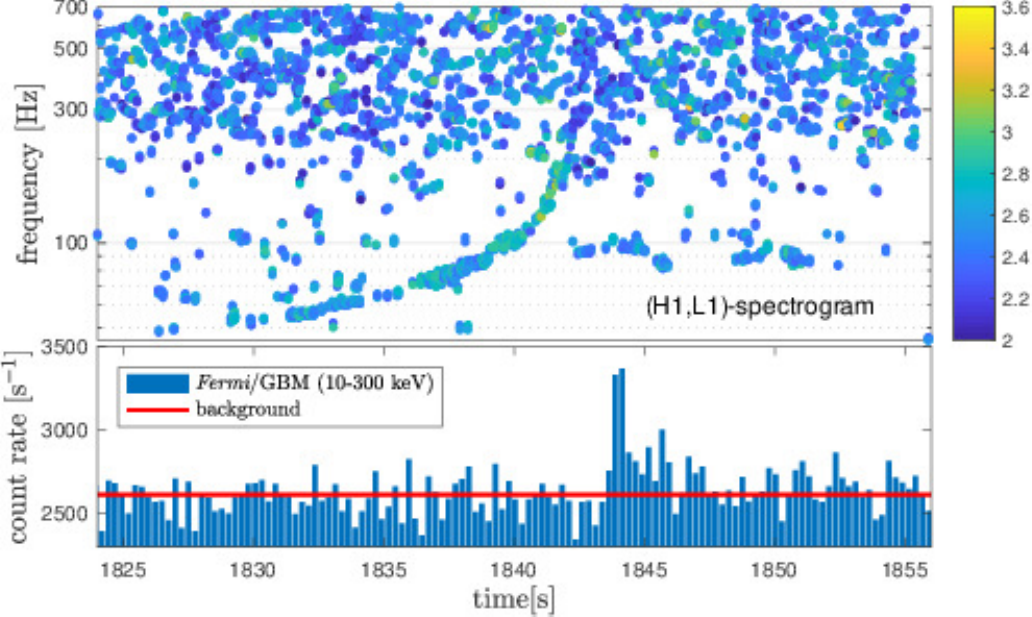}}
\caption{GW170817 features Extended Emission in gravitational-wave radiation post-merger, that appears as an ascending-descending chirp.
The descending chirp has a characteristic time-scale of decay $\tau_s\simeq3\,$s and a duration of about five seconds. 
It is powered by rotational energy $E_J$ in angular momentum, greatly enhanced in a black hole produced in collapse of the hyper-massive neutron star formed 
in the immediate aftermath of the merger. $E_J$ gradually diminishes as the black hole loses its angular momentum $J$ to a surrounding thick torus,
converting it to ${\cal E}_{gw}$ by non-axisymmetric mass-motion in quadrupole gravitational radiation concurrent with GRB170817A. 
Shown are frequency coincidences in butterfly filtering of H1 and L1 with colors indicating strength of
correlations of signal with time-symmetric chirp templates. (Reprinted from \citep{van19a}.) } %van Putten \& Della Valle, 2019, MNRAS, 482, L46.)} 
\label{figGW170817EE}
\end{figure}

GW170817 is of significance to the origin of the most heavy elements inferred from the associated kilonova \citep{kas17,sma17,pia17,dav17}
and the identification of a binary neutron star system as the astronomical progenitor of GRB170817A. 

{Fig. 4 shows Extended Emission post-merger to GW170817 \citep{van19a}. Detailed signal-injection experiments show \citep{van19b}
\begin{eqnarray}
{\cal E}_{gw} \simeq (3.5\pm1)\% M_\odot c^2.
\label{EQN_EE1}
\end{eqnarray}
As a calorimetric constraint, ${\cal E}_{gw}$ offers a unique opportunity to identify the central engine of GRB170817A. 
${\cal E}_{gw}$ is sufficiently large to break the degeneracy between neutron stars and black holes, as it exceeds 
the canonical bound (\ref{EQN_EcEHa}) on $E_J$ of the former. Instead, (\ref{EQN_EE1}) points to $E_J$ of a black hole, i.e., 
enhancement by collapse of the hyper-massive neutron star formed in the immediate aftermath of the merger. %(Fig. \ref{figGW170817EE}).
Accompanying energy emissions appear in magnetic disk winds powering the associated kilonova 
and ultra-relativistic outflows along the black hole spin axis powering the GRB-afterglow emission of GRB170817A. 
Model predictions \citep{van03}, see further (\ref{EQN_Ew}) of \S2.1 and (\ref{EQN_Lj}) of \S3.1 and Figs. \ref{fig:origin}-\ref{fig-T} below,
agree with observations \citep{moo18a,moo18b},
\begin{eqnarray}
Model\,prediction:~~E_w \simeq 4\times 10^{51}\mbox{erg},~~E_j \simeq 5\times 10^{50}\mbox{erg},\\
Observed:~~E_k \simeq 4.5\times 10^{51}\mbox{erg},~~E_j \simeq 10^{49-50}\mbox{erg},
\end{eqnarray}
where we identify total energy output in disk winds with the kinetic energy $E_k\simeq 4.5\times 10^{51}$erg of the kilonova 
and the total energy output $E_j$ in ultra-relativistic jets with the true energy output in GRB-afterglow emission. }

GW170817 is potentially also significant in providing independent measurements of the Hubble parameter $H_0$, in addition to 
the two existing main approaches based on surveys of the Local Universe and $\Lambda$CDM analysis of the CMB. 
This $H_0$ tension problem might indicate, for instance, that the future de Sitter state assumed in 
$\Lambda$CDM) is unstable \citep{van17} pointing to $H_0$ consistent with the outcome of the local distance ladder
\citep{rie19}.
\cite{gui17} estimate that about 50 GW170817 type events will allow an accurate
determination of $H_0$, sufficient to discriminate it from the value obtained
by $\Lambda$CDM analysis of the CMB.

\subsection{Roadmap}

Motivated by the multi-messenger events SN1987A and GW170817EE/GRB170817A, we set out to derive a prospect for
broadband extended gravitational-wave emission from energetic CC-SNe and GRBs {with a focus on central engines harboring
rotating black holes according to the following roadmap:} 

\S2 Observational evidence for black holes as common central engines in LGRBs and SGRBEEs based on BATSE, {\em BeppoSAX}
and {\em Swift}.

\S3 Outlook on multi-messenger powered by $E_J$ of rotating black holes at various stages of accretion \citep{van03a,van17a,lev15}. 

\S4 Gravitational radiation from non-axisymmetric accretion flows onto rotating black holes based on various fluid dynamical instabilities and wave motion.  

\S5 Estimated stochastic background of an astrophysical population of energetic CC-SNe powered by rotating black holes. 

\S6 Searches for broadband extended gravitational-wave emission by butterfly filtering \citep{van14} using heterogeneous computing with {\em Graphics Processor Units} (GPUs).

\S7 Outlook summary on the proposed search for BEGE from CC-SNe in the Local Universe by LIGO-Virgo, soon to be joined by KAGRA.

{For CC-SNe, the above suggests all-sky blind searches for their potential emission perhaps more so than follow-up to triggers from their optical
light curves, given the generic uncertainties their true time-of-onset. For light curves obtained early on, triggered searches might be pursued 
from events ongoing \citep{dro11,li11a} or upcoming all sky optical surveys such as Pan-STARRs \citep{pan11} or the planned Caltech 
Zwicky Transient Facility \citep{kul14,bel15}.} Searches for their contribution to the stochastic background in gravitational waves may be 
pursued by multi-year correlations between two or more gravitational wave detectors (e.g. \cite{sat09}). 
%Existing observations of LGRBs and SGRBEEs justify a vigorous probe of the inner most workings of energetic CC-SNe.

\section{Phenomenology of long GRB-supernovae and SGRBEE's}

Immediately following the serendipitous discovery of GRBs, % in the late 1960s by the nuclear treaty monitoring satellites Vela, 
Stirling Colgate suggested an association to supernovae - gamma-ray flashes from type II supernova shocks \citep{col68,col70,col74} - now seen by association of normal long GRBs to core-collapse of massive stars \citep{woo06}. Indeed, shock breakout in regular CC-SNe is likely to produce high energy emission in UV light \citep{gez08}, X-rays \citep{cam06} up to gamma-rays \citep{wea76,hof09,kat10,nak10,svi12}. While conceivably relevant to {\em low luminosity LGRBs} (LLGRBs), the prompt GRB emission of normal long GRBs is understood to represent dissipation in ultra-relativistic BPJs (below).

BATSE identified short and long GRBs with an observed bimodal distribution in durations $T_{90}$ shown in Fig. \ref{fig4B1} (left panel). 
Short GRBs originate in mergers of NS-NS \citep{eic89}, now confirmed in GW170817, or possibly NS-BH \citep{pac91} for black holes with slow 
spin \citep{van01}, albeit a large overlap between the two populations of short and long durations \citep{bro12,bro13}. $T_{90}$ is defined by the 
time interval covering a 90 percentile in total photon count \citep{kov93}. BATSE identified a mostly non-thermal spectrum, which is typically well described by a smoothly broken power-law Band spectrum \citep{ban93}.

{\em BeppoSAX} seminal discovery of X-ray afterglow emission to GRB 970228 \citep{cos97} allowed rapid follow-up by optical observations \citep{par97}, providing the first cosmological redshift ($z = 0.835$ of GRB 970508) from optical absorption lines of FeII and MgII \citep{met97,ama98}. When detected, afterglow emission of short GRBs tends to be very weak compared to those of LGRBs, consistent with less energy output and burst locations outside star forming regions. Weak X-ray afterglow emission discovered by {\em Swift} in GRB050509B \citep{geh05} and by the {\em High Energy Transient Explorer-2} (HETE II) in GRB 050709 \citep{fox05} was predicted based on energy emitted from rotating black holes \citep{van01}.
Following GRB 970508, {\em BeppoSAX}, HETE II and {\em Swift} provided a growing list of GRBs with measured redshifts. Presently, the total number 
of redshifts identified is about 350 with 287 due to {\em Swift} alone. Fig. \ref{figSwift} shows the distribution of the latter.

For long GRBs, the association with massive stars is now supported by four pieces of evidence: 
\begin{itemize}
\item supernovae (SNe) accompanying a few nearby events \citep{hjo11}; 
\item detection of SN features in the spectra of ``rebrightenings" during GRB afterglow decay, at intermediate redshifts, most recently GRB 130427A ($z$=0.34, \cite{mel13}) up to $z\simeq1$ \citep{del03};
\item the host galaxies are spiral and irregular with active star formation typical for environments
hosting core-collapse SN-Ic's \citep{kel08,ras08}; and
\item a cosmological distribution of redshifts of long GRBs with $z\simeq2$ (Fig. 5). Up to moderate redshifts, this appears to be
         consistent with the cosmic star formation rate \citep{wan10,gri12} (Fig. \ref{figSwift}). 
         At high redshifts, this awaits confirmation by future missions such as the
         planned {\em Transient High Energy Sky and Early Universe Surveyor} (THESEUS) \citep{ama17}
\end{itemize}

\subsection{Hyper-energetic GRB-supernovae}

Aspherical CC-SNe \citep{pap89,hof99,mae08} such as SN1987A derive from relatively massive progenitors. While their explosion mechanism 
remains ill-understood, relatively energetic events are probably powered by magnetic winds from an angular momentum-rich central engine 
\citep{bis70} possibly including relativistic outflows or jet \citep{mac99}. In core-collapse of a massive star, following a drop in thermal pressure at the end of nuclear burning or associated with pair-instability for the most massive stars \citep{bis66,bar67,gal09,cha10}, the central engine will be
a (rotating) neutron star or black hole. 

The kinetic energy $E_k$ in a supernova explosion powered by angular momentum is constrained by the maximal rotational energy of the newly
formed neutron star or black hole and the efficiency in expulsion of the remnant stellar envelope by a putative internal wind or jet. 
The efficiency $\eta=E_k/E_w$ in ejecting an envelope with kinetic energy $E_k$ by a wind or jet with energy $E_w$ depends on the baryon 
loading of the wind, i.e., \citep{van11b}
\begin{eqnarray}
\frac{1}{2}\beta_{ej} < \eta < 1,
\label{EQN_eta}
\end{eqnarray}
where $\beta_{ej} = v_{ej}/c$ denotes the observed velocity of the ejected envelope relative to the velocity of light $c$, that
parameterizes the efficiency by which wind energy $E_w$ released from the central energy is converted to observed kinetic
energy $E_k$ in ejecta. $\eta$ increases with baryon loading, as the outflow velocity becomes less relativistic;
$\eta$ reducing to $(1/2)\beta_{ej}$ in the limit of baryon-poor (Poynting-flux dominated) jets, as the velocity of winds released 
by the central engine approach $c$. By the above, $E_k$ is bounded by $\eta E_J$ of either (\ref{EQN_EcEHa}) or (\ref{EQN_EcEHb}).

\begin{figure}
\centerline{\includegraphics[scale=0.5]{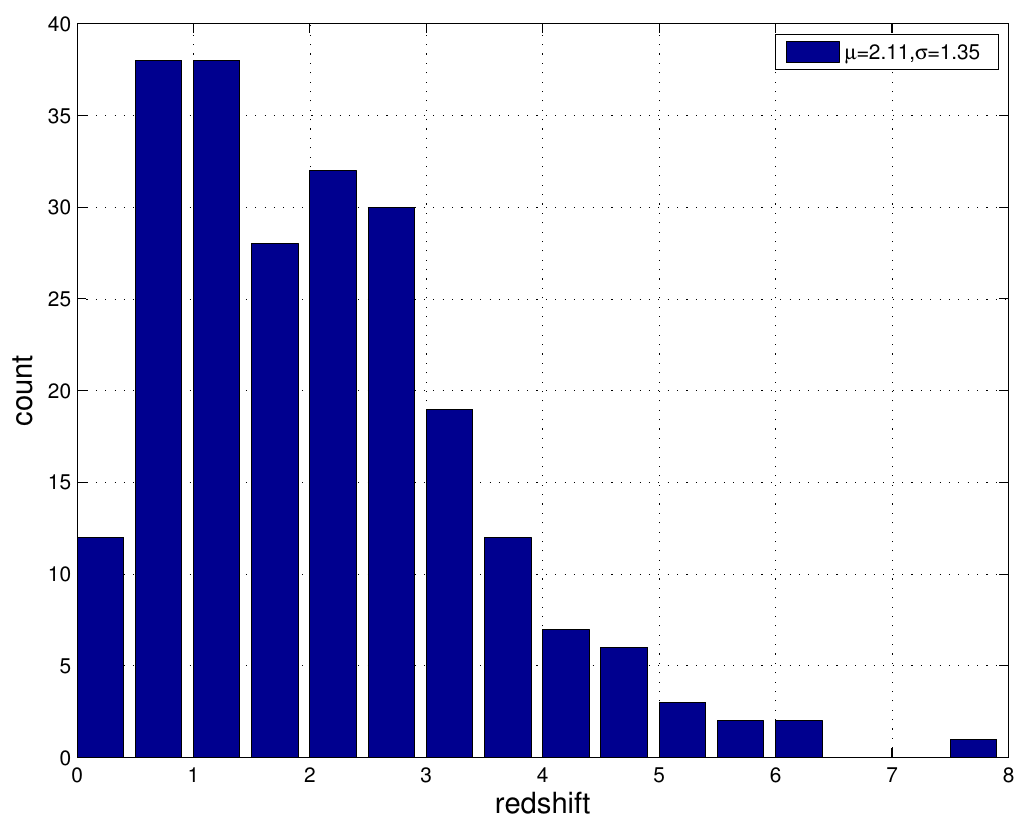}} 
\caption{Observed redshifts of 230 LGRBs in the {\em Swift} catalogue with mean $\mu=2.11$ and standard deviation $\sigma=1.35$. 
This distribution is significantly biased towards low redshifts. (Reprinted from \citep{van12}.)}
\label{figSwift}
\end{figure}

\begin{table*}
\center \textbf{Table 1.}$^\dagger$ Energies \citep{van11b} of GRB and SNe in units of $10^{51}$\,erg. References refer to SNe except for GRB 070125.\\ 
\begin{tabular}{llccrrrrrr}
\hline
GRB & Supernova				& $z$ & $E_\gamma$ & $E_{tot}$ & $E_{SN}$ & $\eta$ & $E_{rot}/E_c$  & Ref.\\
\hline
				& SN2005ap	& 0.283	 	& 			& 			& $>10$ 	&1	& $>0.3$	 & 1\\ 
       				& SN2007bi	& 0.1279 		& 			& 			& $>10$ 	&1	& $>0.3$  & 1\\  
980425		&SN1998bw 	& 0.008 	        	& $<0.001$     	& 			& 50		& 1	& 1.7		& 2\\
031203		&SN2003lw 	& 0.1055	        	& $<0.17$ 	& 			& 60 		&0.25& 10	& 3\\
060218 		&SN2006aj	& 0.033    	 	&$<0.04$ 		& 			& 2		&0.25& 0.25	& 4 \\
100316D		&SN2006aj	& 0.0591		& 0.037-0.06	& 			& 10 		&0.25& 1.3	& 5\\
030329		&SN2003dh	& 0.1685		& 0.07-0.46      & 			& 40		&0.25& 5.3	& 6 \\
050820A		&			& 2.607		&  			& 42			&		&	& 1.4		& 7\\ 
050904  		&			& 6.295	 	& 			&12.9		&		&	& 0.43	& 7\\
070125  		&			& 1.55   		& 			& 25.3	  	& 		&	& 0.84	& 7\\
080319B		&			& 0.937		& 			& 30			&		&	& 1.0 	& 7\\ 
080916C		&			& 4.25		& 			& 10.2		&  		&	& 0.34	& 7\\
090926A		&			& 2.1062		&       		& 14.5 		& 		&	& 0.48	& 8\\
070125  		&   			& 1.55   		& 			& 25.3 		& 	 	&	& 0.84	& 9\\
\hline
\hline
\end{tabular}
\label{TABLE_1}
%}
\mbox{}\\\hskip0.08in
$^\dagger$ \cite{van11b};
1. \cite{gal09,qui11}; 2. \cite{gal98}; 3. \cite{mal04}; 4. \cite{mas06,mod06,cam06,sol06,mir06,pia06,cob06b}; 
5. \cite{cho10,buf11}; 6. \cite{sta03,hjo03,mat03}; 7. \cite{cen10}; 8. \cite{deu11}; 9. \cite{cha08}.
\end{table*}

By (\ref{EQN_EcEHa}-\ref{EQN_EcEHb}), black holes allow $E_k$ to be considerably larger than $\eta E_c$, even at modest efficiencies.
Additionally, black hole-disk systems can produce two-component outflows: an ultra-relativistic jet about their spin axis surrounded by a baryon-rich and possibly collimating disk wind. The first offers potential for gamma-ray emissions upon breakout from the stellar envelope, 
the second for an efficient supernova explosion. In contrast, neutron stars the one-component (magnetic and neutrino-driven) 
outflow of a neutron star may facilitate one but not both.

Exceptional $E_k$'s are observed in GRB 031203/SN2003lw and 030329/SN2003dh with $E_{SN}\simeq 6\times 10^{52}$ erg and $E_{SN}\simeq 4\times 10^{52}$ erg, respectively (Table 1). Like SN1987A, supernovae accompanying LGRBs are aspherical and radio-loud. Hyper-energetic events with $E_k>E_c$ (see (\ref{EQN_EcEHa}-\ref{EQN_EcEHb}) are no exception. In \cite{bis70}, these explosive events are attributed to magnetic winds powered by angular momentum extraction of a compact object, i.e., a proto-neutron star or magnetar (e.g. \cite{woo10,kas10} in SN2007bi \citep{nic13} or a rotating black hole-disk system (BHS, \cite{van03}). Taking into account a finite efficiency for the conversion of angular momentum to a (largely radial) explosion, $\eta<1$, we determined that aforementioned two events require a central energy $E_{rot}$ in angular momentum exceeding the maximal spin energy of a rapidly rotating neutron star by a factor of 10 and, respectively, 5.3 (Table 1). With a total output of about $10^{52}$ erg in optical emission alone, SN2015L \citep{don15} likewise defies the limit $\eta E_c$ in the face of reasonable efficiencies $\eta$ and finite efficiency
in dissipating kinetic energy to electromagnetic radiation, at least not at the same time. 
It appears therefore unlikely that GRB 031203/SN2003lw and 030329/SN2003dh are powered by $E_J$ of a neutron star at any reasonable
efficiency. For this reason, they stand out as potentially being powered by
black holes % engine of these two events stand out as candidate BHS, powering the explosion by the spin energy of a black hole. 
at efficiencies of a few \%. %is brought back to a reasonable few \%. 
In particular, a wind energy $E_w$ may derive from winds from a disk extracting energy of $E_J$, satisfying \citep{van03}
\begin{eqnarray}
E_w = 6\times 10^{52} \left(\frac{\Omega_T}{0.1\Omega_H}\right)^2 \left(  \frac{M}{10M_\odot}\right) \,\mbox{erg}
\label{EQN_Ew}
\end{eqnarray}
for rapidly spinning black hole of mass $M$ parameterized by the ratio of the angular velocity $\Omega_T$ of a torus about the ISCO to the angular velocity $\Omega_H$ of the black hole. 

\subsection{Local event rates of energetic CC-SNe}

As a parent population of long GRBs, energetic core-collapse supernovae in the Local Universe are targets of opportunity (TOOs) to LIGO-Virgo \citep{abr92,ace06,ace07} and KAGRA \citep{som12,kag14}. Current GRB and SN rates show, {based on beaming-corrected event rates,} a branching ratio 
\begin{eqnarray}
{\cal R}=\frac{N(\mbox{GRB-SNe})}{N(\mbox{Type Ib/c})} \simeq 0.2-3 \%
\label{EQN_R1}
\end{eqnarray}
by conservative \citep{van04,ghi13} to optimistic estimates \citep{gue07}. 
The event rate of SN Ib/c exceeds that of GRB-SNe by ${\cal R}^{-1}$.

The origin of the small value of ${\cal R}$ in (\ref{EQN_R1}) is not well understood. Evidently, an observable GRB event requires the succesful formation of an engine sufficiently powerful for outflows to overcome various adverse conditions, namely a high density environment formed in core-collapse of a massive progenitor star, perhaps in a short binary period \citep{pac98}. A possible additional factor is the time of residence of the newly formed black hole in the center of the star, to be sufficiently long only when its kick velocity happens to be low. These successful GRB-supernovae may be rare. Alternatively, it may reflect a small probability of forming nearly extreme Kerr black holes, i.e., about the Thorne limit reached along a modified Bardeen trajectory as the initial condition for long GRBs \citep{van15}.

We estimate SNe of type Ib/c within a distance $D_S$ to have a local event rate
\begin{eqnarray}
\dot{N}(\mbox{Type Ib/c}, D<D_S) \sim 10^{1-2} \,\mbox{yr}^{-1} \left(\frac{D_S}{100\,\mbox{Mpc}}\right)^3
\label{EQN_R2}
\end{eqnarray}
by a weighted volumetric rates from Asiago \citep{cap99,bar99}, Lick surveys \citep{li11a,li11b}, SDSS-II \citep{tay14} and SUDARE VST-OmegaCAM \citep{cap15}.
Broad line events (cf. Fig. \ref{figSN1987A3}) successfully producing a normal LGRBs represent a small fraction of SNe of type Ib/c.
Even so, these energetic BL events are a few per year within 100 Mpc, still more numerous than the true event rate of LGRBs (corrected for 
beaming) by up to one order of magnitude.

Electromagnetic detection of nearby supernovae is remarkably easy, especially in the Local Universe within 100 Mpc. In contrast, electromagnetic counterparts to mergers of double neutron star binaries are generally challenging \citep{kas13,bar13,tan13a} - thusfar detected in GW170817/GRB170817A only.

\subsection{GRBs in the {\em Swift} era}

At present, calorimetry by electromagnetic observations on the kinetic energy in supernovae and the prompt GRB emission offers our strongest observational constraints
on the central engines of GRB-SNe. Indirect constraints may further derive from the MeV-neutrino burst from SN1987A.

We next review the present classification, spectral-temporal properties of GRBs relevant to their putative central engine based on
BATSE, {\em BeppoSAX} and {\em Swift} of long and short GRBs based on Table 2, 
the Amati-relation (Fig. \ref{figAm1}) and associated event statistics (Fig. \ref{figAm2}).

\mbox{}\\
\mbox{}\\
\begin{table*}
%\vskip-0.3in
%\begin{sidewaystable*}
\center {\bf Table 2.}$^\dagger$ {\em Swift} SGRB, SGRBEE$^a$ and LGRBNs. $E_{iso}$ in $10^{52}$ erg, $E_p$ in keV.\\ 
\begin{tabular}{llrrllclc}
\hline
    & $T_{90}$ & $z$ &                   & host$^b$                                    & $E_{iso}$$^c$    & $E_p$$^c$\\
\hline
{{\bf SGRB}}  \\

	 050509B & 0.073 	    &  0.225   &  		&  elliptical galaxy$^2$                                 	& 0.00027$^3$	 			& -  \\ 
	 050813    & 0.45         &  1.8       &  		&  galaxy cluster$^{17,18}$                          & 0.017$^{18}$  	 	 	& - \\
	 051221A & 1.400      &  0.547   &  		&  SF, late type galaxy$^8$                           & 0.25$^9$ 				& \\ 

	 060502B & 0.131      &  0.287   &  		&  massive red galaxy$^4$                          	&  0.022	  				&  193\\ 
	 060801    & 0.49         &  1.131   &     		&           -                              				& 0.027$^{15}$ \\ 

	 061201  & 0.760    &  0.111   &    	&  galaxy cluster$^1$  			                            	& 0.013	 		         & 969 \\ 
	 061217    & 0.210      &  0.827   &     		& faint galaxy$^{12}$                                 	 & 0.008$^{12}$ 			& \\
	 
	 070724A & 0.4           &  0.457   &    		& moderate SF galaxy$^7$                         	&   - 						&  \\ 
	 070429B & 0.47          &  0.904  &    		 & SFR$^{14}$ 1.1$M_\odot$ yr$^{-1}$  		&  - 						& - \\ 
	
	 090426    & 1.2           &  2.609   &       	&  irreg. SF galaxy$^{19}$                		& -  						& - \\ 
	 090510    & 0.3           &  0.903   &    		& field galaxy$^{13}$                       		& 3.8$^{13}$ \\
	 
	100724A & 1.4            &  1.288   &  		&  probably LGRB$^{16}$			       	& -  						& - \\ 
	101219A & 0.6           &  0.718   &      	& faint object$^{11}$                                   	& 0.48	  				& 842 \\ 
        130603B & 0.18         &  0.356   &   		&  SFR$^5$   						&  0.2$^6$                 			&  90$^6$\\ 
        131004A & 1.54         &  0.717   &   		&  low mass galaxy$^{10}$                          	& -  						&   \\ 
\hline
{\bf SGRBEE} \\
         050724  $^d$ $^e$ $^f$ $^g$  	&   69       &  0.258 	&       &  elliptical, weak S$^{22}$ 		&  0.0099$^{23}$      & - \\
         050911  $^d$ $^e$                     	&   16.2   &  1.165  	&       & EDCC493 cluster$^{30}$  		&      0.0019$^{30}$  		& - \\ 
         060614  $^d$ $^e$ $^f$ $^g$&   108.7 &  0.125 	&       &  faint SFR$^{20,21}$   	&  0.21$^{20}$     	& 55$^{20}$ \\
	 061210  $^d$ $^e$ $^f$ $^g$  	&   85.3   &  0.41 	&       & bulge dominated$^{26}$  		&  0.046$^{26}$	& - \\ 
	 061006  $^d$ $^e$ $^f$ $^g$  	&   129.9 &  0.438 	&       & exp. disk profile$^{27}$  		&  0.18   	& 955 \\ 
	 070714B $^d$ $^e$ $^f$ $^g$ 	&   64      &  0.92   	&       & SF galaxy$^{28}$    			&  0.16$^{28,29}$ & -  \\ 
	 071227A $^e$ $^f$                     	&   1.8      &  0.384  	&       &  edge-on S$^{24}$ 			&  0.008$^{25}$       	& - \\ 
	\hline
{\bf LGRBN} \\
 	060505                                    	&   4     	&  0.089   &     & spiral, H$^+$, no SN$^{31}$ 	&  0.0012$^{21}$ & 120  \\ 
	 060614  $^d$ $^e$ $^f$ $^g$  &   108.7 	&  0.125   &     & faint SFR, no SN$^{20}$  	&  0.21$^{21}$     & - \\
	 061021                                          &    46     	& 0.3462  &     & no SN$^{32}$                    	& 0.68                    & 630\\
\hline
\label{TABE2}
\end{tabular}
\mbox{}\\\hskip0.08in
{ $^\dagger$ \cite{van14b}; $^a$ From \cite{HEA}; $^b$ galaxy type, SN association; 
$^c$ Isotropic-equivalent energy and peak energy for events with reliable estimates of the bolometric $E_{iso}$ across a large enough energy band, under the assumption $\Omega_m=0.3$ and a Hubble constant $H_0=70$ km s$^{-1}$ Mpc$^{-1}$; $^d$  \cite{per09}; $^e$  \cite{nor10}; $^f$  \cite{cow12}; $^g$  \cite{gom14};
1. \cite{ber07c} ;2. \cite{fon10,pag06,per09} ; 3. \cite{blo06,blo07}; 4. \cite{blo07}; 5.  \cite{cuc13}; 6.\cite{fre13};
7. \cite{koc10};  8. \cite{ber05,ber07}; 9. \cite{gol05}; 10. \cite{per13}; 
11. \cite{per10}; 12. \cite{ber06b,deu06}; 13. \cite{rau09,gue12}; 14. \cite{cen08}; 15. \cite{cuc06,ber07a}; 
16. \cite{ukw10}; 
17. \cite{blo07,pro06,ber06a,fer07};18. \cite{ber05a};
19. \cite{ant09}; 
20. \cite{del06}; 
21. \cite{fyn06a,cob06} ; 
22. \cite{ber05c,pag06,ber07a,fon10}; 23. \cite{pro05}; 
24. \cite{ber07b}; 25. \cite{ber07d}; 
26. \cite{cen06}; 
27. \cite{fon10,ber07a}; 
28. \cite{gra09}; 29. \cite{gra07};  
30. \cite{ber07}; 
31. \cite{jak07}; 32. \cite{mor06}
%\end{sidewaystable*}
}
\end{table*}

\begin{figure*}
\centerline{\includegraphics[scale=0.35]{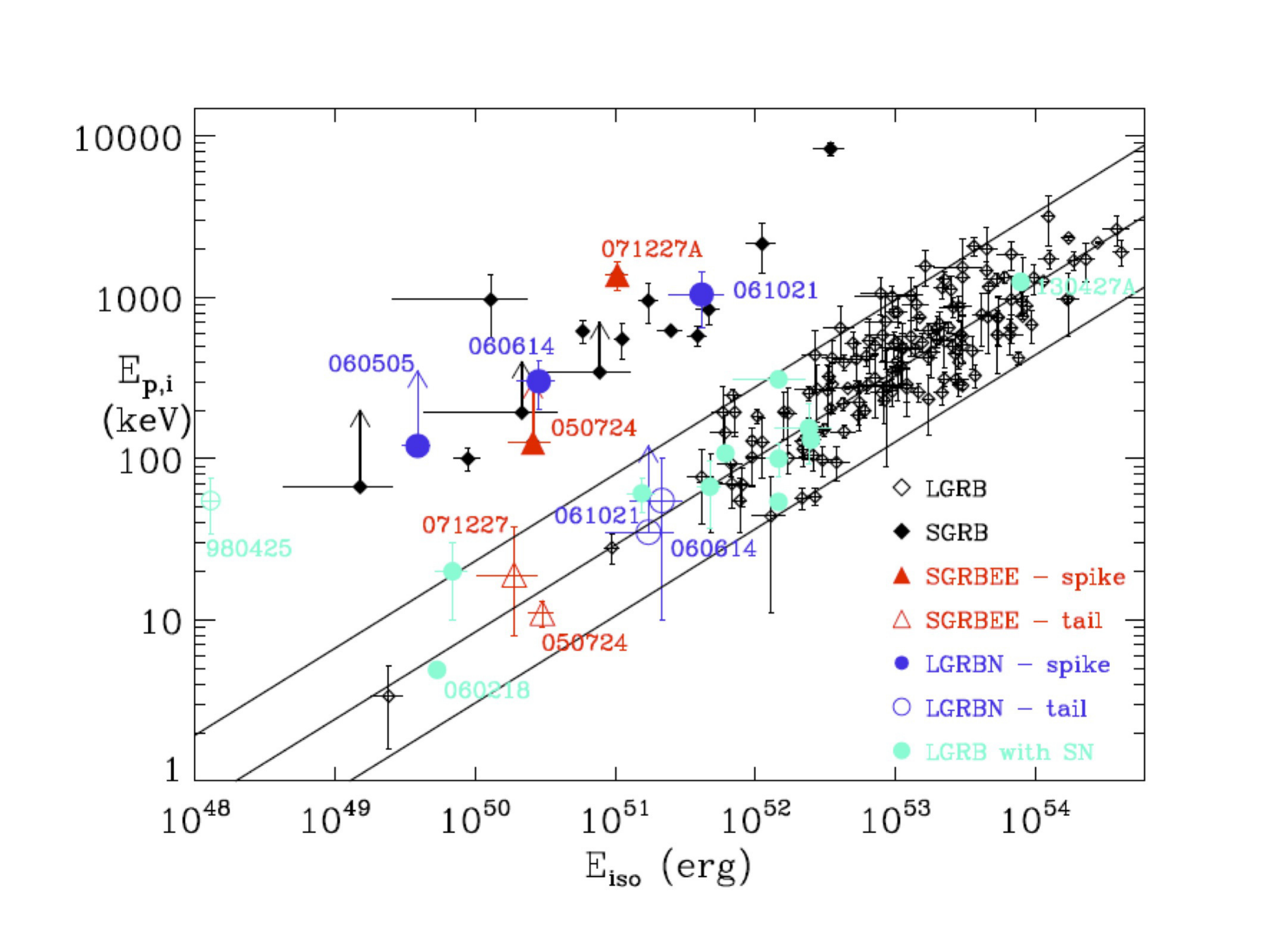}} 
%\centerline{\includegraphics[scale=0.20]{f6}} 
\caption{Short GRBs with Extended Emission (SGRBEE) and long GRBs with and without apparent supernovae (LGRBN) in the $E_{p,i}-E_{iso}$ plane \citep{ama02,ama06}, including GRB-SNe 030329, 050525A, 081007,091127,100316D,101219B. The lines are $E_{p,i}- E_{iso}$ correlatons for normal long GRBs ($\pm 2 \sigma$). Highlights include the sub-energetic GRB980425/SN1998bw and GRBEE 050724 (also a LGRBN). Initial pulses of SGRBEEs (solid triangles, red) fall into the separate group of SGRBs, like initial pulses of LGRBNs (large size filled circle, blue; 90\% confidence limits.) Data mostly from \citep{ama08,ama09,can14,swi14}. (Reprinted from \cite{van14b}.)}
\label{figAm1}
\end{figure*}

{\em Swift} identified the new class of short GRBs with Extended Emission. The SGRBEE GRB060614 ($z=0.125$, $T_{90}$=102 s) has no detectable supernova \citep{del06,fyn06a,gal06} and GRB 050724 is a SGRB with Extended Emission (SGRBEE) with an overall emission time $T_{90}$=69 s in an elliptical host galaxy \citep{ber05c,ber07}. Neither is readily associated with a massive star. Since then, the list of SGRBEEs has grown considerably (Table 2). Table 2 further shows a few long GRBs with no apparent association to supernova (LGRBNs). SGRBEE and LGRBNs challenge the BATSE classification into short and long events. Though both show an initial hard pulse, characteristic of short GRBs, a subsequent long duration soft tail features a spectral peak energy ($E_{p,i}$)-radiated energy ($E_{iso}$) correlation that satisfies the Amati-correlation holding for normal long GRBs. This ``hybrid" structure of observational properties of SGRBEE and LGRBNs suggests that they share the {\em same astronomical origin as short GRBs with the same physics in the central engine as normal long GRBs,} albeit with somewhat smaller values of $E_{iso}$. 

Prompt GRB emission has a characteristic peak energy $E_{p,i}$ at which the $\nu F_\nu$ photon spectrum peaks in the cosmological rest-frame. It typically ranges from tens of keV to thousands of keV. If the isotropic-equivalent energy $E_{iso}$ is the radiation output by a GRB during its whole duration (assuming spherical symmetry), it is found that $E_{iso}$, commonly used in the absence of reliable information on the degree of collimation in individual GRB events, correlates with $E_{p,i}$ (see Fig. \ref{figAm1}). This correlation, now known as the Amati relation, is well established for long GRBs, while short GRBs do not appear to satisfy this Amati relation (\cite{ama06} and references therein).

\begin{figure*}
\centerline{\includegraphics[scale=0.745]{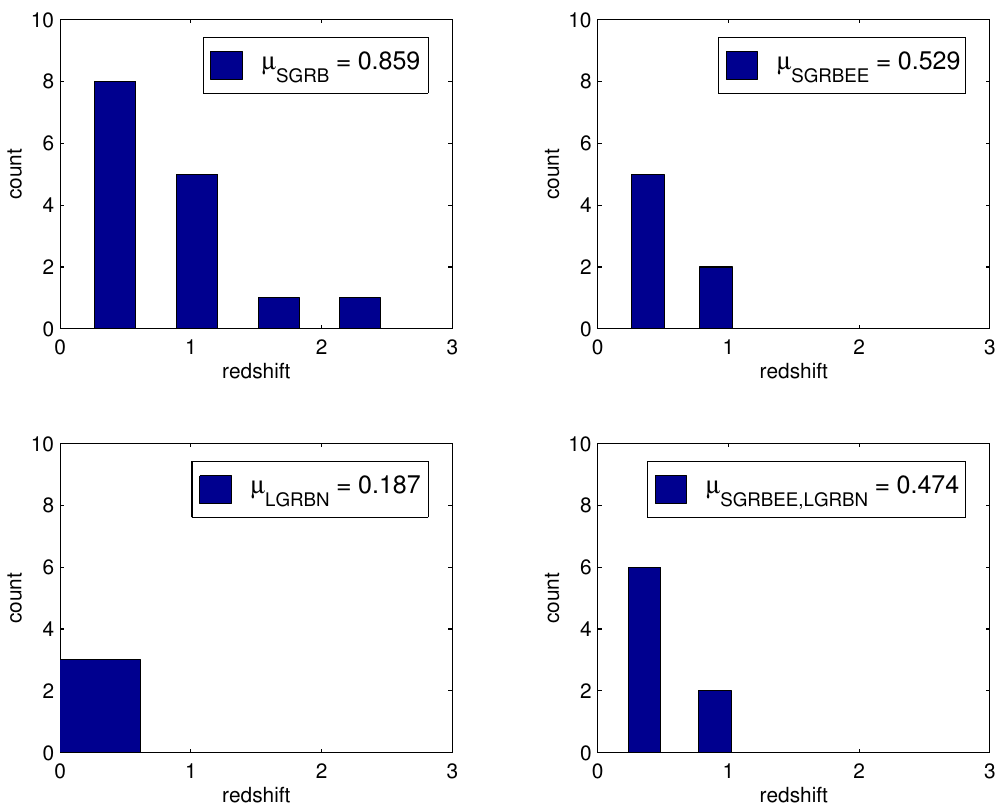}} 
\caption{Redshifts of SGRB, SRGBEE, LGRBN and SGRBEE+LGRBN and their associated mean $\mu$. (Reprinted from \cite{van14b}.)}
\label{figAm2}
\end{figure*}

To quantify our level of confidence in the merger origin of SGRBEE and LGRBNs, we recently considered the mean values $\mu$ of the observed redshifts (Fig. \ref{figAm2}), i.e., $\mu_{L}^{N}$ of LGRBNs, $\mu_{EE}$ of SGRBEEs,  $\mu_{S}$ of SGRBs and $\mu_{L}$ of LGRBs, and concluded that they satisfy
\begin{eqnarray}
\mu_{L}^{N}< \mu_{EE} <\mu_{S}<\mu_{L},
\label{EQN_z}
\end{eqnarray}
where $\mu_{EE}=0.5286$, $\mu_S=0.8587$, $\mu_L^N=0.1870$, and $\mu_L=2.1069$ based on the 
redshifts shown in Table 2. 

By a Monte Carlo test, we determined the probability that, from the mean redshift, the {\em Swift} samples of SGRBEE ($n_1=7$), SGRB ($n_2=15$) and LGRBNs ($n_3=3$) are drawn from the observed distribution of LGRBs ($n=230$). Because of the small $n$ samples and the broad distribution of redshifts of LGRBs (with an observational bias towards low $z$), we proceed with Monte Carlo test by drawing samples of size $n_i$ ($i=1,2,3)$ from the distribution of the $n=230$ redshifts of the latter. Doing so $N$ times for large $N$ obtains distributions of averages $\mu_i$ of the redshifts in these small $n$ samples under the Bayesian null-hypothesis of coming from the distribution of redshifts of LGRBs. We find the following levels of confidence of each \citep{van14b}
\begin{eqnarray}
\begin{array}{rll}
\mbox{SGRBEE} & \not\subset \mbox{LGRB} &: ~~4.6700 \,\sigma \\
\mbox{SGRB}      & \not\subset \mbox{LGRB} &: ~~4.7520 \,\sigma \\
\mbox{LGRBN}   & \not\subset \mbox{LGRB} & :~~ 4.3140\,\sigma \\ 
\end{array}
\label{EQN_s}
\end{eqnarray}
{SGRBs and LGRBS hereby have different astronomical progenitors.} For SGRBs, (\ref{EQN_s}) is consistent with a relatively low redshift origin inferred from identification of host galaxies in the local Universe \citep{tan05}.  At a level of confidence exceeding $4\,\sigma$, SGRBEE and LGRBNs have inner engines originating in mergers in common with normal long GRBs originating in CC-SNe, given that both share the Amati-relation in the long/soft tail. 

Our results (\ref{EQN_s}) show with relatively high confidence that the enigmatic LGRBN GRB060614 is a merger event, suggested earlier based on other arguments \citep{van08,cai09}, whose long durations in soft extended emission can be identified with the lifetime of spin of a rotating black hole (\S2); see \citep{zha06,zha07} for various other explanations of extended emissions from mergers. Baryon-rich jets from accretion disks produced in naked inner engines formed in mergers may dissipate into lower energy emissions, perhaps including a radio burst \citep{van09}. 

The preceding data show a soft/long tail satisfying the same Amati-relation (Fig. \ref{figAm1}) of extended emission in
SGRBEE and normal LGRBNs. 
The association of normal LGRBs with supernovae and their shared spectral-energy properties with Extended Emission to 
short GRBs discovered by {\em Swift} suggests the possibility of a common inner engine, even as SGRBEE's derive from mergers
such as GW17087/GRB170817A. If SGRBEE's indeed derive from mergers, it remains to be determined if these are neutron star-neutron star
or neutron star-black hole mergers. In what follows, we review some of their properties in the electromagnetic spectrum.  

\subsection{Prompt GRB emission of (L)LGRBs}

The three main stages in prompt GRB emission are: (i) formation of outflows from a central energy reservoir, (ii) dissipation of bulk energy in this outflow, and (iii) conversion thereof into electromagnetic radiation. These processes are most likely interrelated. Successful breakout 
of the  jet from the stellar envelope is a necessary condition for producing a GRB. Due to the compactness of the energy source, the largely 
non-thermal electromagnetic emission originates from large radii and, therefore, does not provide a direct probe of the central engine. 
Nevertheless, the central engine possibly leaves an imprint in the light curve of prompt emission on the secular time scale of
accretion and evolution of the black hole \citep{kum08,van09a}. 

Conventional wisdom states that GRB  jets are powered by magnetic extraction of the rotational energy of a magnetar \citep{uso94,met11} or a hyper-accreting black hole, that may be an attractive alternative to account for low baryon-loading \citep{lev93,eic11}. %The following data (\S4.2) 
In some cases \citep{van11b}, evidence from total energy output is tilting towards an association to black holes rather than neutron stars. 
At sufficiently high accretion rates, annihilation of neutrinos that originate from the hot matter surrounding a Kerr black hole can also power a GRB outflow \citep{zal11,lev13b}, although magnetic extraction seems favorable. Efficient conversion of an outgoing Poynting flux into kinetic energy of baryonic contaminants
is not fully understood, yet, it is generally believed to involve gradual acceleration of the flow (e.g., \citep{bog95,chi91,hey89,lyu09}), impulsive acceleration \citep{gra11,lyu11}, magnetic reconnection \citep{gia07,lev97,lyu10,lyu03,zha11,mck12}, and/or current driven instabilities \citep{lev13}. 

The production of high energy emission requires substantial dissipation above or just below the photosphere. 
It most likely results from the formation of internal \citep{mes93,ree92} and/or collimation \citep{bro07,laz09} shocks in cases where the flow is hydrodynamic in the vicinity of the photosphere, or magnetic reconnection \citep{gia07,mck12} if the flow remains highly magnetized at large radii. Dissipation at very large optical depths will merely lead to re-acceleration of the flow, or in case of magnetic extraction to a transition to kinetic energy dominated outflows \citep{gra11,lev13}. It can, nonetheless, help increasing the specific entropy, which seems to be required by the observed SED peaks.

The nature of the prompt emission mechanism remains an open issue. The emitted spectrum, although exhibiting notable variations from source to source, can generally be described by a broken power law Band function \cite{ban93}), with some exceptions, e.g., GRB 090902B. 
It has been originally proposed that the observed spectrum is produced by synchrotron emission of non-thermal electrons accelerated at internal collisionless shocks (for reviews see \cite{pir99,pir04}). However, subsequent analysis (e.g., \cite{bel13,cri97,eic00,pre98}) indicated that the synchrotron model has difficulties accounting for some common properties exhibited by the GRB population, specifically, the clustering of peak energies around 1MeV, the hardness of the spectrum below the peak, and the high efficiencies inferred from the observations. 
At the same time, it has been argued \citep{ryd04,ryd05} that a thermal component appears to be present in some bursts, which may be transient as in the {\em BeppoSAX} event GRB 990712 \citep{fro01}. These developments, and the recent detection of some GRBs with a prominent thermal 
component (e.g., GRB 090902B) or multiple peaks (e.g., GRB 110721A, GRB 120323A) have motivated a reconsideration of photospheric emission \citep{bel13,eic00,gia12,pee06,ryd09,vur13}.

On theoretical grounds, one naively anticipates significant dissipation of the bulk energy of a GRB outflow just below the photosphere, 
either by internal \citep{eic94,bro11b,mor10} or collimation shocks \citep{bro07,laz09}.
They are mediated by radiation and their typical size is on the order of a few Thomson depths \citep{bud10,kat10,lev08,lev12}, larger than any kinetic scale by many orders of magnitudes. Their structure and emission are, therefore, vastly different from those of collisionless shocks above the photosphere, where the Thomson optical depth is well below unity. The large shock width strongly suppresses particle acceleration \citep{lev08,kat10}, yet a non-thermal spectrum can be produced inside the shock via bulk Comptonization \citep{bud10} and formation of a Band-like spectrum is conceivable \citep{ker14}. {Alternatively, sub-photospheric dissipation may be accomplished through shocks or magnetic reconnection. The first can produce non-thermal spectra even without particle acceleration \citep{ito17}. The second holds if the flow remains highly magnetized at mild optical depths, in which case a Band spectrum may follow by particle acceleration \citep{bel13,gia12,vur13}.}

Observationally, a photospheric model (black body plus power law) or a Band function provides a satisfactory fit to most BATSE GRB light curves. However, extending spectra to the low energy range of 2-28 keV (of the {\em BeppoSAX} Wide Field Cameras) poses challenges in a number of cases. For the extended energy range of 2-2000 keV, a Comptonization model appears more robust in producing satisfactory fits \citep{fro13}. The low energy window is consistent with Comptonization of black body background photons by an initially non-relativistic expanding outflow, perhaps representative of the initial launch at stellar breakout of the outflow creating the GRB. High luminosities may particularly
derive from intermittent sources, that may be illustrated by numerical simulations on
the breakout of a striped relativlstic MHD jet \citep{van15a}.

Prompt GRB emission is often followed by afterglow emissions mainly detected at lower energies, especially in X-rays down to radio in some cases. Afterglow emission was anticipated based on the GRB association with ultra-relativistic outflows, further enabling identifying host properties and, in some cases, providing calorimetry in the total energy output. We refer the reader to existing reviews on this subject \citep{pir99,pir04}. {\em Swift} made a key discovery with the identification of long duration X-ray tails, that appear to represent latent activity of the remnant inner engine (e.g. \citep{chi10,ber11}). {A systematic study of temporal, energetic and spectral properties of these X-ray tails to both short and long GRBs points to a similar dissipation and/or emission mechanism of common internal origin \citep{mar11}. }

\begin{figure}
\centering
\includegraphics[scale=0.85]{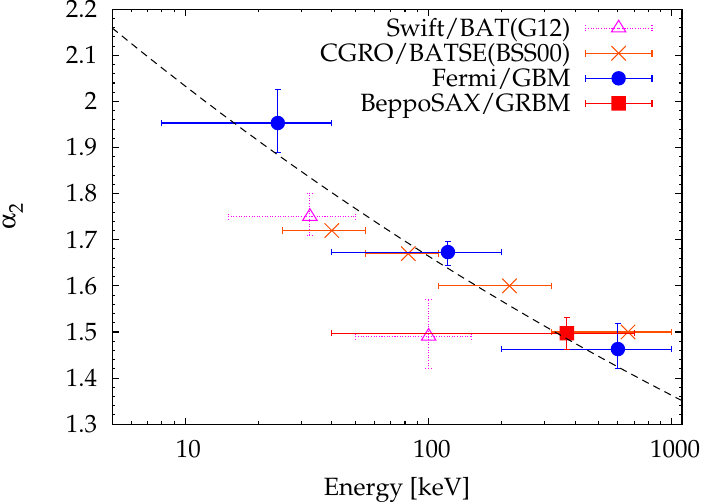}
\caption{Power law index of average PDS over frequencies $10^{-2}<f/{\rm Hz}<1$ by Fourier analysis of different data sets as a function of 
observed energy. Dashed line ($\alpha_2\propto E^{-0.09}$) illustrates $\alpha_2$-dependence on energy estimated from {\em Fermi} data. 
(Reprinted from \cite{dic13a}.)}
\label{fig:alpha}
\end{figure}

\begin{figure}
\centerline{\includegraphics[scale=0.45]{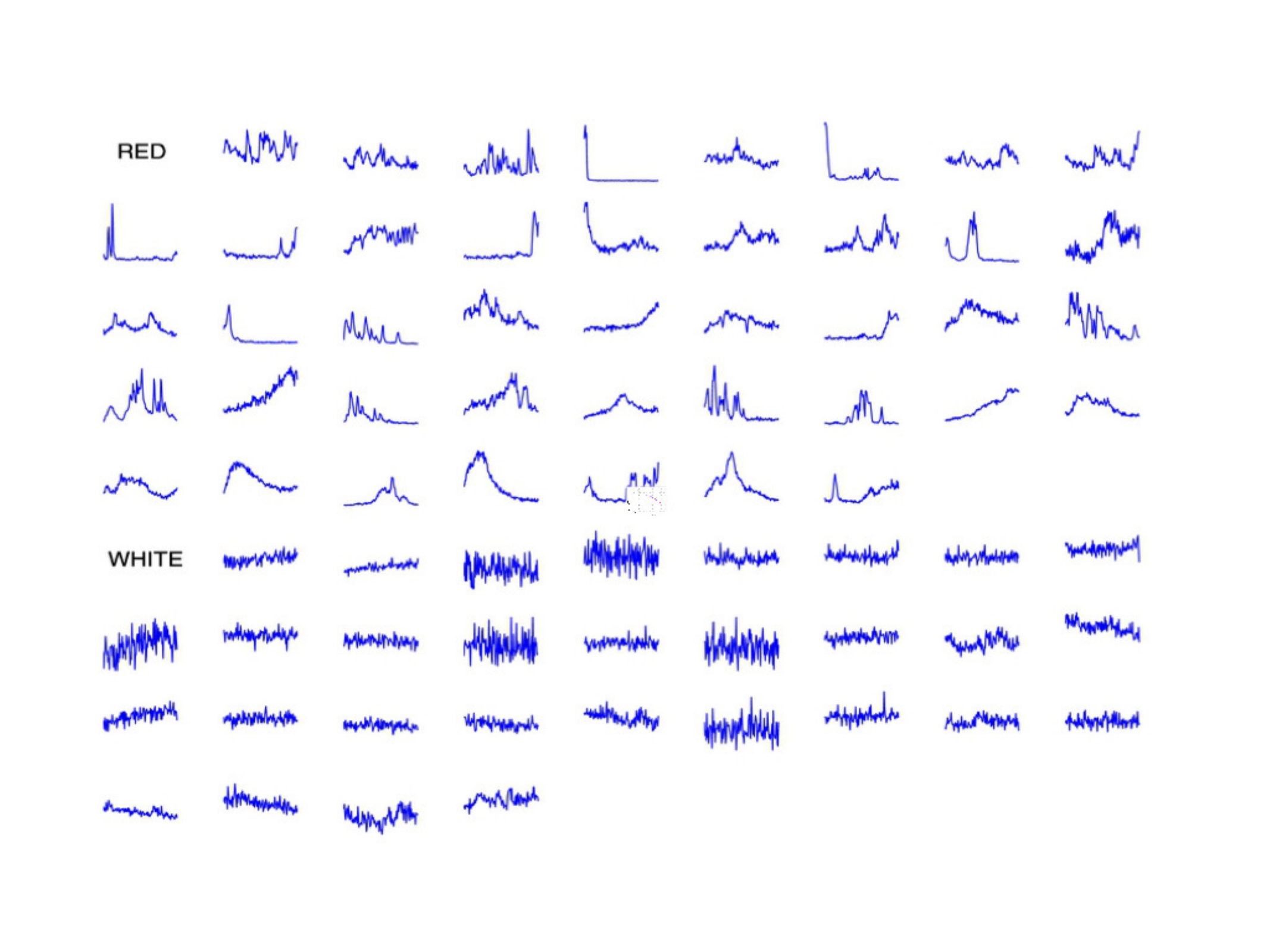}}
\caption{Smoothed light curves of 72 bright long GRBs from {\em BeppoSAX} sampled at 2 kHz for the first 8-10 seconds. 42 have a pronounced autocorrelation (red) with mean photon counts of 1.26 per 500 $\mu$ s bin, while 30 have essentially no autocorrelation (white) with mean 
photon counts of 0.59 per 500 $\mu$ s bin. (Reprinted from \cite{van14}.)}
\label{fig72lc}
\end{figure}

\begin{figure}
\centerline{\includegraphics[scale=0.45]{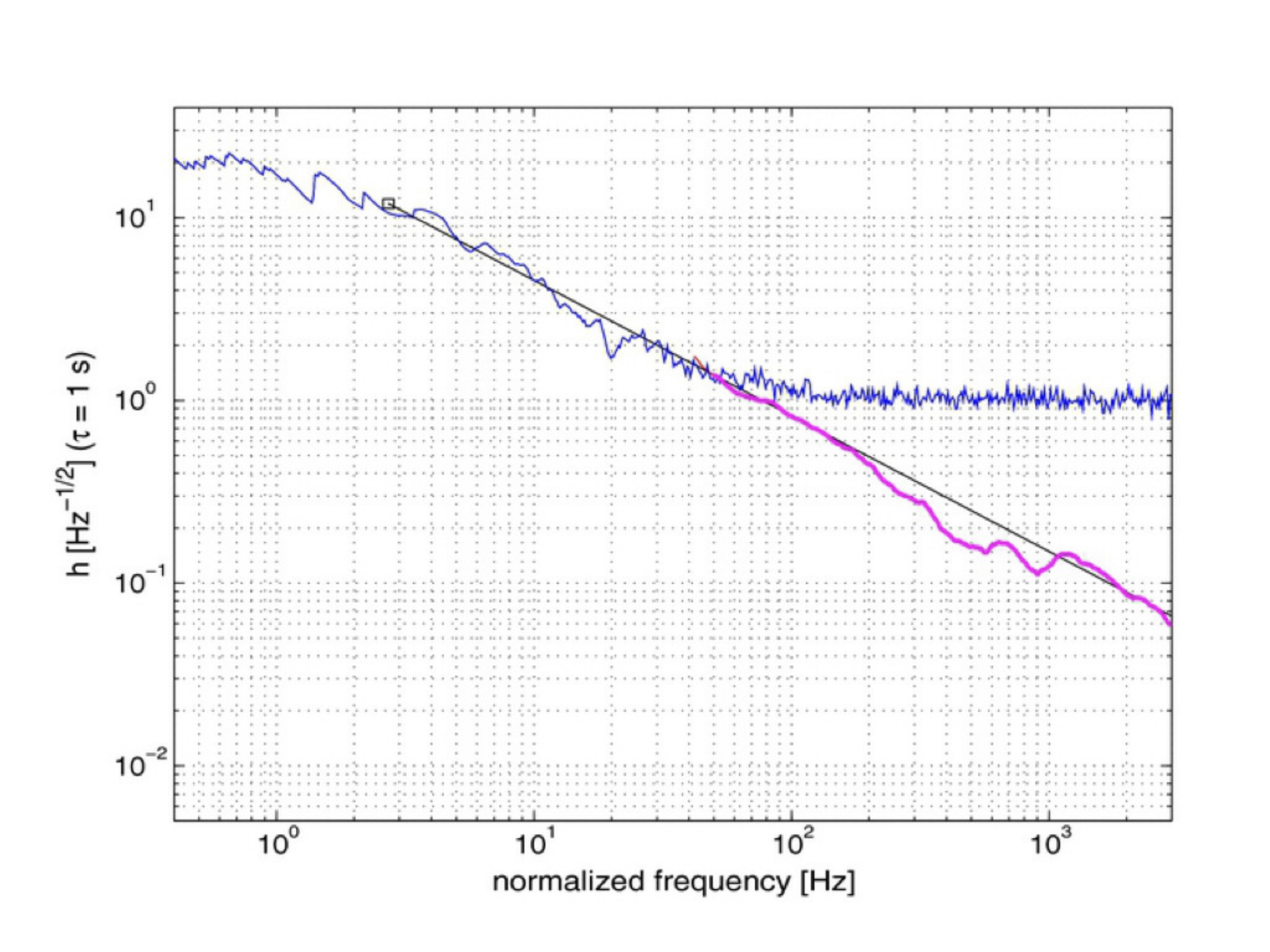}}
\caption{Broadband Kolmogorov spectrum averaged over 42 spectra of (red) bursts with non-trivial autocorrelation functions, extracted by matched filtering using a bank of 8.64 million chirp templates. Results de-redshifted in the source frame show a continuation (black line) to a few kHz (purple, source frame of GRBs) of the Kolmogorov spectrum, extending the same identified at low-frequency by Fourier analysis (blue). {Absent a bump at high frequency expected from canonical mis-alignment of angular momentum and spin in newly formed pulsars} suggests - but does not prove - the absence of magnetars. (Reprinted from \cite{van14}.)}
\label{figKol}
\end{figure}

\subsection{Kolmogorov spectra in LGRBs from {\em BeppoSAX}}

High frequency analysis of prompt GRB emission offers a possible window to intermittency or quasi-periodic behavior in the inner engine, that may be pursued on light curves of the {\em BeppoSAX} catalogue sampled at 2 kHz. As mentioned above, prompt GRB emission probably originates from ultra-relativistic, baryon poor jets, launched from a compact stellar mass object \citep{sar97,pir97} (see also \cite{kob97,nak02}), i.e., neutron star or stellar mass black hole \citep{tho94,met11}. 

Shock induced emission predicts a power law spectrum in energy (e.g. \cite{sar98,tho07}) with turbulent behavior in time. 
Indeed, Fourier analysis of lights curves of long GRBs observed by BATSE and {\em BeppoSAX} reveals a Kolmogorov spectrum in \citep{bel98,bel00,gui12,dic13a,dic13b}. 
The power law index in the observed PSD spectra is broadly distributed about the Kolmogorov 
value of 5/3 with a negative gradient as a function of energy (Fig. \ref{fig:alpha}). Conceivably, this spectral-energy gradient might be due to 
scale dependent dissipation in turbulent flows, which is only beginning to be explored by high resolution numerical simulations in the 
approximation of relativistic hydrodynamics \citep{zra13} (see also \cite{cal14}).

However, Fourier spectra are limited to tens of Hz due to strong Poisson noise in high frequency sampled gamma-ray light curves. 
For long GRBs of {\em BeppoSAX} (Fig. \ref{fig72lc}), Fig. \ref{figKol} shows an extension to the maximal frequency permitted by the Nyquist 
criterion obtained by butterfly filtering, i.e., matched filtering against a dense bank of time-symmetric chirp-like templates \citep{van14}.
A broadband extended Kolmogorov spectrum is found with {no evidence for a bump that might indicate the formation of magnetars}. 
These data point to a {\em common engine producing soft extended emission}, probably so from black holes in mergers and CC-SNe alike. 
Additionally, rotating black holes have ample energy to account for the most energetic GRB-SNe (Table 1). Apart from the low number counts, the only reservation would be extremely sub-luminous CC-SNe (cf. \cite{pas07}), that would be undetectable in our sample of LGRBNs (Fig. \ref{figAm2}). 

Table 3 summarizes our current observational evidence pointing to black hole central engines to LGRBs.

On the premise of black holes unifying the soft extended emission in normal LGRBs (in core-collapse of massive stars) and SGRBEEs (from mergers), we next turn to a model for extended emission from rotating black holes, whose long durations represent the secular time scale of spin down against surrounding high density matter.

\begin{table*}
\textbf{Table 3.} Observational evidence for LGRBs from rotating black holes. (Adapted from \cite{van16}.)
\begin{tabular}{llllll}
\hline
{\sc Instrument} &  {\sc Observation/Discovery} &  {\sc Result} & {\sc Ref.} \\
\hline\hline
\hskip0in {\em Swift}		&	LGRBs with no SN, SGRBEE   	& Extended Emission to mergers 		& (1)	 \\ 
\hskip0.2in 				&      Amati-relation				& Universal to LGRBs and EEs to SGRBs & (1) \\
\hskip0.2in 				&      	X-ray afterglows SGRBs 		& SGRB 050509B 					& (2)\\
\hskip0.in {\em HETE-II}		& 	X-ray afterglows SGRBs  		& SGRB050709 					 & (3) \\
\hskip0.in {\em BeppoSAX}	&      X-ray afterglows LGRBs   		& GRB970228, common to LGRBs, SGRB(EE)s   & (1,4)\\
\hskip0.2in                                 &     Broadband Kolmogorov spectrum & No signature (proto-)pulsars                & (5)\\ 
\hskip0.in BATSE			&	Bi-modal distribution durations. & short-hard and long-soft GRBs                & (6)\\
\hskip0.2in 				&	ms variability				& Compact relativistic central engine          & (7) \\
\hskip0.2in   		      	         &     Normalized light curves LGRBs & BH spin-down against ISCO                    & (8)  \\
\hskip0.in Optical			&	LGRB association to SNe Ib/c   & Branching ratio $<1\%$                           &  (9)  \\
\hskip0.2in                                 &      Calorimetry, $E_k$ of SNe        & $E_{rot}>E_c[NS]$ in some GRB-SNe  & (10)\\\hline
\end{tabular}
\label{TABLE_3}
%}
\mbox{}\\\hskip0.08in
{\bf Note.} (1) Revisited in \cite{van14b}; (2) \cite{geh05}; (3) \cite{vil05,fox05,hjo05}; (4) \cite{cos97}; (5) \cite{van14b}; (6) \cite{kov93}; (7) \cite{sar97,pir97,kob97}, and \cite{van00, nak02}; (8) \cite{van09,van12,nat15}, see further \cite{van08b,sha15}; (9) \cite{van04,gue07}; (10) in the model of \cite{bis70}, $E_{rot}$ exceeds the maximal spin energy $E_c[NS]$ of a (proto-)neutron star in some hyper-energetic events \citep{van11b}.
\end{table*}

\section{Multi-messenger Emission from black hole-torus systems} % $a/M\sim1$ to $a/M\sim0.3$}

{The large energy reservoir $E_J$ in (\ref{EQN_EcEHb}) is predicted by the exact \cite{ker63} solution of rotating black holes of mass $M$, 
angular momentum $J$ and electric charge $Q$.} (For astrophysical black holes of interest here, gravitational contributions by $Q$ can be neglected.)
Rotation is commonly expressed by the dimensionless parameter $a/J$, where $a=J/M$ denotes the specific angular momentum,
$-1\le a/M \le 1$, $\sin\lambda = a/M$ in terms of $|\lambda|\le\pi/2$ \citep{van99}. 
Spacetime about a rotating black hole is dragged into rotation at an angular velocity $\omega=\omega(r,\theta)$, that may be observed as the angular 
velocity of test particles orbiting at a given constant radial distance $r$ and poloidal angle $\theta$ with vanishing specific angular momentum as measured at infinity. 
The angular velocity $\Omega_H$ of the black hole is defined as the limit of $\omega$ as one approaches the black hole $(r\rightarrow  r_H$) that, by the no-hair theorem, reduces to a constant $\Omega_H = \tan(\lambda/2)/2M$ on the event horizon. At a corresponding spin frequency $\nu_H=\Omega_H/2\pi$,
\begin{eqnarray}
\nu_H = 1.6\,\mbox{kHz}\tan(\lambda/2)\left(\frac{10M_\odot}{M}\right),
\label{EQN_k1}
\end{eqnarray}
the total energy $E_J$ in (\ref{EQN_EcEHb}) derives more specifically as $E_J = 2M\sin^2(\lambda/4)$, i.e.,
\begin{eqnarray}
E_J \simeq 6\times 10^{54} \,\mbox{erg} \left(\frac{M}{10M_\odot}\right)\left(\frac{\sin(\lambda/4)}{\sin(\pi/8)}\right)^2.
\label{EQN_k2}
\end{eqnarray}

{$E_J$ is of interest to multi-messenger emission whenever the black hole interacts strongly with surrounding matter in the form if a disk
or possibly thick torus. In addition to accretion, such interaction may be driven by frame dragging, possibly explaining 
(\ref{EQN_alpha1}-\ref{EQN_i2}) when the duration of any such transient emission is identified with the lifetime $T_s$ of black hole spin.}

In core-collapse of a massive star, a black hole grows by Bondi accretion \citep{bon52,sha83}, up to the moment that an accretion disk first forms by angular momentum hang-up about the ISCO of the newly formed rotating black hole. Subsequently, \cite{bar70} accretion is expected to ensue \citep{mck05,kum08b,kin06} - modified by open outflows \citep{van15} - driving the black hole to a near-extremal state, provided there is sufficient mass infall to reach this state. 

{In general time-dependent scenarios, a mass accretion rate $\dot{m}$ in an extended disk is to be distinguished from the accretion flow with mass accretion rate $\dot{M}$ into the black hole of mass $M$. Initially during hyper-accretion the two are expected to be the same, $\dot{M}=\dot{m}$, leading to black hole spin-up by Reynolds stresses in an approximately time-independent accretion flow. When the accretion rate drops sufficiently, Reynolds stresses in the inner accretion flow onto the black hole may become sub-dominant, allowing magnetic stress in Alfv\'en waves to become dominant allowing for suspended accretion mediated by a postive outflow $-\dot{J}$ over an inner torus magnetosphere \cite{van99,van01,van03}. As $\dot{M}\simeq0$, the accretion disk hereby will become time-dependent, that may lead to the formation of a thick torus as matter continues to pile up about and beyond the ISCO by $\dot{m}>0$.}

{In suspended accretion at sub-critical accretion rates in the sense of \cite{glo14}, such near-extremal black hole may experience angular momentum loss 
$\dot{J}$ to surrounding matter, until its angular velocity $\Omega_H$ drops to $\Omega_{ISCO}$ of matter orbiting at the ISCO - a stable fixed point in the equations of suspended accretion \citep{van08}.} 

In GW170817/GRB170817A, the black hole appears to form by collapse of the hypermassive neutron star with mass and angular momentum $(M,J)$ 
in the immediate aftermath of the merger. In this event, the black hole parameters are $(M,J)$, while $E_J$ is significantly enhanced in this collapse
process.

%\subsection{Lifetime of black hole spin}

For core-collapse of massive stars, Fig. \ref{figGrowthe} illustrates the complex sequence of different types of accretion, over the 
course of which the black grows in mass and spins down (Bondi accretion), grows in mass and spins up (Bardeen accretion), and 
loses mass-energy and spins down (suspended accretion). 
This final phase evolves over a secular time-scale \citep{van03,van17a}
\begin{eqnarray}
T_s \simeq 30 \left(\frac{\sigma}{10^{-2}}\right)^{-1}\left(\frac{M}{7M_\odot}\right)\left(\frac{z}{6}\right)^4\,\mbox{s},
\label{EQN_TsA}
\end{eqnarray}
defined by the ratio $\sigma = M_T/M$ and its normalized radius $z=R/M$. 
According to (\ref{EQN_TsA}), the process of losing angular momentum to matter at the ISCO can extend to ultra-long durations 
when $\sigma$ is small, e.g.,  
\begin{eqnarray}
T_s \simeq 100 \left(\frac{\sigma}{10^{-7}}\right)^{-1}\,\mbox{d},
\label{EQN_TsB}
\end{eqnarray}
which time scale can be found in super-luminous supernovae such as SN2015L \citep{van17a}.

%p.29
The process of black holes losing angular momentum to matter at and beyond the ISCO is expected especially from a fully developed turbulent disk. 
Exposed to a finite variance in poloidal magnetic flux, a rapidly rotating black hole develops a finite 
{magnetic moment that preserves maximal magnetic flux through the event horizon at all spin rates, especially so at maximal spin \citep{wal74,dok86,van01p}.} 
Fig. \ref{fig-T} shows the overall efficiency of radiation thus catalytically converted from black hole spin.

In poloidal cross-section, Fig. \ref{figapj15} illustrates the structure of open outflows in a suspended accretion state, allowing the black hole to loose angular momentum $J$ to surrounding matter. Like their supermassive counter parts \citep[e.g.][]{mac97,wal13} or galactic stellar mass black holes in micro-quasars such as GRS 1915+105 \citep{mir94,gre01}, accretion flows in catastrophic events are believed to be likewise magnetized, exposing the central black hole to a finite magnetic flux by accretion \citep{ruf75,bis76,bla77} or the formation of a torus magnetosphere \citep{van99}. 

Strong magnetic fields may derive from the magneto-rotational instability (MRI, \cite{bal91,lub94,glo14}), whose $m=0$ component of the infrared spectrum of MHD turbulence represent a net poloidal flux. In turbulent accretion \citep{bis76} or by forcing \citep{van99}, $\mu_H$ will follow changes in sign in the $m=0$ part of the infrared MHD spectrum on an Alfv\'en crossing time scale. 

%{Exposed to external poloidal magnetic flux, a rapidly rotating black hole develops a finite magnetic moment that preserves maximal magnetic flux through the event horizon at all spin rates, especially at maximal spin \citep{wal74,dok86,van01p}. 
{In the absence of a small angular parameter in the connection of magnetic flux from the latter to the former, an inner torus magnetosphere may hereby mediate angular momentum transport $-\dot{J}$ to surrounding matter by Alv\'en waves. When the black hole spins faster than its surrounding disk or torus,  and accretion rates are sub-critical \citep{lev13b}, $J$ will be transported outwards with gradual expansion of the ISCO as the black hole gradually spins. In this process, matter will be heated and the associated thermal pressures may drive it to develop non-axisymmetric wave instabilities balanced by cooling in gravitational radiation \citep{van12}. }

{The rate of angular momentum transport $-\dot{J}$ outwards is determined by the variance in poloidal magnetic field $B_p^2$. Canonical bounds on $B_p^2$ 
may be derived from a magnetic stability limit $E_{B}/E_k\simeq 1/15$ of energy $E_B$ of the poloidal magnetic field to the kinetic energy $E_k$ in the 
inner disk or torus \citep{van03}. Conceivably, this bound may be circumvented by strongly intermittent inner engines \citep{van15a}, see also \cite{mck12}. 
As a result, a key parameter setting the  lifetime of rapid spin of the black hole is the ratio $\sigma=M_T/M$ of the mass $M_T$ in the torus relative to the mass $M$ of the black hole.}

\subsection{Spin-down in gravitational radiation}

\begin{figure}
\centerline{\includegraphics[scale=0.7]{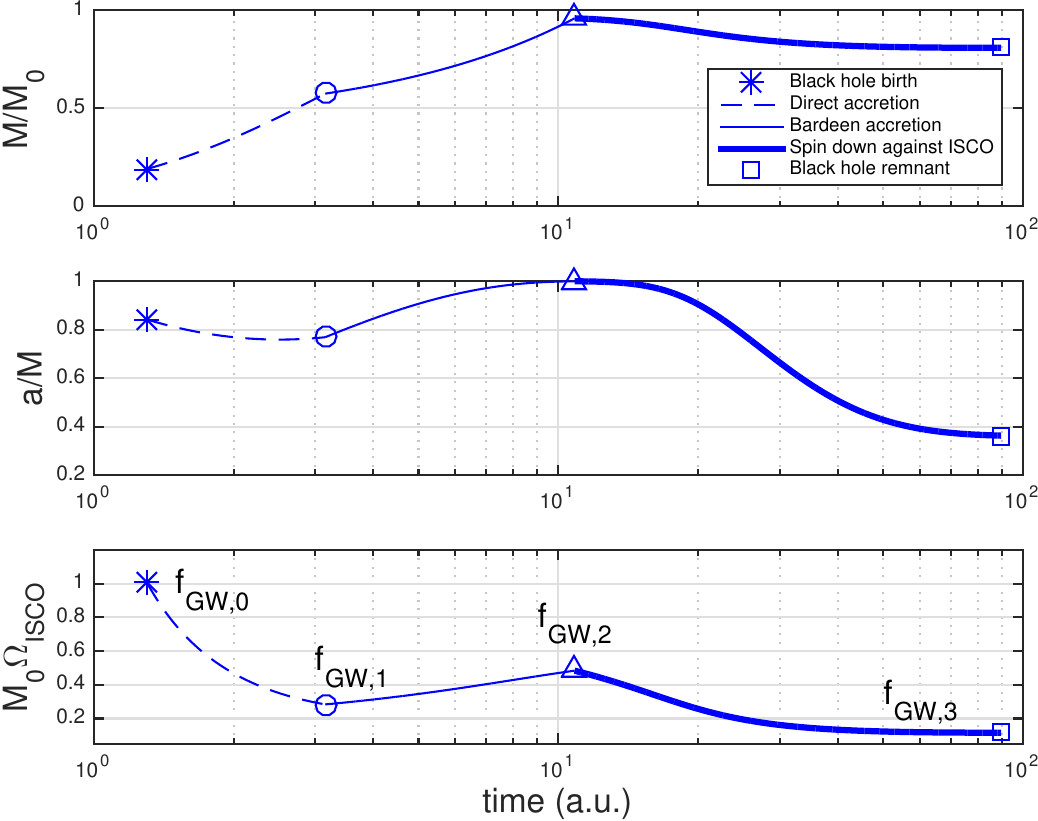}}
\caption{Evolution of a rotating black hole following birth in a progenitor of mass $M_0$ in three phases of accretion: surge in direct accretion ({\em dashed}), growth by Bardeen accretion ({\em continuous}) followed by spin down against matter at the ISCO when accretion becomes subcritical (top and middle panels). Shown is further the associated evolution of any quadrupole gravitational wave signature from matter at the ISCO, marked by frequencies $f_{GW_0}$ at birth, $f_{GW1,}$ at the onset of Bardeen accretion, $f_{GW,2}$ at the onset of spin down and $f_{GW,3}$ at late times, when the black hole is slowly rotating in approximate co-rotation with matter at the ISCO (lower panel.)
(Reprinted from \cite{van17a}.)}
\label{figGrowthe}
\end{figure}

\begin{figure}
\centerline{\includegraphics[scale=0.55]{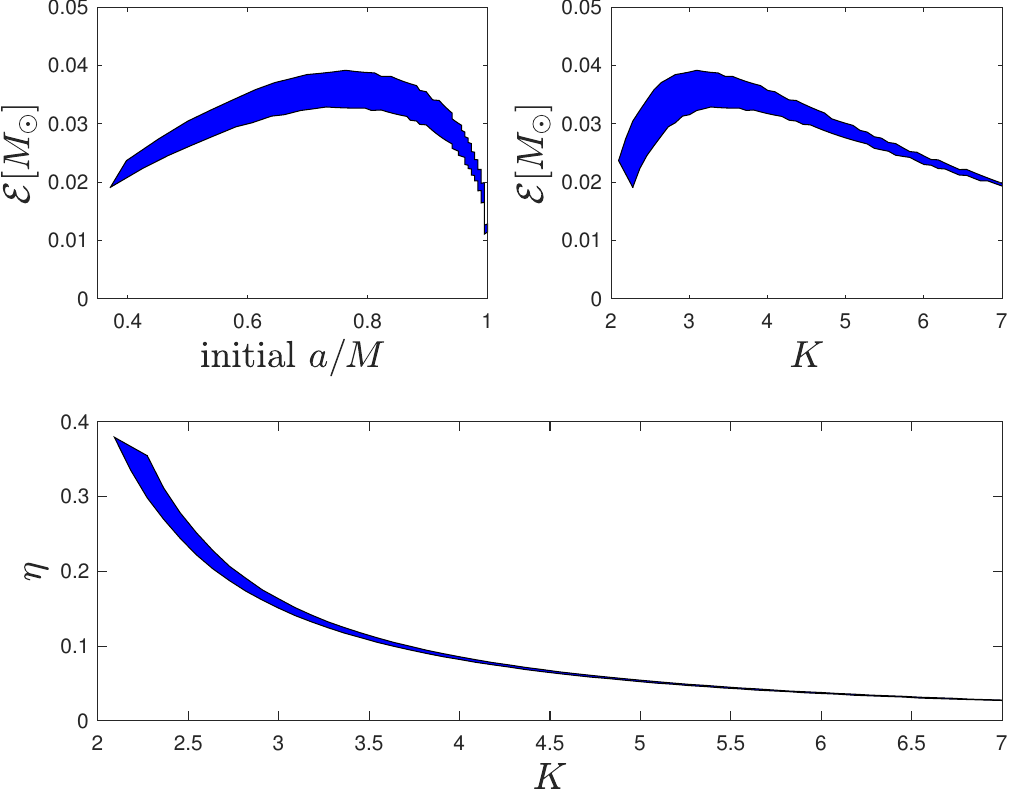}}  
\caption{Model estimate ${\cal E}_{gw}$ of a descending chirp with initial frequency $600\,\mbox{Hz}< f_{gw} < 700\,$Hz,
produced by a non-axisymmetric torus of radius $K$ times the ISCO radius converting spin energy $E_J$ at moderate efficiency
$\eta$ of a rapidly rotating black hole of mass $M=3M_\odot$. 
The graphs show ${\cal E}=3-4\%M_\odot$ for a Kerr black hole with initial 
dimensionless spin parameter $0.7< a/M< 0.8$ corresponding to $E_J\simeq 0.22-0.322 M_\odot c^2$. (Reprinted from \citep{van19b}.)}
\label{fig-T}
\end{figure}

Given the total energy (\ref{EQN_EE1}), Extended Emission to GW170817 derives from $E_J$ of a rotating black hole, observed as a descending
chirp satisfying (Fig. 4)
\begin{eqnarray}
f_{gw}(t)=(f_s-f_0)e^{-(t-t_s)/\tau_s}+f_0~~(t>t_s)
\label{fig_exponential}
\end{eqnarray} 
with $\tau_s=3.01\pm0.2\,$s, $t_s=1843.1$\,s, $f_s=650\,$Hz and $f_0=98$\,Hz.
At a source distance $D$, here scaled to $40\,$Mpc for GW170817 ({\em 1}), the dimensionless strain
$h=L_{gw}^{1/2}/(\Omega D)$ produced by a dimensionless quadrupole mass-moment $\zeta=\delta m/M$ satisfies
$h(t) \simeq 2.7 \times 10^{-23} \left({\zeta}/{3\%}\right) \left({D}/{40\,\mbox{Mpc}}\right)^{-1} \left({f_{gw}}/{650\,\mbox{Hz}} \right)^\frac{2}{3}$,
$L_{gw}$ = $(32/5) \zeta^2 (M/r)^5 \simeq 2\times 10^{52}$ $\left({\zeta}/{3\%}\right)^2 \left({10M}/{r} \right)^5 \mbox{erg~s}^{-1}$ $\simeq$ 
$1\% M_\odot c^2/$s$^{-1}$.

A model estimate ${\cal E}_{gw}$ in (\ref{EQN_EE1}) derives for catalytic conversion of $E_{J}$ 
operating at an efficiency $\eta$ given by the ratio $\Omega_T/\Omega_H$.
The estimated initial frequency of about 744\,Hz at the time of coalescence $t_c$ inferred from our Extended Emission
starting at $t_s<1$\,s is below the orbital frequency at which the stars approach the ISCO of the system as a whole, 
about 1100\,Hz at $r\simeq16$ km according to the Kerr metric. At this point, a binary system of two equal mass neutron stars 
has a dimensionless specific angular momentum $a/M = 0.72 < 1$ consistent with the values found in numerical simulations of \cite{bai08}, 
allowing for prompt collapse to a $\sim3M_\odot$ Kerr black hole with $E_J \simeq 24\%M_\odot c^2$, i.e., about 
one-third of its maximal spin energy. For a torus radius $K$ times the ISCO radius well beyond the ISCO, Fig. 11 shows the 
result of numerical integration of the equations describing spin-down, predicting a maximum output ${\cal E} \simeq 3-4\%M_\odot$ 
over $0.7< a/M< 0.8$ subject to the observed initial gravitational-wave frequency $600\,\mbox{Hz} < f_{gw} < 700\,$Hz 
at $t_s<1$\,s post-merger. % - consistent with (\ref{EQN_E}).

\subsection{Ultra-relativistic black hole jets} 

\begin{figure*}
\centerline{\includegraphics[scale=0.4]{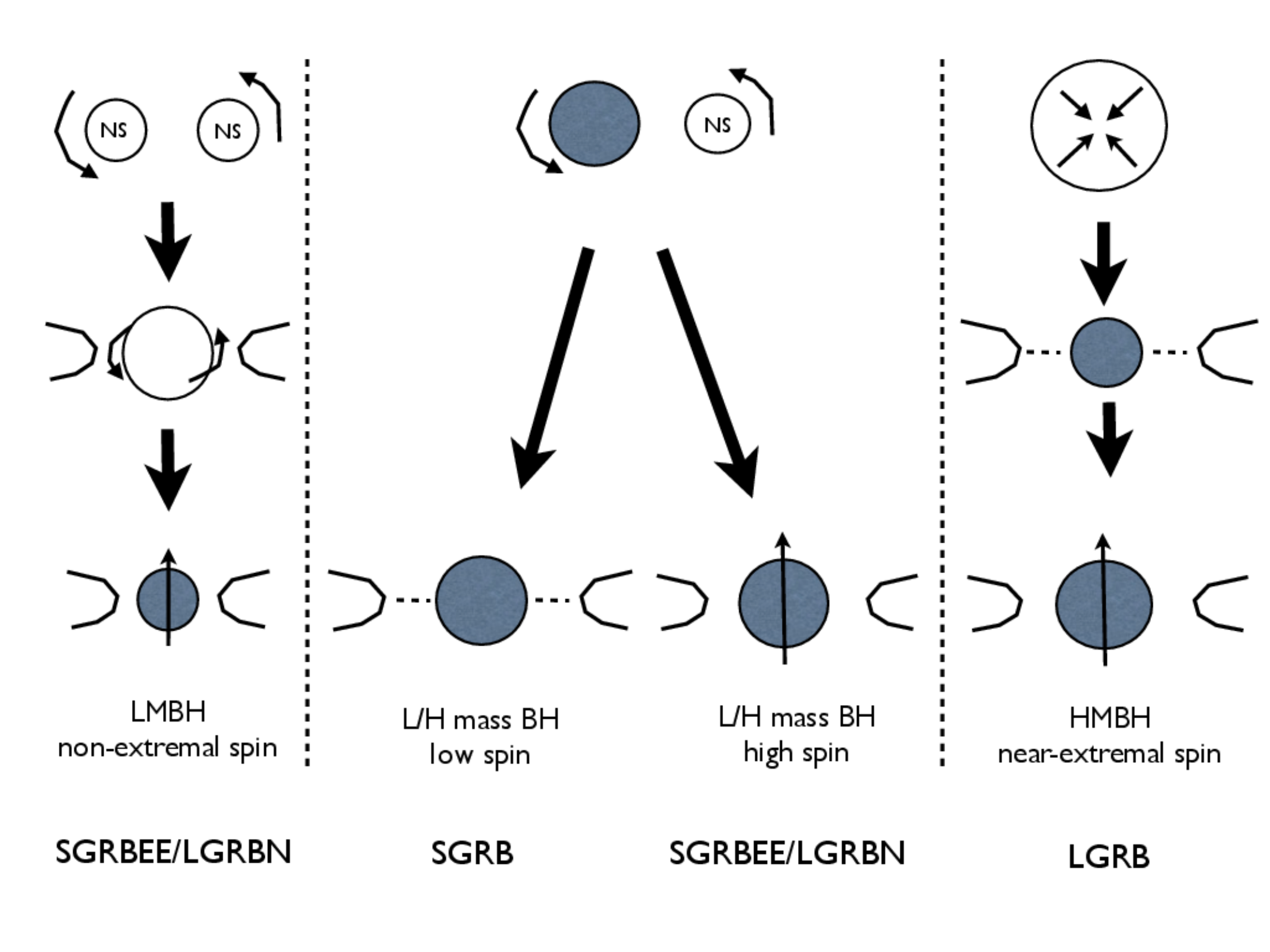}}
\caption{Scenarios of GRBs from rotating black holes from double neutron star mergers (NS-NS, left), mergers of a neutron star with another black hole (NS-BH, middle) and core-collapse of a massive star (CC-SN, right). A black hole-disk or torus system will be of relatively low (LMBH) or high mass (HMBH). NS-NS produces a LMBH with high but non-extremal spin after passing through a super-massive near-extremal neutron star phase \citep[e.g.][]{bai08}.
Spin of the HMBH following NS-BH mergers depends largely on the BH in the progenitor binary, unless its mass and spin are extremely low. Black hole formation in core-collapse passes through Bondi and modified Bardeen accretion, causing the black hole to surge to high mass with near-extremal spin. 
We identify soft Extended Emission (EE) with the spin down phase of rapidly rotating black holes producing SGRBEEs or LGRBs with no supernovae (LGRBN) from mergers (involving rapidly rotating black holes) and normal LGRBs from core-collapse of massive stars. This common physical origin is supported by a common Amati-correlation {of EE and LGRBs shown in Fig. 6.} (Reprinted from \cite{van15}.)}
\label{fig:origin}
\end{figure*}

In (\ref{EQN_TeT90}), black holes are envisioned to be losing $J$ mostly to surrounding matter about and beyond the ISCO, leaving a 
minor release about the spin axis in open outflows powering the baryon-poor ultra-relativistic jets (BPJ) seen in gamma-rays with luminosity \citep{van03,van09}
\begin{eqnarray}
L_j \simeq 1.4\times 10^{51}\,\mbox{erg}\,\mbox{s}^{-1}\,\left(\frac{M}{7M_\odot}\right) \left(\frac{T}{20\,\mbox{s}}\right) \left(\frac{\theta_H}{0.5}\right)^4,
\label{EQN_Lj}
\end{eqnarray}
where $\theta_H$ refers to the half-opening angle of the open magnetic flux-tube on the event horizon of the black hole, considered in its lowest energy
state. For rapidly rotating black holes, $L_j$ is produced by frame dragging along open magnetic flux tubes, supported by Carter's magnetic moment $\mu_e^H$ 
of the black hole \citep{car68,coh73,wal74} in equilibrium with an external magnetic field, supported by the surrounding matter at sub-critical
accretion rates schematically indicated in Fig. \ref{figapj15} \citep{van01,van03}.

\begin{figure}
\centerline{\includegraphics[scale=1.00]{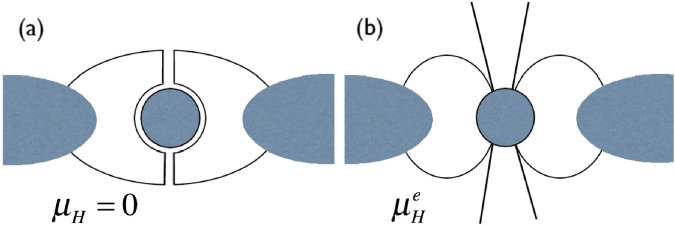}}
\caption{Extremal black holes with vanishing Carter's magnetic moment are out-of-equilibrium with vanishing horizon flux (a). It rapidly settles down to an equilibrium state with essentially maximal horizon flux (b), allowing an open magnetic flux tube to form along the axis of rotation supported by the equilibrium value ${\mu }_{H}^{e}$. At sub-critical accretion rates, frame dragging may produce BPJs while accretion from the ISCO is suspended by feedback from the black hole by Alv\'en waves via an inner torus magnetosphere (b). Major dissipation ($E_D > 0$) by forced MHD turbulence about the ISCO implies ${R}_{{jD}}={E}_{j}/{E}_{D}<< 1$. By slip and no-slip boundary conditions on the event horizon and, respectively, matter at the ISCO, the black hole. Gradually loss of angular momentum gives rise to a finite lifetime $T_s$ of rapid spin. Reprinted from \cite{van15}. 
}
\label{figapj15}
\end{figure}

The output (\ref{EQN_Lj}) is robust as a consequence of differential frame-dragging. ($\omega$ decays to zero at infinity from $\Omega_H$ on the black hole.) 
Locally, frame dragging is formally manifest in Papapetrou forces \citep{pap51,pir56}, where, for charged particles, frame dragging acts on the canonical angular 
momentum supporting a Poynting flux-dominated outflow. The line-integral thereof is a potential energy  $E=\omega J_p$ \citep{van09a}. For charged particles, $J_p$ is defined by total magnetic flux on the flux-surface at hand (which is an adiabatic invariant). In super-strong magnetic fields typically considered in models of GRB inner engines, $E$ assumes energies on the scale of Ultra-High Energy Cosmic Rays (UHECRs). This may be processed downstream to gamma-ray emission in relativistic shocks or, for intermittent sources, into UHECRs by acceleration of ionic contaminants ahead of outgoing Alfv\'en fronts \citep{van09a}.

{The jet luminosity (\ref{EQN_Lj}) may be compared with open model of force-free flux surfaces rotating at one-half the angular velocity of the black hole envisioned in 
\cite{bla77} (with $\theta_H=\pi/2$).
Channeling an outflow in $\theta_H<<\pi/2$ limits the output to be a very small fraction of the total black hole luminosity $L_H$, leaving most $L_H$ to be incident to the surrounding matter mediated by the inner torus magnetosphere. Upon integration in time, (\ref{EQN_Lj})
predicts the same small fraction of $E_J$ \citep{van03} (applied to Eq. 27)
\begin{eqnarray}
E_j\simeq \frac{1}{4z^4}E_J,
\label{EQN_EJj}
\end{eqnarray}
where $z=r/M$. If, on the other hand, $L_j$ would represents an opening outflow supported by horizon flux over the {\em full hemisphere} of the black 
hole event horizon, $E_j$ would be a substantial fraction of $E_J$. The canonically small fraction $1/4z^4$ is in quantitative agreement with observations 
indicating $E_j$ to be a relatively small fraction of $E_J$ in (\ref{EQN_EcEHb}), equivalently, (\ref{EQN_k2}) \citep{van03,van15}.}

\subsection{Spin-down in long GRBs} 

In the electromagnetic spectrum, indirect observational evidence for black hole spin down may be seen by time domain analysis of long GRBs from
the BATSE catalogue. As a proxy for the BATSE durations $T_{90}$ of prompt emission of long GRBs (Fig. \ref{fig4B1}) \citep{van99,van01,van03} (Fig. \ref{fig:origin}), consider $T_{engine}\simeq T_{90}$ with
\begin{eqnarray}
T_{s} \simeq T_{90}
\label{EQN_TeT90}
\end{eqnarray}
while $\Omega_H> \Omega_{ISCO}$, where $\Omega_{ISCO}= ( M (z^{3/2}\pm\sin\lambda))^{-1}$
for prograde (+) and retrograde (-) orbital motion \citep[e.g.][]{sha83}
The inequality $\Omega_H>\Omega_{ISCO}$ is readily satisfied, whenever their rotational energy 
exceeds about 5.3\% of their maximal spin energy (for a given black hole mass-at-infinity $M$).

%{\bf Indirect evidence for the evolution of black hole spin may be sought in a time-domain analysis by suitable averaging of BATSE light curves of long GRBs.
{BATSE Light curves may be normalized in time and count rates by matched filtering using a mode light curve $L_j(t)$ from 
(\ref{EQN_Lj}), defined by numerical integration of the equations of suspended accretion (with fixed points at extremal and slow spin). 
The resulting normalized light curves may then be averaged. The results compare well the model light curve, showing consistency especially
for relatively long duration GRBs. Such model light curve can be created by considering $\theta_H=\theta_H(t)$ to be positively correlated 
to the ISCO: $L_j\propto \Omega_H^2$ and, since $L_j$ is dimensionless in geometrical units, $L_j\propto (r_{ISCO}\Omega_H)^2$. 
In this set-up, $L_j\propto z^2\Omega_H^2$ in a Taylor series expansion in $z=r_{ISCO}/M$, where $r_{ISCO}$ denotes the ISCO radius.}

Normalized light curves (nLC) extracted from the BATSE catalogue of LGRBs allow a confrontation with model templates of spindown, 
of black holes against high density matter at the ISCO and (proto-)neutron stars by magnetic winds. 
In making a connection to the observed GRB emission, we consider a linear correlation between ultra-relativistic baryon-poor outflows and the 
observed prompt GRB emission. Fig. \ref{figmf} shows a match between nLC and model templates. The results favour the first,
especially so for very long duration events with $T_{90}$ exceeding tens of seconds. Similar results obtain for normalized light curves 
extracted from {\em Swift} \citep{gup12}.
\begin{figure*}
\centerline{\includegraphics[scale=0.4]{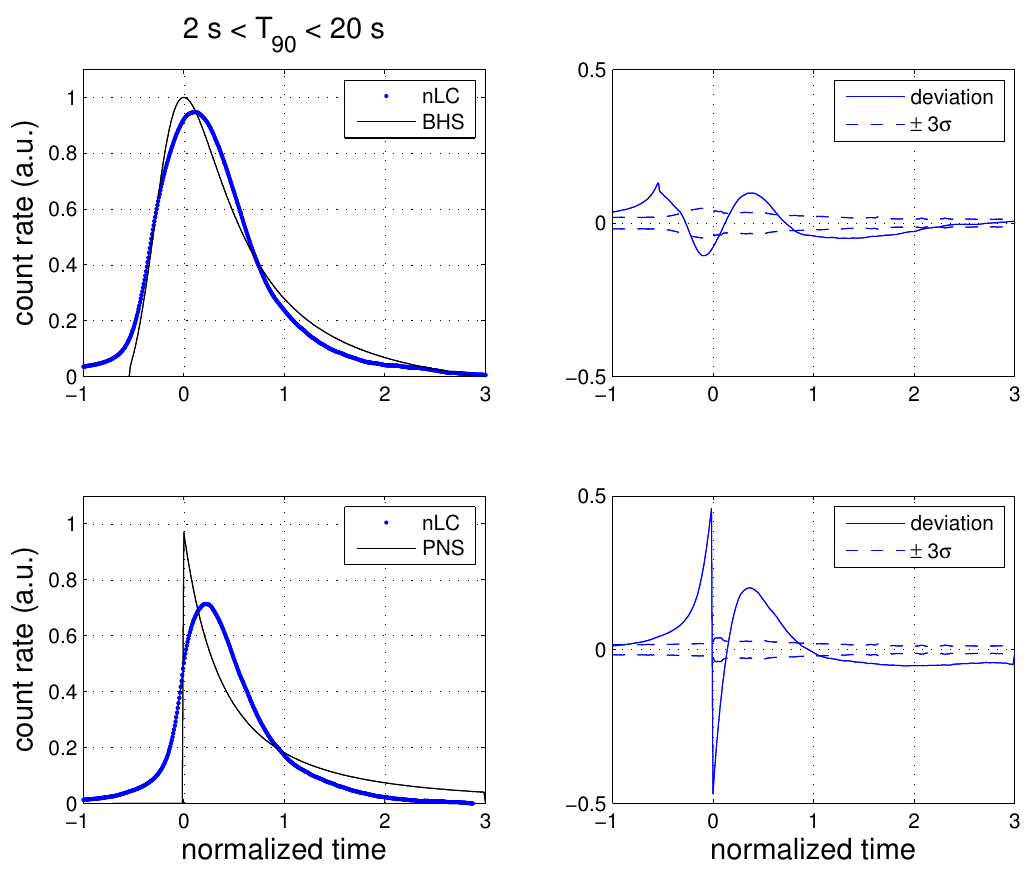}\includegraphics[scale=0.4]{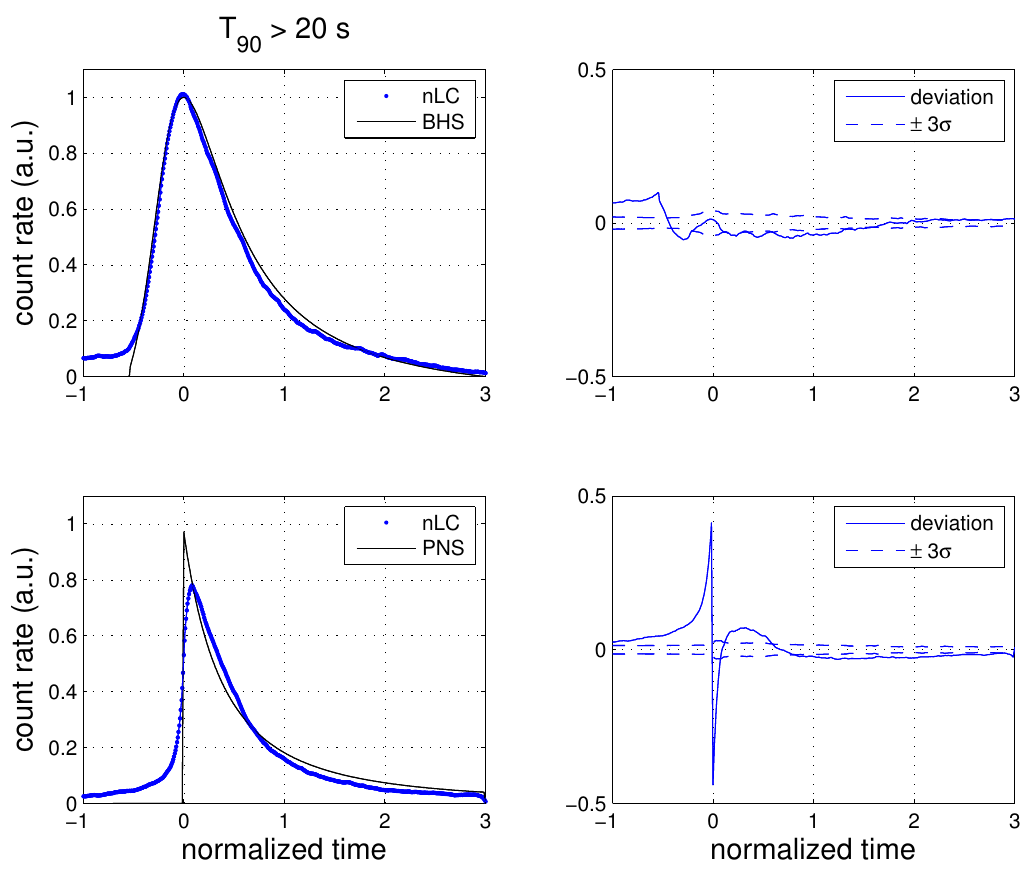}}
\caption{Normalized GRB light curves (nLC, thick lines) extracted from the BATSE catalog, by matched filtering on model templates (thin lines) for 
outflows from spin-down of rotating black holes (BHS, A) and proto-neutron stars (C). Consistency is relatively better for the former, especially 
so for durations greater than 20\,s. We attribute this time scale to that of jet breakout of a stellar remnant envelope. (Adapted from \cite{van12}.)}
\label{figmf}
\end{figure*}
Noticeable also is an improvement in the fit for relatively long duration events with $T_{90}> 20$ s \citep{van09}. We attribute this to the time scale of jet breakout in a remnant stellar envelope for the majority of long GRBs originating in CC-SNe \citep{van12}, possibly further in association with the most rapidly spinning black holes \citep{van09}. Based on different observations related to the relatively flat distribution of $T_{90}$ below 20 s, a similar conclusion obtains \citep{bro13}. 

Eq.(\ref{EQN_TeT90}) may be contrasted with various time scales of accretion, generally associated with growth and spin-up of the central black hole.
In \cite{kum08b}, a distinction is outlined between fall-back at high and low (down to zero) accretion rates. 
The first implies an initial $\dot{m}\propto t^{-1/2}$ for the first ten seconds or so in transition to $\dot{m}\propto t^{-3}$  on the time scale of one hundred seconds. 
The second implies $\dot{m}\propto t^{-n}$ with $4/3\le n \le 2$, depending on the detailed radial profile of mass-loss in winds. 
Assuming a linear correlation between black hole luminosity $L_H$ powering the prompt GRB emission and accretion rate \citep{kum08b}, 
the same procedure applied to these accretion models shows matches vastly below par to those shown in Fig. \ref{figmf} \citep{van17a}.

For power law accretion profiles, discrepant behavior appears notably in a spike between the nLC and the model light curves, characterized by a prompt switch-on (cf. short GRBs). In contrast, Fig. \ref{figmf} shows a satisfactory match the nLC with the model light curve of black holes losing angular momentum against matter at the ISCO across the full duration of the bursts. The results for $T_{90}>20\,$s provide some support for very long GRBs commencing from near-extremal rotating black holes, perhaps in the Thorne limit of the Bardeen trajectory of evolution \citep{van15}. Furthermore, extreme luminosities in GRBs can derive from a nonlinear response to intermittent accretion about the ISCO \citep{van15a}, perhaps stimulated by feedback from the black hole starting from aforementioned nearly extremal spin.

\subsection{Slowly rotating remnants}

A principle outcome of black hole evolution shown in Fig. \ref{figGrowthe} is a {slowly rotating remnant}, whose angular velocity has reached
the fixed point $\Omega_H=\Omega_{ISCO}$ in the equations of suspended accretion, satisfying
\begin{eqnarray}
a/M \simeq 0.36
\label{EQN_FP}
\end{eqnarray}
as a black hole gradually lost most of its angular momentum, and the surrounding Kerr space time relaxed to the space time of a slowly spinning black hole. 
(Just such slow spin appears to be present in the progenitor binary estimates (\ref{EQN_GW15}) of GW150914, but this may also be attributed to 
canonical isolated stellar evolution prior.)

As a stable fixed point, the black hole luminosity will reach a plateau with finite luminosity in open outflows, provided there is a continuing, latent accretion. 
%In attributing SN2015L to black hole spin down following (\ref{EQN_TsB}), just such plateau is seen at late times in the optical light curve.
 The plateau in the optical light curve at late times appears natural when attributing SN2015L to black hole spin down following (\ref{EQN_TsB}).
Following (\ref{EQN_TsA}), it may signal X-ray tails (XRT) over time scales of thousands of seconds, discovered by {\em Swift}, 
that are remarkably universal to LGRBs and SGRBs alike, pointing to a common remnant \citep{eic09}. 

We identify this remnant to be a slowly rotating black hole about the stable fixed point (\ref{EQN_FP}), as a sure outcome regardless of prior formation 
and evolution history and progenitor. % (Fig. \ref{fAPC1}). 
In the present context, XRT's may possibly be accompanied by long lasting low luminosity gravitational 
wave emission. As a common endpoint, this may appear both to normal LGRBs and SGRBs originating in mergers, following messy break-up of neutron stars 
in the tidal field of a companion black hole \citep{lee98,lee99} or in the merger of two neutron stars \citep{ros07}. 

Any misalignment between its angular momentum and that of the black hole would further lead to Lense-Thirring precession and 
hence line-broadening of the trajectories shown \citep{van04b} and thick or extended disks will produce frequencies lower than those shown.
In particular, Fig. \ref{figGW170817EE} shows Extended Emission to GW170817 appears at $< 700$\,Hz in the first quadrant, identified with
emission from mass-quadrupole moment at about three times the radius of the ISCO. 

At late times, a further release in X-rays may appear with luminosity $L_X\simeq 0.25\,\dot{m}$ for an accretion rate $\dot{m}$. 
Accompanying gravitational wave emissions should be very weak with negligible increase in black hole mass and angular momentum in view 
of the observed X-ray luminosities, e.g., $L_X\simeq 10^{41}$ erg s$^{-1}$ in GRB060614 \citep{man07}. If unsteady, large amplitude flaring may 
occur in, e.g., GRB050502B \citep{geh09} by fluctuations between feedback of the black hole $(\Omega_H> \Omega_T)$ 
or accretion $(\Omega_H< \Omega_T)$. See also \citep{lei08}. % \citep{eik03,lei08}.

\section{Non-axisymmetric accretion flows } 

The relatively high densities anticipated in accretion flows in catastrophic events such as mergers and core-collapse of massive stars forms a 
promising starting point for broadband extended gravitational-wave emission. In essence, we expect gravitational radiation derived from
accretion flows and, possibly, waves at or about the ISCO excited by input from the black hole, converting angular momentum in orbital
motion and, respectively, spin of the central black hole. A key pre-requisite for this outlook is the onset of non-axisymmetric waves.

\subsection{Alpha-disk model}

The {\em alpha-disk} model gives a general frame work for mass-inhomogeneities in accretion flows with 
the following properties:

\begin{enumerate}
\item A kinematic viscosity $\nu$ expressed in terms of a dimensionless $\alpha$ parameter given by
\begin{equation}
\nu=\alpha c_s H=\frac{\alpha \Omega H^2}{\sqrt{2} }=\frac{\alpha H^2}{\sqrt{2}M}\left(\frac{M}{r}\right)^{3/2},
\label{EQN_alpha}
\end{equation}
using $c_s=\Omega H/\sqrt{2} $ for the sound speed in terms of orbital angular velocity $\Omega$. Here,
$r$ is the radial distance to the black hole of mass $M$ and, typically, $0.001 < \alpha < 0.1$. Where accretion flows are governed by viscous torques,  the surface density of the disk satisfies $\Sigma(r)=\dot{m}/(3\pi\nu)$ \citep{pri81} for an accretion rate $\dot{m}$ with 
asymptotic radial migration velocity $v_r^\nu$, i.e.,
\begin{eqnarray}
%\begin{array}{ll}
\Sigma =  \frac{M\dot{m}\sqrt{2}}{3\pi \alpha H^2}\left(\frac{r}{M}\right)^\frac{3}{2},~~
v_r^\nu =  \frac{3\nu}{2r} = \frac{3\alpha H^2}{2\sqrt{2}M}\left(\frac{M}{r}\right)^\frac{5}{2}.
%%\end{array}
\label{EQN_sigma}
\end{eqnarray}
Shown below, under certain conditions, there exists a critical radius $r_\nu$ % in (\ref{EQN_rnu2}) 
within which angular momentum loss is dominated by gravitational radiation;
\item 
A Lagrangian disk partition, given by annular rings of radius $r$, radial width $l(r)$ and mass $\Delta m(r)\equiv \sigma(r) M$, 
here in the approximation that $l(r)$ is similar to the vertical scale height  $H(r)$ of the disk. A ring is parametrised 
by mass inhomogeneity, total energy and gravitational wave luminosity 
\begin{eqnarray}
%\begin{array}{ll}
\delta m  = \xi \Delta m,~~
\Delta E  = \frac{\sigma M}{2} \left(\frac{M}{r}\right),~~
\Delta L_{gw}  =\frac{32}{5}\xi^2\sigma^2 \left( \frac{M}{r}\right)^5.
%\end{array}
\label{EQN_dm}
\end{eqnarray}
in terms of the dimensionless parameter $0\le \xi<1$ (Appendix A),  where $\xi$ is not necessarily small, in a local
Keplerian approximation $q=3/2$ in the angular velocity distribution 
\begin{eqnarray}
\Omega(r)=M^{-1}\left(\frac{M}{r}\right)^q.
\label{EQN_Omq}
\end{eqnarray}
\end{enumerate}

Gravitational radiation from mass-inhomogeneities in accretion flows down to the ISCO may appear as instabilities driven by cooling, wave-like or as fragments, when cooling times are on the order of the orbital period \citep{gam01,ric05}. In accretion flows onto black holes, a crucial condition is that such instabilities set in at a radius outside the ISCO. In this event, accretion may be driven by angular momentum loss in gravitational radiation rather than viscous torques across some critical radius greater than $r_{ISCO}$. Although details on the origin and structure of mass-inhomogeneities are uncertain, we shall, for illustrative purposes, discuss these in the quadrupole approximation. In this approximation, migration of mass-inhomogeneities is described by (\ref{EQN_hchar0}-\ref{EQN_hchar2}) in Appendix A. 

Spiral in of inhomogeneities by gravitational radiation dominated angular momentum loss may commence at radii large compared to $r_{ISCO}$. In this event, the luminosity in gravitational waves at a given mass accretion rate 
\begin{eqnarray}
\dot{M} = \dot{m} \dot{m}_0,~~\dot{m}_0 = \frac{c^3}{G}=4\times 10^{38}\,\mbox{g}\,\mbox{s}^{-1},
\end{eqnarray}
satisfies
\begin{equation}
\frac{L_{gw}}{\dot{m}}\simeq \frac{M}{2r_{ISCO}}.
\label{EQN_AISCO2}
\end{equation}
For illustrative purposes, we express (\ref{EQN_AISCO2}) in a Newtonian approximation of the gravitational binding energy at the ISCO to the central black hole. 
A more precise estimate involves $1-e$, where $e$ denotes the specific energy of orbiting matter in the Kerr metric given. 
Our aim here is to develop leading order estimates within a factor of a few. Accretion onto the ISCO may be followed by a plunge into the black hole
or mass ejection in the form of a disk wind.  

The scale (\ref{EQN_AISCO2}) points to a potentially substantial energy output $E_{GW}$ in gravitational waves, provided that a window $r_c> r_{ISCO}$ or $r_b  > r_{ISCO}$ for gravitational radiation dominated angular momentum transport exists. As the following two sections show, this depends on cooling, viscous transport by random walks of large scale eddies describes by the alpha disk model above and mass-inhomogeneities parameterized by $\xi$. The $\alpha$ and $\xi$ are probably inversely correlated, although a detailed description thereof is not known. For instance, small $\alpha$ disks have relatively high density and/or low temperature, by which they are prone to a variety of self-gravity and wave-like instabilities that may produce $\xi$. 

We introduce
\begin{eqnarray}
P = \frac{\xi}{\alpha}
\label{EQN_P}
\end{eqnarray}
to reflect the ratio of gravitational radiation-to-viscous mediated angular momentum transport. $P$ effectively acts as an efficiency parameter in the gravitational wave output from the extended disk $r>r_b$ in the alpha model, assumed to hold for $r\ge r_b> r_{ISCO}$. In this region, the efficiency in gravitational radiation 
is relatively low and the approximation $l\simeq H$ in (\ref{EQN_sigma}) gives the mass density profile
\begin{equation}
\sigma =2\pi r\, l \Sigma =\frac{2\sqrt{2}\,M}{3\alpha H}\left(\frac{r}{M}\right)^\frac{5}{2}\dot{m}~~(r> r_b).
\label{EQN_sig}
\end{equation}
Accordingly, (\ref{EQN_dm}) implies 
\begin{eqnarray}
\Delta L_{gw} = \frac{256}{45h_0^2} P^2\left(\frac{M}{r}\right)^2 \,\dot{m}~~(r>r_\nu).
\label{EQN_DLGW}
\end{eqnarray}  
Adopting a scaling $H=h_0 r$, the total disk luminosity $L_{gw} = \int_{r_b}^\infty \Delta L_{gw} dn$, $dn=dr/l$, satisfies
\begin{equation}
\frac{L_{gw}}{\dot{m}}=\frac{256}{90h_0^3}\frac{P^2}{z^{2}_b}  \simeq 
1.4\,  \frac{P_1^2}{z^{2}_\nu}\left(\frac{h_0}{0.1}\right)^{-3}\, \left(\frac{\dot{M}}{M_\odot\,\mbox{s}^{-1}}\right),
\label{Lgw-ineffic}
\end{equation}
where $z_b= r_b/M$ and $P = 10\,P_1$ associated with a fiducial value $\alpha=0.1$. 

Based on these preliminaries, we next turn to some specific estimates of $r_c$ and $r_b$.

\subsection{Fragmentation chirps}

In self-gravitating accretion flows, a possible origin of mass inhomogeneities is fragmentation when the cooling time of the accreted matter in the instability zone is on the order of or shorter than the orbital time  (e.g., \citep{gam01,ric05,mej05}). The disk is unstable to axisymmetric perturbations if \citep{too64,gol65}
\begin{equation}
Q= \frac{c_s\Omega}{\pi G \Sigma}<1,\label{Qtoomr}
\end{equation}
and to non-axisymmetric perturbations at slightly larger values, $Q\lesssim  2$ (e.g. \citep{gri11}).  

For our $\alpha$-disk model, (\ref{Qtoomr}) yields a characteristic radius beyond which the instability may be generated \citep{pir07}:
\begin{eqnarray}
\frac{r_c}{M}  > \left(\frac{3\alpha h_1^3}{\sqrt{2}\dot{m}}\right)^{2/3} \simeq 
300 \alpha_{-1}^{2/3} \left( \frac{h_1}{0.5}\right)^2  \left(\frac{\dot{M}}{M_\odot \,\mbox{s}^{-1}}\right)^{-\frac{2}{3}} 
\label{EQN_rc}
\end{eqnarray}
adopting $H=h_1r$ at this radius \citep{pop99,che07} with the fiducial scale $h_1=0.5$ for the relatively colder disk flow further out. The characteristic wavelength of the fastest growing mode is of the order of $QH$, and its mass is $(QH)^2\Sigma$.

Cooling may derive from several channels. Among electron-positron pair annihilation to neutrinos, URCA process, and photo-disintegration of $^4$He, it has been argued that the latter may be most effective one in the instability zone \citep{pir07}. Rapid cooling may thus lead to fragmentation into a gravitationally bound clumps of mass $m_f$ up to a few percent of the mass of the black hole, i.e. \citep{pir07}: 
\begin{equation}
m_f = M \sigma_f\simeq M\Sigma(QH)^2\simeq \frac{M}{\pi}h_1^3\simeq 0.04\, M.\label{sig_frag}
\end{equation}
It is unclear how many fragments are produced in this process. In \cite{pir07}, it is suggested that if multiple fragments form, they may merge into a mass of $0.1-1\,M_\odot$. Fragments thus produced will subsequently migrate inwards, initially so by viscous stresses. Any gravitational wave emission hereby derives its energy from the accretion flow. The characteristic strain amplitude hereby scales with the instantaneous strain amplitude, i.e., $h_{char}(f)\propto f^{2/3}$ (cf. \ref{EQN_h1987A}). When it reaches small enough radius with associated transition frequency $f_{e}$ in gravitational waves, angular momentum loss may be overtaken by gravitational wave emission. In this event, the gravitational wave luminosity effectively derives from gravitational binding energy of the inhomogeneities to the central object, as opposed to the accretion flow, until complete disruption by tidal forces.

At the fragment's Roche radius $r_d\simeq 1.26 \eta M (m_f/M)^\frac{1}{3}$ for a black hole size $\eta M$, where $1\le \eta\le 2$ parametrizes uncertainty in black hole spin \citep{fis72,lat74,lat76}, the orbital frequency is roughly $\Omega_{d}\simeq 1/M (M/r_d)^\frac{2}{3}$. With (\ref{sig_frag}) adopted for $m_f$, the corresponding gravitational wave frequency window is
\begin{eqnarray}
W_c:~~ f_c< f< \min\{f_d,2f_{ISCO}\},
 \label{EQN_WC}
 \end{eqnarray}
 where where $f_{ISCO}$ denotes the orbital frequency at the ISCO  and
\begin{eqnarray}
f_c\simeq 1.2 \,M_1^{-1} \frac{1}{\alpha_{-1}} \left( \frac{h_1}{0.5}\right)^{-3} 
   \left(\frac{\dot{M}}{M_\odot \,\mbox{s}^{-1}}\right) \,\mbox{Hz},~~f_{d}\simeq 300-900 \,M_{1}^{-1}\left(\frac{h_1}{0.5}\right)^\frac{3}{2}\,\mbox{ Hz},
\label{EQN_fc1}
\end{eqnarray}
where $M=M_{1}\,10M_\odot$ and the range in $f_d$ refers to the uncertainty in $\eta$.  The broad bandwidth in $W_c$ essentially covers the full operational bandwidth of sensitivity of LIGO-Virgo and KAGRA. The associated characteristic strain amplitude satisfies the canonical scaling of binary coalescence. For a mass fragment (\ref{sig_frag}), we have, adapted from (\ref{EQN_hchar2}) of Appendix A:
\begin{eqnarray}
h_{char}(f>f_e)  = 
1.7 \times 10^{-22} D_{100}^{-1} \,M_1{^\frac{1}{3}} \left(\frac{\sigma_f}{0.04}\right)^\frac{1}{2}  \left(\frac{f}{100\,\mbox{Hz}}\right)^{-\frac{1}{6}}~(f\epsilon W_{c})
\label{EQN_hchar2b2}
\end{eqnarray}
with $\sigma_f = m_f/M$, $D=D_{100} \,100$ Mpc and $W_{c}=[f_{e},\min\{f_d,2f_{ISCO}\}]$. A plot of the fragmentation chirps is exhibited in Fig. \ref{fig:0}. %\ref{fig:hchar}. 

\subsection{Wave patterns in accretion disks}

We shall consider continuous accretion in a non-fragmented disk with deformations $\xi>0$ originate at large radii in wave motion, such as spiral waves by self-gravity in Lin-Shu type wave instabilities \citep{gri11}. These deformations may evolve as matter accretes inwards, though this is poorly understood at present. 

In what follows, we consider a general framework of steady-state accretion and assume, for simplicity, that $\xi$ is a constant. Given the definition of $\delta m$ in (\ref{EQN_dm}), this means that $\delta m$ varies in proportion to $\sigma=\sigma(r)$. This model approach is hereby distinct from ordinary constant mass inhomogeneities.

Once more we focus on quadrupole emission, although any Jeans type self-gravitating instability, as in fragmentation, is inherently local and need not couple to the large scale structure of the disk to ensure that the lowest order modes are the most unstable. For illustrative purposes, we nevertheless focus on the $l=m=2$ instabilities.

Fig. \ref{fig:0} illustrates our identification of quadrupole mass-moments in a spiral wave, in rings following discretization in polar coordinates. Each ring is has finite annular width. By the underlying spiral wave, each ring has a quadrupole
mass-moment of over-dense regions, here represented by a pair of mass-inhomogeneities $\delta m$. Due to rotation, each pair of $\delta m$ emits quadrupole gravitational wave emission at twice the Keplerian frequency, defined by the radius of each ring. In accretion flows, the spiral wave gradually tightens. As $\delta m$ migrates inwards, their gravitational radiation broadens in frequency, each imprinting their own stamp in the gravitational wave spectrum. 
The total luminosity is the sum of the luminosity in each ring by Parseval's theorem.

Our focus on quadrupole wave emissions gives a leading order approximation, that ignores possibly further emissions by higher order mass moments. The full spectrum of gravitational wave emission from all multiple mass moments 
can be calculated, but likely so the total luminosity is dominant in $m=2$  \citep{van02,bro06}.

\begin{figure}[h] %[C]
\centerline{\includegraphics[scale=0.6]{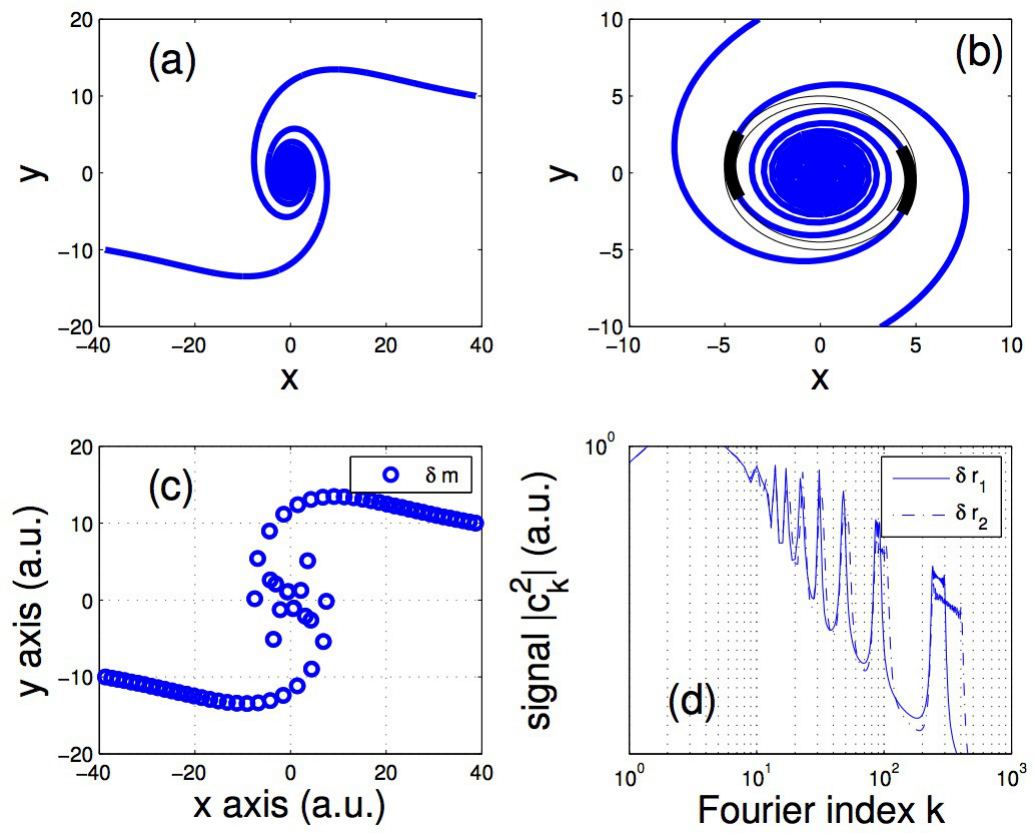}}
\caption{(a) A spiral density wave pattern in a disk. (b) Over-dense regions (thick black) in an annular region $4.5 < r < 5$ (thin black circles) of finite angular extend $\delta \varphi/2\pi<<1$. (a) Approximation of the over-dense regions in (b) by local mass-inhomogeneities $\delta m$, here $2\times 32$ on a grid with 32 annular rings. (d) Keplerian rotation implies distinct quadrupole emission spectra of each $\delta m$ in (c) that are non-overlapping in frequencies. The Fourier coefficients $|c_k|^2$ show broadening due to accretion, shown in (d) for two accretion rates, corresponding to radial migrations $\delta r_2<\delta r_1<0$. (Reprinted from \cite{lev15}.)}
\label{fig:0}
\end{figure}

We consider a transition radius $r_b$, beyond which accretion is by viscous angular momentum transport following standard thin disk accretion theory. Within $r_b$, accretion is driven by gravitational radiation losses in the inspiral in region
\begin{eqnarray}
r_{ISCO} < r < r_b.
\label{EQN_AISCO1}
\end{eqnarray}
Provided that $r_b$ is sufficiently larger than $r_{ISCO}$, (\ref{EQN_AISCO1}) implies maximal efficiency in gravitational radiation with a luminosity satisfying (\ref{EQN_AISCO2}). We now show that $r_b>r_{ISCO}$ is satisfied at hyper-accretion rates.
 
By $\Delta L_{gw}$ in (\ref{EQN_dm}), the ring shrinks, thereby reducing its total energy by $d\Delta E = M^2\sigma/2r^2dr = -\Delta L_{gw}dt_{gw}$  $(dr<0)$, that is 
\begin{equation}
dt_{gw} = \frac{5}{64} \left(\frac{M}{r}\right)^{-3} \xi^{-2}\sigma^{-1}dr.
\label{EQN_tgw}
\end{equation}
The time for a radial drift $dr$ by viscous torques alone is
\begin{eqnarray}
dt_b = \frac{2r}{3\nu} dr = \frac{\sqrt{8}M^2}{3\alpha H^2} \left(\frac{M}{r}\right)^{-\frac{5}{2}}dr.
\label{EQN_tnu}
\end{eqnarray}
We define the transition radius $r_b$ below which gravitational driven migration dominates over viscous
transport by equating (\ref{EQN_tgw}) and (\ref{EQN_tnu}), i.e., $dt_{gw}=dt_b$ at the transition radius
\begin{eqnarray}
\frac{r_b}{M} =\left( \frac{128\sqrt{2}}{15} \frac{\xi^2 \sigma M^2}{\alpha H^2}\right)^2. 
\label{EQN_rnu}
\end{eqnarray}

To be specific, consider a vertical scale height $H(r)/r$ slowly increasing with $r$ from about 0.1 within a few gravitational radii to about 0.5 at about 100 $M$, based on numerical results for neutrino cooled disk models \citep{pop99,che07}. These numerical results show that $H/r$ depends relatively weakly on $\alpha$.  For illustrative purposes, we shall therefore adopt $H(r)=h_0 r$,  $h_0\simeq0.1$, in the inner region. Our $(\alpha,\xi)$ model (\ref{EQN_alpha}-\ref{EQN_dm}) thus obtains
\begin{equation}
\frac{r_b}{M}=\frac{512\xi^2}{45\alpha^2 h_0^3}\dot{m}\simeq 5.7\,P_1^2\,\left(\frac{h_0}{0.1}\right)^{-3}\left(\frac{\dot{M}}{M_\odot\,\mbox{s}^{-1}}\right),
\label{EQN_rnu2}
\end{equation}
where $P=10\,P_1$, $P_1=\xi / \alpha_{-1}$, $\alpha=0.1\alpha_{-1}$. 
Integration of (\ref{EQN_DLGW}) then yields
\begin{eqnarray}
L_{gw}(r)   = \frac{\dot{m}Mr_\nu}{4r^2}~~(r>r_b),~~L_{gw}(r_b)  = 2.2\times 10^{-7} P_1^{-2}\left(\frac{h_0}{0.1}\right)^3.
\label{Lgw-ineffic2}
\end{eqnarray}

In the region $r_{ISCO}<r<r_b$ the accretion flow is driven by gravitational radiation losses, where the mass profile of the enclosed number of $dn=dr/l$ rings is determined from the relation $(\sigma M)dn=\Delta m \,dn = \dot{m} \,dt_{gw}$. 
Using (\ref{EQN_tgw}) with $l=H$, $H=h_0r$, we obtain
\begin{equation}
\sigma =\frac{1}{5.7\xi} \sqrt{{5h_0\dot{m}}} \left(\frac{r}{M}\right)^2.
\label{sig-gw}
\end{equation}
Note the relatively steep radial dependence compared to (\ref{EQN_sig}) due to a modified surface density $\Sigma$. Following substitution of (\ref{sig-gw}) into (\ref{EQN_dm}), it is seen that $\Delta L_{gw}\propto \dot{m}\, r^{-1}$ is strongest at the smallest radii. The total luminosity is  roughly the sum over the inner rings,  $L_{gw}\simeq (r_{ISCO}/l)\Delta L_{gw}(r_{ISCO})\simeq h_0^{-1}\Delta L_{gw}(r_{ISCO})$. In the high efficiency regime $r_b>r_{ISCO}$, we thus obtain the total disk luminosity (\ref{EQN_AISCO2}). This shows that a major fraction of the accretion flow is converted to gravitational radiation, {\em independent} of $P$. 

According to (\ref{EQN_rnu2}), $r_{b}>r_{ISCO}$ holds at hyper-accretion rates 
\begin{eqnarray}
\frac{\dot{M}}{M_\odot\,\mbox{s}^{-1}} > \frac{1}{5.7} z\,P_1^{-2}
\label{EQN_hz}
\end{eqnarray}
with luminosity (\ref{EQN_AISCO2}) provided that mass inhomogeneities originating at $r>r_b>r_{ISCO}$ survive all the way to $r_{ISCO}=zM$. If they dissipate at $r_{ISCO}<r_{diss}<r_b$, then $r_{ISCO}$ should be replaced by $r_{diss}$ in (\ref{EQN_AISCO2}) , and the net result is a relatively lower $L_{gw}$ emitted at lower frequencies. It will also be appreciated that for (\ref{EQN_hz}) to proceed on a long duration time scale of tens of seconds, a large progenitor remnant stellar envelope mass is required or $\alpha$ is small.  

We express the outlook above in terms of the frequency window associated with quadrupole emissions at $r_b$ 
and $r_{ISCO}$:
\begin{eqnarray}
W_b:~~ f_b < f< \max\{f_{diss},2f_{ISCO}\},
 \label{EQN_Wnu}
 \end{eqnarray}
where
\begin{equation}
f_b\simeq 468\,M_1^{-1} P_1^{-3}\left(\frac{h_0}{0.1}\right)^{\frac{9}{2}} \left(\frac{\dot{M}}{M_\odot\,\mbox{s}^{-1}}\right)^{-\frac{3}{2} }\,\mbox{Hz}.
\label{EQN_fnu}
\end{equation}

The observable relevant to $W_b$ is the associated characteristic strain amplitude.
Over a time period $\tau$, an accretion rate $\dot{m}$ implies a mass migration $\dot{m}\tau$ in the annular region down
to $r$ from  $r_b$ by the associated energy $\Delta E_{rad}=({1}/{2})\left({M}/{r}\right)\dot{m}\tau$ in gravitational
radiation. For $r_{ISCO}\le r << r_b$, it is emitted over a bandwidth given by the difference between the gravitational 
wave frequency at $r$ and $f_b$,
\begin{eqnarray}
\Delta f = f(r) - f_b \simeq f(r) = (\pi M)^{-1} \left(\frac{M}{r}\right)^\frac{3}{2}
\label{EQN_A4}
\end{eqnarray}
in the present Keplerian approximation. Consequently, 
\begin{eqnarray}
\frac{\Delta {E}_{rad}}{\Delta f} = \frac{\pi\dot{M}\tau M}{2}\left(\frac{M}{r}\right)^{-\frac{1}{2}},
\label{EQN_A5}
\end{eqnarray}
and hence the characteristic strain for a source at a distance $D$ is
\begin{eqnarray}
h_{char}(f) = \frac{\sqrt{2}}{\pi D}\sqrt{\frac{\Delta { E_{rad}}}{\Delta f}} \simeq
  \frac{M}{\sqrt{\pi} D}\, \sqrt{\frac{\dot{M}\tau}{M }} \left(\pi M  f \right)^{-1/6}=\sqrt{2}\kappa \left(f/f_b\right)^{-\frac{1}{6}} 
\label{EQN_A6A}
\end{eqnarray} 
for $f\epsilon W_b$ with
\begin{eqnarray}
\kappa =  \kappa_0 \left(\frac{f_b}{f_0}\right)^{-\frac{1}{6}},~~\kappa_0= \frac{M}{\sqrt{2\pi }D}\, \sqrt{\frac{\dot{M}\tau}{M}} \left(\pi M  f_0 \right)^{-1/6}.
\label{EQN_A6B}
\end{eqnarray} 
Numerically, we have 
\begin{eqnarray}
\kappa_0 = 8.3 \times 10^{-22}\,\frac{M_1^\frac{1}{3}}{D_{100}} \sqrt{\frac{M_{acc}}{M_\odot}} \left(\frac{f_0}{1000\,\mbox{Hz}}\right)^{-\frac{1}{6}},
\label{EQN_A6C}
\end{eqnarray} 
where we put $M_{acc}=\dot{M}\tau$. This shows that $h_{char}(f)$ is maximal in $W_b$ at $f\simeq f_b$. Observationally, $f_b$ is the most relevant frequency in $W_\nu$ provided that it falls in the thermal or shot noise dominated
region of the LIGO-Virgo and KAGRA detectors.  

By the strong dependence of $f_b$ on $\dot{M}$ in (\ref{EQN_fnu}), it will be appreciated that $W_b$ opens up a window
$f_b< 2f_{ISCO}$ only at large hyper-accretion rates, when 
\begin{eqnarray}
\frac{\dot{M}}{M_\odot\,\mbox{s}^{-1}} > \left( \frac{M_1f_{ISCO} }{216 \,\mbox{Hz}}\right)^{-\frac{2}{3}} \alpha_{-1}^2\simeq 
\frac{1}{6} \left(z^\frac{3}{2}+\hat{a}\right)^\frac{2}{3}\alpha_{-1}^2\ge 0.26\,\alpha_{-1}^2,
\label{EQN_m0}
\end{eqnarray}
where $z=r_{ISCO}/M$, $\hat{a}=a/M$ denotes the dimensionless spin rate of the black hole and the lower bound on the right hand side refers to an extremal Kerr black hole $(z=\hat{a}=1)$. The inequality on the right hand side of (\ref{EQN_m0}) provides a necessary but not sufficient condition. Around non-extremal black holes, the required $\dot{M}$ is larger.
Furthermore, the frequency windows $W_c$ and $W_b$ satisfy different scalings with accretion rate $\dot{M}$. Around a nearly extremal Kerr black hole, we note that $f_c = f_b$ only when
\begin{eqnarray}
\dot{m} \simeq 10 \,\left(\frac{h_1}{0.5}\right)^\frac{6}{5} \alpha_{-1}^\frac{8}{5}.
\end{eqnarray}
Unless $\alpha_{-1}$ is small, i.e., $0.001 < \alpha < 0.01$, we expect
\begin{eqnarray}
W_b \subset W_c.
\label{EQN_WW}
\end{eqnarray}

In the external region $r>r_b$, the emission in gravitational waves is at frequencies $f<f_b$. In this region, the emission is relatively inefficient and satisfies (\ref{Lgw-ineffic2}). We can estimate the effective strain $h_{eff}=h(f) \sqrt{f\tau}$ from the instantaneous strain $h=L_{gw}^{1/2}/\Omega D$, $\Omega = \pi f$, and the number of wave periods $n=f\tau$.  By $h_{char}\sqrt{f^{-1} \Delta f}=\sqrt{2}h_{eff}$, we thus estimate the characteristic strain associated with a one-sided frequency spectrum as
\begin{eqnarray}
h_{char}(f) = \frac{M}{\sqrt{2 \pi} D} \sqrt{\frac{M_{acc}}{M}} \left(\pi M f_b\right)^{-\frac{1}{6}} (f/f_b)^{\frac{1}{6}}
\,(f < f_\nu).
\label{EQN_inefficD}
\end{eqnarray}

A matching of the expressions (\ref{EQN_A6A}) and (\ref{Lgw-ineffic2}) obtains by noting that the luminosity in the inner radiatively efficient region $[r,r_b]$, $r_{ISCO}\le r << r_b$, satisfies  
\begin{eqnarray}
L_{gw}^i([r,r_\nu])= \frac{M\dot{m}}{2r} - \frac{M\dot{m}}{4r_b} = \frac{M\dot{m}}{4r}\left[2-(f_b/f)^\frac{2}{3}\right] .
\label{EQN_LGWi}
\end{eqnarray}
With (\ref{EQN_A6C}), we therefore have \citep{lev15}
\begin{eqnarray}
h_{char}(f) = \kappa  \left\{
\begin{array}{lr}
(f/f_b)^\frac{1}{6}  &  (f<f_b)\\
(f/f_b)^{-\frac{1}{6}}\left[ 2 - (f_b/f)^\frac{2}{3}\right]^\frac{1}{2} & (f\ge f_b).
\end{array}\right.
\label{EQN_A6D}
\end{eqnarray} 
This broadband spectrum increases with $\dot{m}$. Due to (\ref{Lgw-ineffic2}), the increase in $f<f_b$ is entirely due a shrinking of the bandwidth, as $r_\nu$ increases and $f_b$ decreases with $\dot{M}$. In contrast, the increase in $f>f_b$ is due to an increase in the luminosity (\ref{EQN_LGWi}). 
A plot of the proposed broadband emission (\ref{EQN_A6D}) is exhibited in Fig. \ref{fig:hchar} for various choices of $M_{acc} = \dot{m}\tau$ and $f_b$. 

\subsection{Non-axisymmetric waves in a torus}

Following (\ref{EQN_TeT90}), we consider sub-critical accretion allowing for the formation of a torus in suspended accretion, which
catalytically converts most of the input from the black hole into gravitational radiation at frequencies about 
\begin{eqnarray}
f_{gw} \simeq \pi^{-1} \Omega_{T} %ISCO}
\label{EQN_fgw2}
\end{eqnarray}
with $\Omega_T\lesssim \Omega_{ISCO}$ (inequality for an extended disk or thick torus) 
with lesser emission at higher frequencies \citep{bro06}, accompanied by a minor output (\ref{EQN_Lj}) in BPJs. The former features a 
distinctive descending chirp, due to expansion of the ISCO as the black hole spins down \citep{van08}. This emission is subsequent to any
gravitational burst associated with the initial formation of the black hole and continuing accretion \citep{ree74,due04,lip83,leu97,fon01,nag07,lip09,fry02} and separate from any quasi-normal mode ringing (QNR) of the event horizon \citep{tho86,kok99} with relatively high frequencies ($l=m=2$) \citep{ech98}
\begin{eqnarray}
f_{QNM}=3200 \,\mbox{Hz}\,\left[1-0.63\left(1-\frac{a}{M_H}\right)\right]^\frac{3}{10} \left(\frac{M_H}{10M_\odot}\right)^{-1}.
\label{EQN_fQNR}
\end{eqnarray}

Quite generally, gravitational radiation from ISCO waves is described by mass moments $I_{lm}$ in the quantum numbers $l$ and $m$ of spherical harmonics
has a luminosity (Appendix A)
\begin{eqnarray}
L_{gw} \propto \Omega_T^{2l+2}I_{lm}^2.
\label{EQN_LQ1}
\end{eqnarray}
In practice, it appears that most of the emission comes from $m=l$ \citep{bro06}. 
To leading order $\Omega_T^2\propto r^{-3}$, whereby (\ref{EQN_LQ1}) satisfies
\begin{eqnarray}
L_{gw} \propto \frac{1}{r^{m+3}}.
\label{EQN_Q2}
\end{eqnarray}
Consequently, gravitational wave luminosity tends to be maximal at the ISCO, such as exemplified in the previous section.

Susceptibility to non-axisymmetric instabilities at the ISCO is due to a variety of processes. 
In particular, energetic feedback by black hole sets in at $E_{rot}^H$ in excess of 5.3\% of $E_{rot}^{max}$, when $\Omega_H>\Omega_{ISCO}$. 
Black holes losing angular momentum to surrounding matter enforce turbulence with associated heating and enhanced (thermal and magnetic) pressure \cite{van02,van12}.
An improved criterion for the black hole to be rapidly spinning, is when its forcing onto the surrounding matter exceeds the luminosity in magnetic winds from the disk, taking advantage of $\Omega_H/\Omega_{ISCO}>1$ (the ratio extends up to 1.4396 in the Kerr metric) and the substantial surface area of the event horizon. 
Using the Shakura-Sunyaev solution \citep{sha73} as a leading order approximation of the inner disk, the time integrated output $E_H$ from the black hole luminosity onto the inner disk exceeds the energy loss $E_w^*$ of this disk in magnetic winds whenever $a/M\ge 0.4433$ \citep{van12}. Equivalently, 
$E_J$ initially exceeds $E_J^{max}$ by about 9\%, which is still small. The excess $E_H - E_w^*$ can be radiated off in gravitational waves and MeV neutrinos. 
Here, we attribute the excitation of non-axisymmetric wave instabilities to enhanced thermal and magnetic pressures induced by such feedback.

%\subsubsection{Mass moments about the ISCO}

Papaloizou-Pringle \citep{pap84} considered (\ref{EQN_Omq}) with non-Keplerian rotation index $q>3/2$. Applied to a torus, the associated surface gravity allows surface waves to be excited at the inner and outer surface. In the approximation of an incompressible fluid, a detailed analysis shows that the coupling between waves on these two opposite faces allows for angular momentum transport outwards on a dynamical time scale. In the singular limit of an infinitesimally slender torus considered in \citep{pap84}, these surface waves become unstable to all of $m\ge 1$ modes at the same critical rotation index $q_c=\sqrt{3}$. In what follows, we again highlight results for the idealized limit $\Omega_T \simeq \Omega_{ISCO}$, where
$\Omega_T < \Omega_{ISCO}$ for an extended disk or thick torus.

We envision $q>3/2$ induced by feedback from the central rotating black hole. Extending the Papaloizou-Pringle instability to tori of arbitrary width \citep{van02}, non-axisymmetric instabilities are found to set in consecutively, starting at $m=1$, as a function of the rotation index of a torus (Appendix B). 
When $q>q_{cm}(\delta)\ge \sqrt{3}$, %exceeds a critical value, 
the azimuthal modes $m=1,2,\cdots$ become successively unstable (as a function of $q$, Fig. \ref{fig:p6a}), where \citep{van02}
\begin{eqnarray}
%\begin{array}{ll}
q_{c1}(\delta)= \sqrt{3} + 0.27 \left(\frac{\delta}{0.7506}\right)^2,~~
q_{c2}=\sqrt{3} +0.27\left(\frac{\delta}{0.3260}\right)^2,\cdots
%\end{array}
\label{EQN_APB4}
\end{eqnarray}
as a function of the ratio $\delta=b/R$ of minor-to-major radius of the torus. Here,
$\sqrt{3}$ refers to the singular limit $\delta=0$ of infinitesimally slender tori in \citep{pap84}. The curves (\ref{EQN_APB4}) are analytic approximations to semi-analytical results shown in Fig. \ref{fig:p6a}. Gravitational radiation from such non-axisymmetric wave instabilities in a torus of finite aspect ratio anticipated in \citep{van01b,van02,van03} is now 
also seen in more recent numerical simulations \citep{kiu11,tos19}.

From heating alone, a minimum value of $q$ can be associated with the temperature $T=T_{10}10^{10}$ K \citep{van03}
\begin{eqnarray}
T_{10} \simeq 2L_{\nu,52} \left(\frac{M_T}{0.1M_\odot}\right)^{-\frac{1}{6}}.
\label{EQN_APB2}
\end{eqnarray}
For a torus of minor-to-major radius $\delta = b/R$, the resulting thermal pressure enhances $q$ according to
\begin{eqnarray}
q = 1.5 + 0.15 \left(\frac{R}{M}\right) \left(\frac{\delta}{0.2} \right)^2 T_{10}.
\label{EQN_APB3}
\end{eqnarray}
Thermal pressures and magnetic pressures will be similar at temperatures of about 2 MeV \citep{van09a}. 

A mode that becomes unstable is generally strengthened by gravitational radiation back reaction \citep{van02}. As a result, inducing gravitational wave emissions by (\ref{EQN_APB4}) is reminiscent of a Hopf bifurcation \citep{van12}. The result is a characteristic descending chirp during black hole spin down (Fig. 4). 
%(Fig. \ref{fig:p6a}). 

Ultimately, the luminosity is determined by a nonlinear saturation amplitude, balancing {\em heating} in dissipation and {\em cooling} in gravitational waves, MeV neutrino emission and magnetic winds \citep{van12}, unless aforementioned feedback is intermittent or the torus as a whole is unstable. The first allows for an analytic estimate in the case of a flat infrared spectrum of MHD turbulence \citep{van01b}, possibly of interest to the most luminous sources resulting from hyper-accretion onto the ISCO in core-collapse supernovae \citep{van15a}.

\begin{figure}[h] %[B]
\centerline{\includegraphics[scale=0.31]{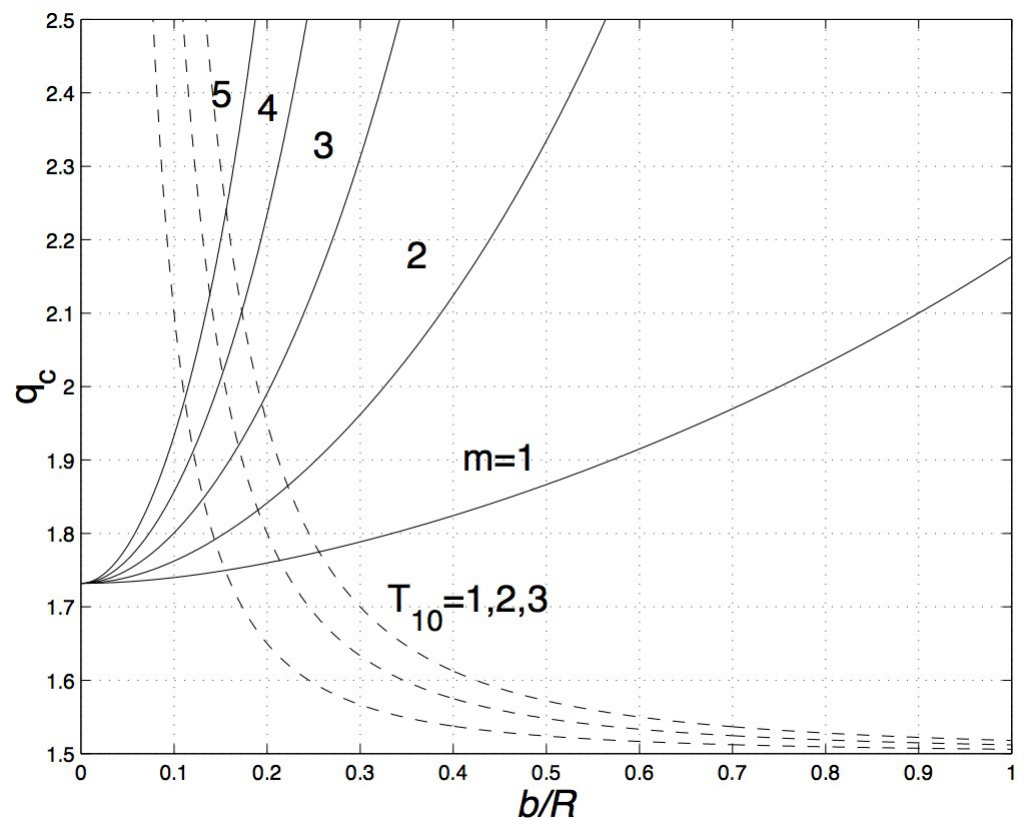}\includegraphics[scale=0.31]{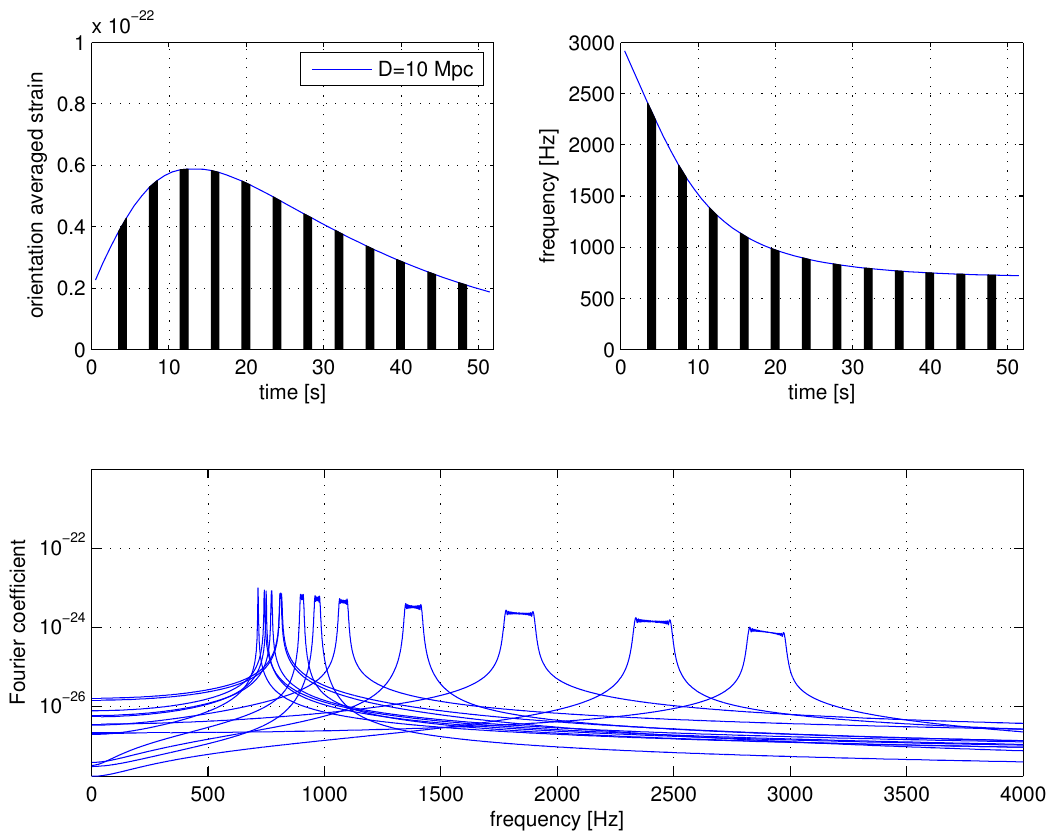}\includegraphics[scale=0.31]{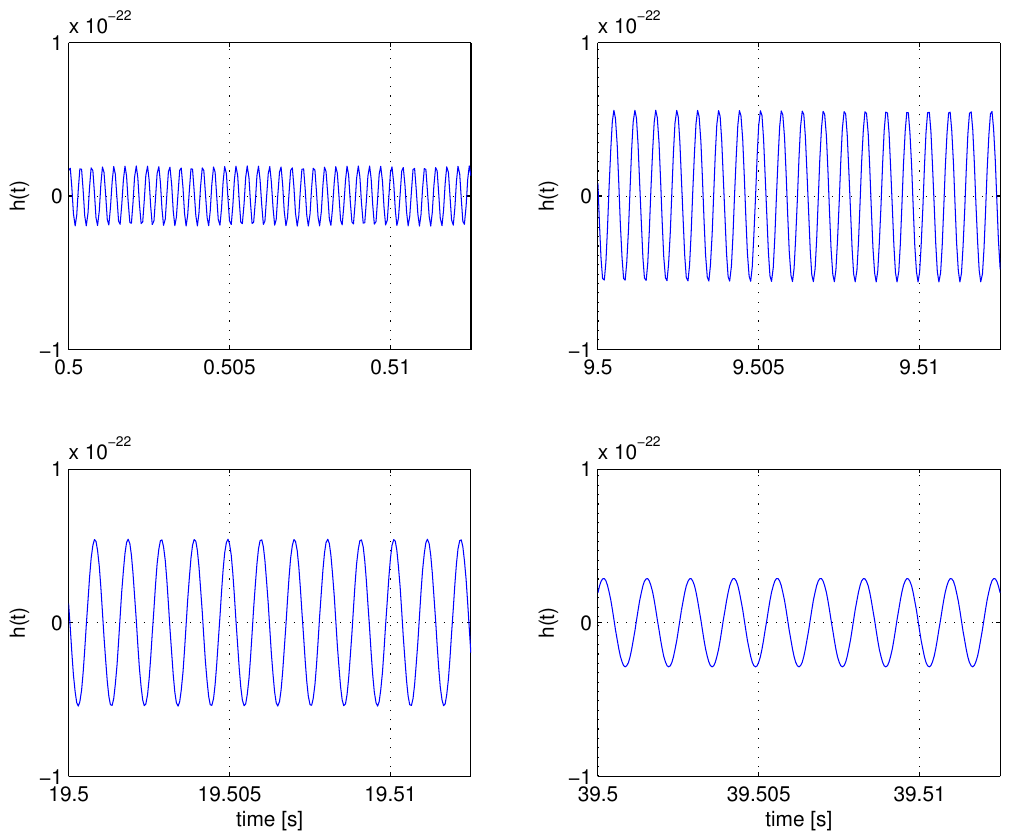}}
\caption{(Left panel.) The neutral stability curves for buckling modes expressed in terms of the critical rotating index $q_c$ as a function of minor-to-major radius $\delta=b/a$. Labels refer to azimuthal quantum numbers $m=1, 2, \cdots$, where instability sets in above and stability sets in below. Of particular interest is the range $q \le 2$, where $m=0$ is Rayleigh-stable. For $q = 2$, instability sets in for $b/a < 0.7385$ ($m = 1$), 0.3225 ($m = 2$) and, asymptotically, for $b/a< 0.56/m$ ($m \ge 3$). Included are the curves of $q_c$ as a function of $\delta$ at various temperatures $T=T_{10}\,10^{10}$ K. (Reprinted from \citep{van02,van12}.) (Middle and right panels.)
Theoretical wave form of ISCO waves induced around an initially extremal black hole obtained by integration of the equations of suspended accretion in the strong interaction limit $L_{gw}\simeq L_H\simeq -\dot{M}$. Shown is the orientation averaged strain amplitude $h(t)/\sqrt{5}$. Due to a turbulent background accretion flow, phase-incoherence is anticipated to be limited intermediate time scales. The wave form is sliced into chirp templates of intermediate duration $\tau=1$\,s for use in matched filtering. (Reprinted from \citep{van08}, working in the idealized limit
$\Omega_T \simeq \Omega_{ISCO}$.)}
\label{fig-fsc1}
\label{fig:p6a}
\end{figure}

\subsection{Gravitational strain frequency and amplitude}

The wave instability (\ref{EQN_APB4}) has the desirable property that the associated mass-moments are predominantly at the lowest quantum numbers, which ensures that most of the gravitational wave output is at the lowest quadrupole emission frequency. Around a stellar mass black hole of $10M_\odot$, the resulting frequency is broadly in the range of 500-3000 Hz for emissions from quadruple mass moments within the sensitivity wave band of the upcoming advanced ground based detectors. Emissions from $m=3$ and higher are unlikely to be detectable by these detectors. (Emissions from $m=1,2$ in a disk or torus produce quadrupole emissions at the same frequency from the combined black hole plus disk or torus system). Even so, their output may be of interest to future, next generation detectors. The same conclusion holds for gravitational wave spectra produced by magnetic pressure induced multipole mass moments based on a numerical simulation \citep{bro06}. 
{Accordingly, the frequency of gravitational wave emissions is fixed by the Kerr metric for a given mass and angular momentum of the black hole with those of quadrupole emission satisfying}
\begin{eqnarray}
f_{gw} < 2f_{ISCO},
\label{EQN_fgw}
\end{eqnarray}
where $\Omega_T\le \Omega_{ISCO}$ denotes the angular velocity of the torus formed about or beyond the ISCO. 
At late times, when the angular velocity of the black hole approaches that of the ISCO, we have for an initial black hole mass $M$ the frequency range \citep{van11}
\begin{eqnarray}
f_{gw} <  595 - 704 \,\mbox{Hz}\,\left(\frac{M}{10M_\odot} \right)^{-1},
\label{EQN_APB5}
\end{eqnarray}
where the 15\% frequency range 595-704 Hz refers to different choices of black hole initial spin. 

%In (\ref{EQN_APB5}), the minimum value of 595 Hz corresponds to an initially maximally spinning black hole in 
%the idealized limit $\Omega_T \simeq \Omega_{ISCO}$.
%For stellar mass black hole systems, this ensures that frequencies are within the LIGO-Virgo and KAGRA bandwidth of sensitivity. According to Figs. (\ref{figGW170817EE}) and (\ref{fig-T}), in reality we may expect frequencies to be considerably lower by quadrupole mass-moments at radii a few times that of the ISCO - by a factor of about three in case of GW170817EE. 

{The limits (\ref{EQN_fgw}-\ref{EQN_APB5}) serve as strict upper bounds (for quadrupole emission). For thick tori, $f_{gw}$ will be 
considerably lower, when the quadrupole mass-moment giving rise to the emission develops over an extended radius $r=Kr_{ISCO}$.
For relatively low stellar mass black holes, such {\em fortuitously moves $f_{gw}$ closer to the minimum of detector strain noise of the present ground 
based detectors} (cf. Fig. 19). 
With $K\simeq 3$ estimated in (\ref{fig-T}), {this appears to be the case in the Extended Emission from the black hole-torus system following delayed 
collapse of a hyper-massive neutron star post-merger to GW170817} (Fig. 4).}

\begin{figure}
\centerline{\includegraphics[scale=0.6]{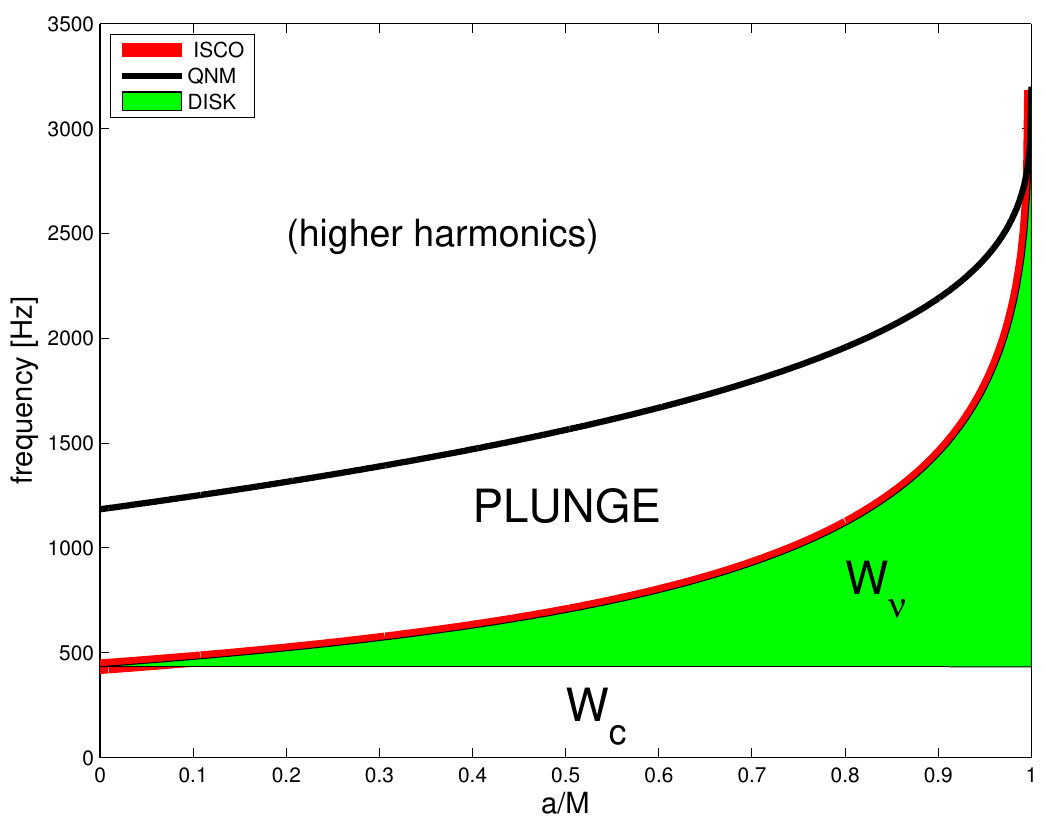}}
\caption{Schematic overview of quadrupole gravitational radiation derived from accretion flows onto rotating black holes in non-axisymmetric wave patterns ($W_b$) or spiral in of fragments ($W_c$), or ISCO waves (red curve). The first produces one or more ascending chirps at sufficiently high accretion rates, the second descending chirps. Fragmentation may occur at sufficient cooling in the extended accretion disk. Wave instabilities from a torus about the ISCO arise from heating and magnetic pressure by feedback from the central black hole if $\Omega_H > \Omega_{ISCO}$. High frequency radiation can be produced by Quasi-Normal Mode ringing of the event horizon that may be exciting by matter plunging in (black curve). The spectrum may contain additional radiation from
higher order modes (not shown), e.g., by high $m$ multipole mass moments in the disk or torus, as well as fragments in elliptical orbits (not shown).(Reprinted from \cite{lev15}.)}
\label{fig:fisco}
\end{figure}

\begin{figure*} %[A]
\centerline{\includegraphics[scale=0.65]{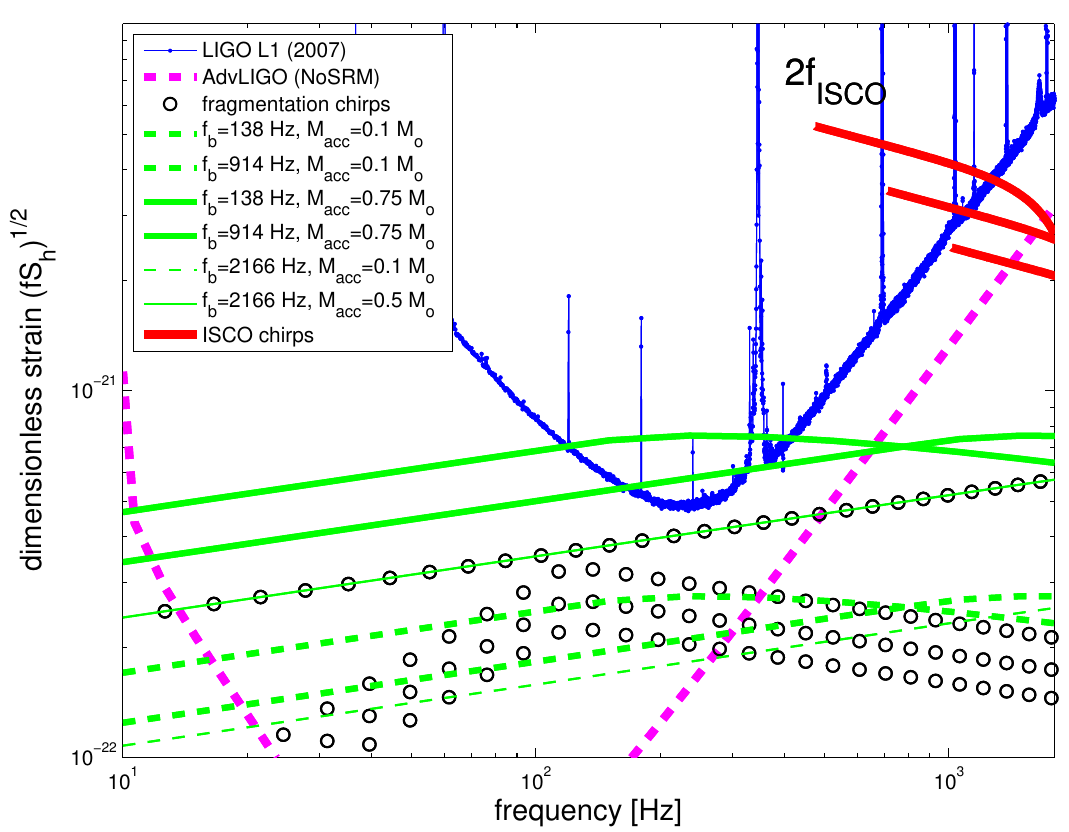}}
%\centerline{\includegraphics[scale=0.35]{f26}}
\caption{Overview of the characteristic strain $h_{char}(f)$ of quadrupole gravitational radiation from accretion flows around rotating black holes formed in core-collapse of massive stars at $D=100$ Mpc. The vertical distance to the dimensionless strain $h_n=\sqrt{fS_h}$ in LIGO S5 represents the maximal attainable S/N ratio obtainable by filtering. The model curves shown are broadband emission from non-axisymmetric accretion flows (green), fragmentation chirps of \citep{pir07} (circles, 
$\sigma_f=0.1$, $f_{e} = 120$ Hz) and ISCO waves induced by feedback from a central black hole (red). 
The curves shown refer to a black hole mass $M=10M_\odot$ (black), $M=7,10$ and $15M_\odot$ (green, red). (Reprinted from \cite{van16}.)}
\label{fig:hchar}
\end{figure*}

The instantaneous strain implies an {\em effective} strain $h_{eff}=\sqrt{n}h$, where $n\simeq fT$ at a characteristic frequency $f$ for a burst duration $T$. 
Thus, $h_{eff} \simeq \pi^{-1} D^{-1} \sqrt{E/f}$ represents the strain that can be recovered by matched filtering in the ideal limit of matching the entire signal 
with a model. Similar to the derivation of (\ref{EQN_inefficD}), we have, in the frequency domain \citep{fla98a,fla98b}, 
the corresponding characteristic amplitude
\begin{eqnarray}
h_{char}(f) \simeq \frac{\sqrt{2}}{\pi D} \sqrt{ \frac{dE}{df} } \simeq 3.5\times 10^{-21}\,D_{100}^{-1} \,M_1
\label{EQN_heff}
\end{eqnarray} 
based on $dE/df \simeq 4.5\times 10^{12} M_1^2$ cm$^{2}$ across a frequency range of 600 - 1500 Hz $(10M_\odot/M)$.% calculated in the next section.

Figs. \ref{fig:fisco}-\ref{fig:hchar} summarize the frequencies and strain amplitude of quadrupole emissions from accretion flows and a torus about the ISCO. Concluding this section, we anticipate, further to (\ref{EQN_TeT90}), that for the gravitational wave emissions most likely observable by LIGO-Virgo and KAGRA with durations
$T_{gw}\simeq T_{engine}$, where $T_{gw}$ represents either the time scale of hyper-accretion or the lifetime of spin of the central black hole. 

\section{Stochastic background from core-collapse supernovae}

In a homogeneous isotropic universe, the contribution to the stochastic background in gravitational waves from a source population locked to the cosmic star formation rate (SFR) can be expressed in terms of the spectral energy density $\epsilon_B^\prime$ per unit volume per unit frequency. It is common to express the same per unit logarithmic frequency relative to the closure density, i.e., $f\epsilon_B^\prime$ relative to $\rho_c$, where $f$ denotes the observed frequency in gravitational waves and $\rho_c = 3H_0^2/8\pi$ in geometrical units for a present-day Hubble constant $H_0$.  It can be calculated \citep{van05b} based on (a) Einstein's adiabatic relationship $E=hf$ for the energy $E$ of a graviton of frequency $f$, where $h$ is Planck's constant; (b) conservation of radiation energy within a co-moving volume during cosmological evolution; and (c) a scaling of the SFR over the cosmic evolution described by the Hubble constant $H(z)=H_0h(z)$ \citep[e.g.][]{por01}. In three-flat $\Lambda$CDM at late times, $h(z)=\sqrt{1-\Omega_{m,0}+\Omega_{m,0}(1+z)^3}$ with present-day matter (dark and baryonic) density $\Omega_{m,0}$ expressed relative to $\rho_{c,0}$.

The event number density per unit redshift $N(z)$ and the event rate volume density $R(z)$ satisfy $N(z)dz=R(z)dt_e$ where $dz$ refers to the redshift interval with a corresponding locally measured time interval $dt_e$ (measured in the source frame). Consequently,
\begin{eqnarray}
\begin{array}{l}
N(z) = R(z) \frac{dt_e}{dz} dz = R(z) \frac{dt_e}{dt} \frac{dt}{dr} \frac{dr}{dz} = \frac{R(z) dz}{(1+z)H_0h(z)},
\end{array}
\label{EQN_B1}
\end{eqnarray} 
where $dt_e/dt = 1/(1+z)$ is the cosmological dilation in time, $dr/dt=1$ is the velocity of light measured at $z=0$ and $dr/dz=(H_0h(z))^{-1}$ denotes the change in proper distance with respect to redshift. According to (c) derived in \cite{por01}, the cosmic SFR expressed in terms of a rate per unit volume $R_{SF2}(z,\Omega_\Lambda)$ in a three-flat cosmology parametrized by $\Omega_\Lambda$ satisfies
\begin{eqnarray}
R_{SF2}(z,\Omega_\Lambda) = R_{SF2}(z,0)\frac{h(z)}{h_0(z)},~~h_0=(1+z)^\frac{3}{2}.
\label{EQN_B2}
\end{eqnarray}

By Einstein's adiabatic relationship (a), $dE/df$ is redshift invariant as a function of $(1+z)f$ for a given $f$. Hence, by conservation of radiation (b), we have (cf. \citep{phi01} for a closely related expression)
\begin{eqnarray}
\epsilon_B^\prime(f) = \int_0^{\infty} \frac{dE}{df}((1+z)f) N(z) dz,
\label{EQN_B3}
\end{eqnarray}
where $z_{max}$ denotes the maximal redshift in the cosmic SFR model rate (\ref{EQN_B2}). For a gravitational wave source locked to the cosmic SFR, 
\begin{eqnarray}
N(z)=N_0 \frac{R_{SF2}(z,\Omega_\Lambda) }{ R_{SF2}(0,\Omega_\Lambda)},
\end{eqnarray}
where $N_0$ denotes the observed rate volume density at $z=0$. By (\ref{EQN_B2}), it follows that
\begin{eqnarray}
\begin{array}{l}
\epsilon_B^\prime(f) = n_0 \int_0^{\infty} \frac{dE}{df}\left[(1+z)f\right] \frac{\hat{R}_{SF2}(z)}{(1+z)^\frac{3}{2}} \,dz, n_0=\frac{R_0}{H_0},
\end{array}
\label{EQN_B4}
\end{eqnarray}
where $\hat{R}_{SF2}(z)=R_{SR2}(z,0)/R_{SF2}(0,0)$ denotes the normalized cosmic SFR satisfying $\hat{R}_{SF2}(0,0)=1$.

A concrete example of a cosmic SFR model rate is \citep{por01}
\begin{eqnarray}
R_{SF2}(z,0) = \frac{0.16 h_{73} }{1+660 e^{-3.4(1+z)}}\,M_\odot \,\mbox{yr}^{-1}\,\mbox{Mpc}^{-3},
\label{EQN_B5}
\end{eqnarray}
where $H_0=h_{73} \times 73$ km s$^{-1}$ Mpc$^{-1}$, whereby
\begin{eqnarray}
\begin{array}{l}
\hat{R}_{SF2}(z) = \frac{23}{1+660 e^{-3.4(1+z)}},~~
N(z)= \frac{23N_0}{(1+660 e^{-3.4(1+z)})(1+z)^\frac{3}{2}}
\end{array}
\label{EQN_B5b}
\end{eqnarray}
is an approximation to the observed cosmic SFR over the redshift range $0\le z \le 5$. For a given observational parameter $N_0$ and source model $dE/df$, $\epsilon_B^\prime(f)$ can thus be evaluated by numerical integration. In what follows, we shall write $E_f = dE/df$. Thus, the Schwarz inequality provides an a priori bound on $\epsilon_B^\prime(f)$, given by
\begin{eqnarray}
\begin{array}{l}
\epsilon_B^\prime(f) \le n_0A_2E_2,~E_2= \sqrt{  \int_0^\infty  E_f^2(x) \frac{dx}{x} },~~
A_2=\sqrt{ \int_0^{\infty} \frac{\hat{R}^2_{SF2}(z)}{(1+z)^2} \,dz}
\end{array}
\label{EQN_B6}
\end{eqnarray}
with $A_2=12.72$.% for (\ref{EQN_B5}). 

For a source effectively described by a constant $E_f=E_f^0$ over a finite bandwidth $B=f_2-f_1$ between two cut-off frequencies $f_{1,2}$, i.e., 
we have
\begin{eqnarray}
\begin{array}{l}
\epsilon_B^\prime(f) = n_0E^0_f \int_{1+z=f_1/f}^{1+z=f_2/f} \frac{\hat{R}(z)}{(1+z)^\frac{3}{2}}\,dz \le n_0E^0_f A_\frac{3}{2},~~
A_\frac{3}{2}=\int_0^\infty \frac{\hat{R}(z)}{(1+z)^\frac{3}{2}}\,dz,
\end{array}
\label{EQN_B7a}
\end{eqnarray}
where $A_\frac{3}{2}=5.8$. The maximum of $\epsilon_B^\prime(f)$ attains at $\epsilon_B^{\prime\prime}(f) = 0$, i.e.,
\begin{eqnarray}
\hat{R}\left(\frac{f_1}{f}-1\right) = \sqrt{\frac{f_1}{f_2}} \hat{R}\left(\frac{f_2}{f}-1\right).
\label{EQN_B7b}
\end{eqnarray}
For a cosmic SFR that is asymptotically constant, i.e., $\hat{R}(z)\simeq \hat{R}_*$ at large $z$, e.g., $\hat{R}_*=23$ in (\ref{EQN_B5}),
(\ref{EQN_B7b}) reduces to the implicit equation
\begin{eqnarray}
\hat{R}\left(\frac{f_1}{f}-1\right) = \sqrt{\frac{f_1}{f_2}}\hat{R}_*
\label{EQN_B7c}
\end{eqnarray}
whenever $f_2>>f_1$. To exemplify, $f_1=600$ Hz and $f_2=3000$ Hz associated with a black hole mass $M=10\,M_\odot$ imply a maximum at 
\begin{eqnarray}
f_{B,peak}=272\,\mbox{Hz}
\label{EQN_B7d}
\end{eqnarray}
as the root of (\ref{EQN_B7c}) for the model rate (\ref{EQN_B5}) (with a corresponding $z_c=0.84$). Note that $f_2/f_c=11.0$, which is still within the redshift range of star formation. 
At this frequency, (\ref{EQN_B7a}) gives the maximum $\epsilon_B^\prime(f) = n_0E^0_f A_*$,
where $A_*=3.52$. In this approximation, we have, consequently,
\begin{eqnarray}
\Omega_B = \frac{B\epsilon_B^\prime(f)}{\rho_c} \simeq 10^{-8}\, \left(\frac{k}{10\%}\right)
\label{EQN_B7f}
\end{eqnarray}
for a bandwidth $B\simeq 1000$ Hz in gravitational waves from SN Ib/c. Here, we consider a branching ratio $k$ of SN Ib/c into broad line events that may successfully produce long gravitational wave bursts.

\section{Searches for broadband extended gravitational radiation}

To search for broadband extended gravitational radiation from transient events with a finite energy reservoir $E_J$, we focus on
chirps - ascending or descending in frequency over a duration $T$ of seconds or more. To this end, we use {\em butterfly filtering}: matched 
filtering against a dense bank of time-symmetric chirp-like templates. It differs from Fourier-based spectrograms \citep{sut10,pre12,thr13,thr14,cou15,abb15,gos15} 
by bandpass filtering of signals with a finite slope $\left|df(t)/dt\right|\ge\delta > 0$ for some $\delta >0$ \citep{van17b}. %(Fig. \ref{figKol}, Fig. \ref{figmf}). 
It permits searches for un-modeled signals of long duration, e.g., seconds to tens of seconds, different from transients of sub-second durations 
\citep[e.g.][]{moh12} without clustering or path algorithms \citep[e.g.][]{cha07,cha17} using templates with relative bandwidths of less 
than 10\% \citep{van16,van17b} (Fig. 20). } 
As phase coherence in gravitational-wave emission from (magneto-)hydrodynamic sources such as accretion flows onto compact objects
is expected to be limited to intermediate time scales $\tau$, the templates used are of duration 
\begin{eqnarray}
P<<\tau<<T
\end{eqnarray} 
for periods $P$ and total
burst durations $T$. Thus, $\tau$ is one of the search parameters in a probe for gravitational-wave emission.
Butterfly filtering produces single detector output in terms of spectrograms, that may be merged across two detectors by frequency coincidences.

\begin{figure}
\centerline{\includegraphics[scale=1.00]{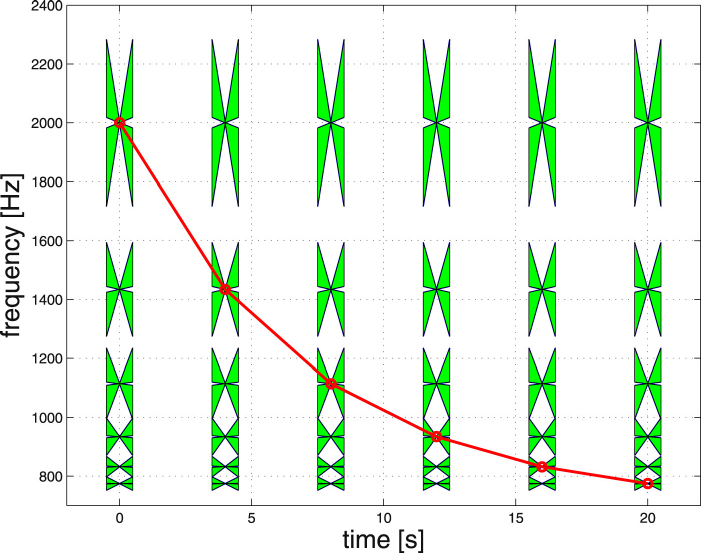}}
\caption{Butterfly filtering is a bandpass filter of trajectories of long-duration chirps with finite slope $0<\delta \le df(t)/dt$ in frequency $f(t)$, suppressing signals with essentially constant frequencies. Butterfly filtering is realized by matched filtering against a bank of chirp templates, here of intermediate duration $\tau = 1$\,s and covering of bandwidth of 350-2000 Hz in frequency. (Reprinted from \cite{van16}.)} 
\label{figWb}
\end{figure}

While LIGO data in the shot-noise dominated frequency range 350-2000 Hz is largely Gaussian over intermediate bandwidths 
of our templates over $\tau = 0.5$\,s or $\tau=1\,$s, it features frequent spurious signals, commonly referred to as {\em glitches}, some of which
may be triggered by earthquakes \citep{acc10,bis18}, e.g., a seven minute long descending chirp EQ170223 in H1 triggered by a $4.1\sigma$
earthquake 80s prior in Belfair \citep{van19c}. 
Furthermore, the LIGO data sets are large. LIGO S6 alone covers well over one year of observations in 1 TB of data; LIGO O1(O2)
offers 586 (3113) frames of 4096\,s of simultaneous H1 and L1 data.

Butterfly filtering derives its sensitivity (Eq. 26) by {linear amplification} of candidate signals according to the theory of matched filtering, 
{\em prior} to correlating the output of two detectors (by frequency coincidences \citep{van19a,van19b}). It permits
deep searches in LIGO data (Fig. \ref{figGW170817EE}) similar to that in the time-series of GRB light curves (Fig. \ref{figKol}). 
Acceleration on {\em Graphics Processor Units} (GPU) is required
to produce broadband spectrograms using a large bank of templates and detailed image analysis thereof. A heterogeneous
compute platform comprising an appreciable number of GPU-CPU nodes permits {\em blind searches} over an entire observational run
such as S6, O1-2 or, upcoming, O3 with no triggers from electromagnetic radiation or neutrinos. This is ideally suited to search
for un-modeled signals including descending chirps from newly formed black holes post-merger to mergers of a neutron star
with a companion neutron star (Fig. \ref{figGW170817EE}) or black hole including those formed in nearby core-collapse supernovae. 

\section{Summary and future prospects}

We reviewed some novel prospects for multi-messenger emission from energetic CC-SNe forming black holes and their associated
long GRBs, hinted at by current theory and observations. We are led to this outlook by prospects for non-axisymmetric inner disks
around black holes (Fig. 2), the large energy reservoir $E_J$ in angular momentum of rapidly rotating Kerr black holes, the formation of 
high-density matter in SN19817A (Fig. 3), and calorimetric evidence for black hole spin-down during GRB170817A (Fig. 4) 
at frequencies emitted from a thick torus extending beyond the ISCO (Fig. \ref{fig-T}). 
A similar outcome might be produced by SGRBEE's, and mergers of neutron stars with a black hole companion.

While GRBs are rare, their parent population of CC-CNe are far more numerous, including supernovae of type Ib/c. At a fraction of these
share the same central engine, it appears opportune to search for multi-messenger emission relatively energetic CC-SNe that may
harbor newly formed black holes, derived from non-axisymmetric accretion flows similar to those considered for post-merger emission 
from GW170817, summarized in Table 4. In particular, signals from non-axisymmetric accretion flows onto rotating black holes may appear as 
\begin{enumerate}
\item {\em Ascending chirps} from accretion flows with fragments or non-axisymmetric wave patterns (Fig. 16), formed within a critical radius where cooling conspires with self-gravity or where angular momentum loss by gravitational radiation is dominant over angular momentum loss by viscous transport;
\item {\em Descending chirps} from waves in a disk or torus by catalytic conversion of black hole spin energy $E_J$ via an inner torus magnetosphere. Non-axisymmetric waves are expected from heating and enhanced magnetic pressure, balanced by cooling in gravitational radiation, MeV neutrino emission and magnetic winds. In this process, frequencies descend in time as the ISCO expands with diminishing angular momentum of the black hole.
\end{enumerate}
These emissions may be preceded by QNM ringing and the random black hole kicks in black hole formation, during a surge in black hole mass prior to the formation of an accretion disk, possibly continued during further growth to a nearly extremal black hole leading up to an accompanying LGRB. 

\begin{table*}
\textbf{Table 4.} Parameter estimates of broadband extended gravitational radiation from accretion flows onto rotating black holes.
\begin{tabular}{llllll}
\hline
{\sc Quantity} &  {\sc Scale} &  {\sc Comment} & {\sc Ref.} \\
\hline\hline
{\sf BH-disk}\\
\hskip0.2in $M$			&	$10^1$ $M_\odot$   			& & \citep{bai98,woo06}	 \\ 
\hskip0.2in $E_{rot}^H$		&       $1\, M_\odot c^2$			& & \citep{ker63} \\
\hskip0.2in $f_{ISCO}$		&       $10^{2-3}$ Hz	 			& & (\ref{EQN_APB5})\\
\hskip0.2in $M_{acc}$		& 	$ 10^{-1}-10^0$ $M_\odot$  	& $\dot{M}\tau$ & (\ref{EQN_A6C}) \\
\hskip0.2in $M_{T}$			&       $10^{-2}$ $M$   			& & (\ref{EQN_g4b})\\
\hline
{\sf Long GW-burst} 		&							& & \\
\hskip0.2in efficiency			&	$< 50\%$		 		         & & Fig. \ref{fig-T} \\
\hskip0.2in ${\cal E}_{gw}$	&	$< 0.1\,M_\odot c^2$			& & Fig. \ref{fig-T} \\
\hskip0.2in $T_{gw}$  	        &      $10^1$ s	 				& $T_{90},t_{ff}, T_{spin}$ & Figs. 11-12, (\ref{EQN_TsA})  \\
\hskip0.2in $L$				&	$0.1M_\odot c^2$ s$^{-1}$	& & Fig. \ref{fig-T} \\
\hskip0.2in $\tau_s$                  &      1 s                                             & & (\ref{fig_exponential})
\\
\hline
{\sf Accretion} \\
{\hskip0.08in  Fragments} 		&										& 						& \citep{pir07} \\	
 \hskip0.2in $f_c < f_{gw} < 2f_{ISCO}$	&   $10^0-10^3$ Hz 							& 						& (\ref{EQN_WC})\\
\hskip0.2in $h_{char}(f)$			&   $1.7\times10^{-22}(f/f_{e})^{\frac{2}{3}}$ 		& $f_c<f<f_{e}$ 			&  cf. (\ref{EQN_h1987A}) \\
\hskip0.2in $h_{char}(f)$			&   $1.7\times10^{-22}(f/f_{e})^{-\frac{1}{6}}$ 		& $f_{e} < f < 2f_{ISCO}$		& (\ref{EQN_hchar2b2}) \\
\hskip0.08in { Disk waves} 		& 										& 						& \citep{lev15} \\
\hskip0.2in   $f_{gw}<2f_{ISCO}$		&  $10^0-10^3$ Hz 							& 						& \\
\hskip0.2in $h_{char}(f)$			&   $1.2\times10^{-21}(f/f_b)^{\frac{1}{6}}$ 		&$f<f_b$ 				&  (\ref{EQN_A6D}) \\
\hskip0.2in $h_{char}(f)$			&   $1.2\times10^{-21} (f/f_b)^{-\frac{1}{6}}$ 		&$f_b < f < 2 f_{ISCO}$		& (\ref{EQN_A6D}) \\
\hskip0.08in { ISCO waves}  		& 										& 						& \citep{van02,van03}\\
\hskip0.2in $f_{gw}<2f_{ISCO}$ 	&  $10^{2-3}$ Hz 							& 						&  (\ref{EQN_APB5}) \\
 \hskip0.2in $h_{char}(f)$			&  $3.4\times10^{-21}$  						& 		& (\ref{EQN_heff}) \citep{van01b} \\
\hline
\hline
\end{tabular}
\label{TABLE_4}
%}
\mbox{}\\\hskip0.08in
\end{table*}

While emanating in the proximity of black holes well-described by the Kerr metric, these signals are to be searched for as 
essentially un-modeled due to a gradual loss of phase coherence over an extended time of emission. 
Observations of un-modeled signals will be based on features indicating correlated behavior across two or more gravitational-wave detectors.
The latter may appear in merged spectrograms produced by butterfly filtering over dense banks of time-symmetric chirp-like 
templates of intermediate duration (capturing phase-coherence) by heterogeneous computing, for an individual event in the Local Universe 
or the stochastic background of their astrophysical source population. 
Significance will critically depend on the sharpness at which physical properties of the central engine - $(M,J)$ and $(K,\tau_s)$ - are resolved, 
by image analysis of candidate features in merged spectrograms.

A major scientific objective is true calorimetry on all emission channels to identify the true nature of the inner engine of energetic CC-SNe,
furthering our observational evidence of black hole spin-down in the Extended Emission post-merger to GW170817.
Upcoming LIGO-Virgo and KAGRA observations hereby promise to decisively identify the physical mechanism of the most extreme transients
in the Universe and provide new insights in the diversity of CC-SNe and GRBs, long and short including SGRBEE's, and their relation to
mergers of neutron stars with another neutron star or black hole. Based on GW170817EE, they may be found in all-sky blind searches in 
gravitational-wave data up to distances of about 100Mpc in LIGO-Virgo O3 \citep[e.g.][]{heo15}.

\begin{acknowledgments} 
This work was performed in part at Aspen Center for Physics, which is supported by National Science Foundation grant PHY- 1607611.
The first author gratefully acknowledges support from the National Research Foundation of Korea under grants 2015R1D1A1A01059793,
2016R1A5A1013277 and 2018044640. This work made use of LIGO O2 data from the LIGO Open
Science Center provided by the LIGO Laboratory and LIGO Scientific Collaboration. LIGO is funded by the U.S. National Science Foundation. Additional support
is acknowledged from MEXT, JSPS Leading-edge Research Infrastructure Program, JSPS Grant-in-Aid for Specially Promoted Research 26000005, MEXT
Grant-in-Aid for Scientifc Research on Innovative Areas 24103005, JSPS Core-to-Core Program, Advanced Research Networks, and the joint research program of
the Institute for Cosmic Ray Research. AL acknowledges supported by a grant from the Israel Science Foundation no. 1277/13. 
The BeppoSAX mission was an effort of the Italian Space Agency ASI with the participation of The Netherlands Space Agency NIVR. 
F. Frontera, C. Guidorzi and L. Amati acknowledge financial support from the Italian Ministry of Education, University, and Research through the PRIN-MIUR 2009 project on Gamma-Ray Bursts (Prot. 2009 ERC3HT). 
\end{acknowledgments}

\appendix
\setcounter{secnumdepth}{0}
\section{{\bf Appendix A.}  Gravitational radiation}

The Einstein equations are a hyperbolic-elliptic system of equations for a four metric $g_{ab}$ with signature $(-,+,+,+)$ in the line element 
\begin{eqnarray}
ds^2 = g_{ab}dx^adx^b.
\label{EQN_line1}
\end{eqnarray}
This can be made explicit following a foliation of space-time in Cauchy surfaces \citep{arn62} or in a Lorentz gauge on SO(3,1) connections in the Riemann-Cartan formalism \citep{van96}. The resulting gravitational wave motion is subject to elliptic constraints given by conservation of energy and momentum. 

\subsection{Hyperbolicity}

The linearized equations of motion about Minkowski space-time reveal the two modes of
transverse gravitational waves of spin two. Traditionally, the derivation is given in 3+1 in a special choice of coordinates, that exploits gauge covariance in the choice of coordinates. Here, we note a derivation utilizing the constraints of energy-momentum conservation.

To start, consider slicing of space-time into space-like hypersurfaces of constant coordinate time $t=x^0$. 
Let $h_{ij}$ denote the three-metric intrinsic to these hypersurfaces, that are coordinated by the 
remaining $x^i$ $(i=1,2,3)$. The line-element can be equivalently expressed as (e.g. \citep{tho86})
\begin{eqnarray}
ds^2 = -\alpha^2 dt^2 + h_{ij}\left( dx^i + \beta^i dt\right) \left( dx^j + \beta^j dt\right),
\label{EQN_AP1}
\end{eqnarray}
where $(\alpha,\beta^i)$ denote the lapse function and, respectively, shift functions. In foliating space-time into three-dimensional 
hypersurfaces, the $(\alpha,\beta^i)$ are a gauge, and are not dynamical variables. The three-metric $h_{ij}$ defines parallel
transport of vectors in the hypersurfaces of constant coordinate time $t$ and, as such, comes with a covariant three-derivative $D_i$
and associated Christoffel symbols $\Gamma_{ij}^k$. 

Following (\ref{EQN_AP1}), the Ricci tensor $^{(4)}R$ of $g_{ab}$ expands to $^{(3)}R$ of $h_{ij}$ and quadratic terms of the extrinsic 
curvature tensor $K_{ij} = -1/(2\alpha) L_t h_{ij}$, where $L_t$ denotes the Lie derivative of $h_{ij}$ with respect to the coordinate time $t$. The Hilbert action for $g_{ab}$ hereby reveals explicit contributions from ``potential" and ``kinetic" energies in
\begin{eqnarray}
S= \frac{1}{16\pi} \int \left( ^{(3)}R + K_{ij}K^{ij} - K^2\right) \alpha \sqrt{h} dx^3 dt.
\label{EQN_AP2}
\end{eqnarray}
Following \citep{arn62}, variations with respect to the non-dynamical variables $(\alpha, \beta^i)$ obtain the conservation laws of
energy and momentum, given by the constraints
\begin{eqnarray}
R - K_{ij}K^{ij} + K^2 = 0,~~D^iK_{ij} - D_j K = 0,
\label{EQN_AP3c}
\end{eqnarray}
where $R=R_{i}^i$ denotes the trace of the Ricci tensor of $h_{ij}$.
Variation with respect to $h_{ij}$ gives the first-order evolution equation 
\begin{eqnarray}
L_t K_{ij} =\left(R_{ij} - D_iD_j\right) \alpha + \left( K K_{ij} - 2K_i^m K_{jm}\right) \alpha,
\label{EQN_AP3a}
\end{eqnarray} 
where $L_t K_{ij} = \partial_t K_{ij} - \left( K_i^m D_m\beta_j + K_j^m D_m \beta_i + \beta^m D_m K_{ij} \right)$. Similar to the latter, we may expand $L_t h_{ij}$ to obtain the first-order evolution equation
\begin{eqnarray}
\partial_t h_{ij} = D_i \beta_j + D_j \beta_i - {2\alpha} K_{ij}.
\label{EQN_AP3b}
\end{eqnarray} 
Combined, (\ref{EQN_AP3c}-\ref{EQN_AP3b}) define a constraint Hamiltonian system of equations for the
dynamical variables $(h_{ij},K_{ij})$. 

The existence of gravitational waves represents the hyperbolic structure of (\ref{EQN_AP3c}-\ref{EQN_AP3b}). 
This becomes explicit by analysis of harmonic perturbations in the curvature driven gauge \citep{van10,van12b}
\begin{eqnarray}
\partial_t \alpha = -K, ~~\beta^i = 0.
\label{EQN_AP4a}
\end{eqnarray}
We next consider small perturbations about Minkowski space-time $(h_{ij} = \delta_{ij},\alpha=1,K_{ij}=0$), where
$\delta_{ij}$ denotes the Kronecker delta symbol, that is,
$h_{ij} = \delta_{ij} + \delta h_{ij}$, $\alpha = 1 + \delta \alpha$ and $K_{ij} = \delta K_{ij}$. In the gauge (\ref{EQN_AP4a}),
(\ref{EQN_AP3a}-\ref{EQN_AP3b}) imply
\begin{eqnarray}
\partial_t^2 h_{ij} = - 2 \left( R_{ij} - D_iD_j \delta \alpha\right) ,~~\partial_t^2 K = \Delta K,
\label{EQN_AP4b}
\end{eqnarray}
where we used that $^{(3)}R$ is of second order according to the Hamiltonian energy constraint in (\ref{EQN_AP3c}),
Here, we recall the perturbative expansion \citep{wal84}
\begin{eqnarray}
R_{ij} = - \frac{1}{2} \Delta \delta h_{ij} + \frac{1}{2} \partial_i \partial^e \delta \bar{h}_{ej} + \frac{1}{2} \partial_j \delta \bar{h}_{ej},
\label{EQN_AP4c}
\end{eqnarray}
where $\bar{h}_{ij} = \delta h_{ij} - \frac{1}{2} \delta_{ij} \delta h$, $\delta h = \delta^{ij} \delta h_{ij}$. 
A harmonic plane wave $\delta h_{ij} = \hat{h}_{ij} e^{-i\omega t} e^{ik_ix^i}$ (similarly for $K_{ij}$) of angular frequency $\omega$ with wave vector $k_i$ can be applied to (\ref{EQN_AP3a}) and (\ref{EQN_AP4a}), giving $k_i\hat{K}_{ij} = k_j \hat{K}$, $-i\omega \hat{\alpha} = - \hat{K}$, $\delta \hat{h}_{ij}=-2i\omega^{-1} \hat{K}_{ij}$ and, for $\partial_i\partial^e\bar{h}_{ej}$, 
\begin{eqnarray}
%\begin{array}{ll}
k_i k^e \hat{h}_{ej} - \frac{1}{2} k_i k_j \delta \hat{h} =  
i \omega^{-1} \left( - 2k_i k^e \hat{K}_{ej} + k_i k_j \hat{K}\right) = - i\omega^{-1} k_i k_j \hat{K}.
%\end{array}
\label{EQN_5a}
\end{eqnarray}
By (\ref{EQN_AP4c}), there follows that
$\hat{R}_{ij} - \partial_i\partial_j \hat{\alpha} = \frac{1}{2} k^2 \hat{h}_{ij} - i\omega^{-1} k_i k_j \hat{K} + i\omega^{-1} k_i k_j \hat{K},$
whereby the first evolution equation in (\ref{EQN_AP4b}) reduces to the dispersion relation
\begin{eqnarray}
\omega^2 = k^2
\label{EQN_5b}
\end{eqnarray}
of propagation along light cones in a local Minkowski background space-time. 

The Einstein equations, $G_{ab} = 16\pi T_{ab}$ describe the response of space-time curvature to a stress-energy
tensor $T_{ab}$ of matter and fields, where $G_{ab} = ^{(4)}R_{ab} - \frac{1}{2}g_{ab}^{(4)}R$ is the Einstein tensor.
Following (\ref{EQN_AP4c}) and (\ref{EQN_5b}), the covariant wave equation for perturbations 
$g_{ab} = \eta_{ab} + \delta g_{ab}$ in the four-metric on a fixed background space-time with metric $\eta_{ab}$ is
\begin{eqnarray}
\Box_\eta \delta g_{ab} = -16\pi T_{ab},
\label{EQN_Gab}
\end{eqnarray}
where $\Box$ denotes the d'Alembertian associated with $\eta_{ab}$. The coefficient $-16\pi$ in (\ref{EQN_Gab}) results 
from the factor $-\frac{1}{2}$ in (\ref{EQN_AP4c}). 

Searches for contemporaneous emission in electromagnetic and gravitational from cosmological GRBs have been suggested to test for gravitons to be massless as described by (\ref{EQN_Gab}). However, these tests only serve to identify differences in masses of gravitons and photons, as may be seen by expressing wave motion of both in terms of four vector fields. Let $\omega_{a\mu\nu}$ denote the Riemann-Cartan connection of four-dimensional space-time in the SO(3,1) tetrad formalism and $A_a$ denote the vector potential of the electromagnetic field. In the Lorentz gauge to both \citep{van96}, propagation in vacuum satisfies
\begin{eqnarray}
%\begin{array}{ll}
\hat{\Box} \omega_{a\mu\mu} - R_a^c\omega_{c\mu\nu} - [\omega^c,\nabla_a\omega_c]_{\mu\nu} = 0,~~
\Box A_a - R_a^cA_c =0,
%\end{array}
\label{EQN_APWE}
\end{eqnarray}
where $\hat{\Box}$ denotes the d'Alembertian associated with the SO(3,1) gauge covariant derivative 
$\hat{\nabla}_a = \nabla_a + [\omega_a,\cdot]$. In the presence of a cosmological constant $\Lambda>0$
\citep{rie98,per99}, $R_{ab} = \Lambda g_{ab}$, whereby (\ref{EQN_APWE}) becomes
\begin{eqnarray}
%\begin{array}{ll}
\hat{\Box} \omega_{a\mu\mu} - \Lambda\omega_{c\mu\nu} - [\omega^c,\nabla_a\omega_c]_{\mu\nu} = 0,~~
\Box A_a - \Lambda A_c =0,
%\end{array}
\label{EQN_APWEb}
\end{eqnarray}
showing gravitons and photons of the same effective mass $m=\sqrt{\Lambda}\simeq 10^{-29}$ cm$^{-1}$ in
geometrical units, as defined by the dispersion relation of (\ref{EQN_APWEb}). 

\subsection{Luminosity in gravitational radiation}

Consider the transverse traceless perturbations \citep{mis73}
\begin{eqnarray}
\delta h^{TT}_{ij} = \left( \begin{array}{ccc} h_+ & h_\times & 0 \\ h_\times & -h_+ & 0 \\ 0 & 0 & 0 \end{array} \right) =
h_+ e_{ij}^+ + h_\times e_{ij}^\times,
\label{EQN_AP6a}
\end{eqnarray}
decomposed in the two linear polarization tensors $e_{ij}^+$ and $e_{ij}^\times$ of gravitational waves
in the $(x,y)$ plane orthogonal to the direction of propagation along the $z-$axis. The perturbed line-element (\ref{EQN_line1}) 
now assumes the form
\begin{eqnarray}
ds^2=\eta_{ab} dx^a dx^b + h_+(dx^2-dy^2) + 2h_\times dx dy.
\label{EQN_line2}
\end{eqnarray}
With rotational symmetry over an angle $\pi$ about the $z-$axis, gravitational waves are of spin-2 \citep{fie39}.
In the far field region away from a source region, the Hilbert action (\ref{EQN_AP2}) reduces
to the kinetic term $K_{ij}K^{ij}$, and hence by (\ref{EQN_AP3b}) to 
\begin{eqnarray}
S = \frac{1}{16\pi} \int \frac{1}{2}\left[ \left(\partial_a h_+\right)^2 + \left( \partial_a h_\times\right)^2 \right] dx^3 dt.
\label{EQN_AP2r}
\end{eqnarray}
We can now read off the stress-energy tensor of gravitational wave motion:
\begin{eqnarray}
t^{00} = t^{0z} = t^{zz} = \frac{1}{16\pi} \left < \dot{h}_+^2 + \dot{h}_\times^2 \right> 
\label{EQN_AP2s}
\end{eqnarray}
The two polarization wave modes (\ref{EQN_AP6a}) and the associated gravitational wave stress-energy tensor (\ref{EQN_AP2s})
are characteristic properties of general general relativity. Alternative theories may have additional degrees of freedom \citep{sat09}. 
 
For a source described by a stress-energy tensor $T_{ab}$, the gravitational wave emission results from the associated
time harmonic perturbations in the tidal gravitational field. Following (\ref{EQN_Gab}), we have in response to a distance source
over a region $V$
\begin{eqnarray}
\delta h_{ij}(r,t)= \frac{4}{r} \int_V T_{ij}(t-r,x^i) d^3x.
\label{EQN_AP2w}
\end{eqnarray}
Explicit evaluation for a circular binary of point masses $M_1$ and $M_2$, orbital frequency $\Omega$ and orbital
separation $a$ shows the quadrupole gravitational wave formula \citep{wal84,tho02}
\begin{eqnarray}
L_{gw} = \frac{32}{5} \Omega^6 a^4 \mu^2 = \frac{32}{5} \left(\Omega \mu\right)^{\frac{10}{3}}
\label{EQN_AP2gw}
\end{eqnarray}
in units of $L_0 = c^5/G$, where $\mu = M_1^{\frac{3}{5}}M_2^\frac{3}{5}/(M_1+M_2)^\frac{1}{5}$ denotes the chirp mass,
$c$ is the velocity of light and $G$ is Newton's constant.   

To see (\ref{EQN_AP2gw}), we first recall the following identity for a mass distribution with velocity four-vector $u^b$ (e.g. \citep{tho02}) 
\begin{eqnarray}
\int_V T^{ij}d^3x = \frac{1}{2}\partial_0^2 I_0^{ij} 
\label{EQN_idT}
\end{eqnarray} 
between $\int_V T^{ij} d^3x = \int_V u^iu^jdm$, $dm=\rho d^3x$ and the moment of inertia tensor
\begin{eqnarray}
I^{ij}_0 = \int_VT^{00}x^ix^j d^3x \simeq \int_V x^ix^j dm,
\label{EQN_Iij}
\end{eqnarray}
where the latter refers to the non-relativistic limit $u^0\simeq 1$. The identity (\ref{EQN_idT}) follows from the
conservation of energy-momentum, $\nabla_aT^{ab}=0$. About a flat space-time background, $\nabla_a\nabla_bT^{ab}=0$ 
implies $\partial_0^2T^{00} + 2\partial_0\partial_i T^{0i} + \partial_i\partial_jT^{ij}=0$, i.e.,
$\partial_0^2T^{00}-\partial_i\partial_jT^{ij}=0$
using $\partial_0T^{00}+\partial_iT^{0i}=0$. Integration by parts twice of $\partial_0^2 \int_V x^ix^j T^{00} d^3x$ = $\int_V x^ix^j \partial_k\partial_l T^{kl}$ obtains (\ref{EQN_idT}). 

Consequently, (\ref{EQN_AP2w}) gives for the {\em traceless} metric perturbations
\begin{equation}
h_{ij}^{T}(t,r)=\frac{2}{r}\frac{d^2I_{ij}(t-r)}{dt^2},~~{I}^{jk}=I^{jk}_0-\frac{1}{3}\delta^{jk}\delta_{lm}I^{lm}_0.
\label{qmom}
\end{equation}
By (\ref{EQN_AP2r}), we arrive at the gravitational wave luminosity 
\begin{equation}
L_{gw}=\frac{dE_{GW}}{dt}=\frac{1}{5}
\langle{ \frac{d^3I_{jk}}{dt^3} \frac{d^3I^{jk}}{dt^3}} \rangle,
\label{dEgw/dt}
\end{equation}
taking into account and reduction factor 2/5 as only two of the five degrees of freedom in the traceless metric perturbation $\delta h^T_{ij}$ are physical degrees of freedom representing outgoing gravitational radiation \citep{tho02}.

\subsection{Radiation from multipole mass moments}

Consider a ring having cross-sectional radius $b$ and  density $\rho$, rotating around a central object
in the $(x,y)$ plan at angular velocity $\Omega$ in a circular orbit of radius $r$.  Let $m=\int_V{\rho d^3x}$
denote the total mass of the ring, where the integration is over the ring's volume $V$.  We restrict the analysis to a thin ring, $b<<r$,
and compute $I^{jk}$ to order $O(b^2/r^2)$.  Let $(x',y')$ denote a Cartesian coordinate system rotating with the ring.  One can always 
choose the axis such that  
\begin{eqnarray}
 I^{x'x'}_0=\frac{1}{2}mr^2(1+\xi),~~
 I^{y'y'}_0=\frac{1}{2}mr^2(1-\xi),~~
 I^{x'y'}_0=0.
\end{eqnarray}
to order  $O(b^2/r^2)$.  Here, $\xi$ quantifies the mass quadrupole inhomogeneity, with $\xi=0$ for an axi-symmetric ring.
For example, for a ring having a density  $\rho=\rho_0+\rho_2\cos^2\theta$ in cylindrical coordinates, with $\rho_0$ and $\rho_2$ being constants,
one obtains  $\xi=m_2 /4m$, where $m=\int_Vd^3x\rho$ is the total mass, and $m_2=\int_Vd^3x\rho_2$.  
Now, transforming to the non-rotating frame, 
\begin{equation}
\left(\begin{array}{c} x\\ y
\end{array}\right)
=\left(\begin{array}{cc}
 x'\cos\Omega t & y'\sin\Omega t\\
-x'\sin\Omega t & +y'\cos\Omega t
\end{array}\right)
\end{equation}
yields 
\begin{eqnarray}
\begin{array}{l}
 I^{xx}_0=I^{x'x'}_0\cos^2\Omega t+I^{y'y'}_0\sin^2\Omega t=\frac{1}{2}mr^2(1+\xi\cos2\Omega t),\nonumber\\
 I^{yy}_0=I^{x'x'}_0\sin^2\Omega t+I^{y'y'}_0\cos^2\Omega t=\frac{1}{2}mr^2(1-\xi\cos2\Omega t),\label{Ijk}\\
 I^{xy}_0=\frac{1}{2}(I^{x'x'}_0-I^{y'y'}_0)\sin2\Omega t=-\frac{1}{2}mr^2\xi\sin2\Omega t.\nonumber
\end{array}
\end{eqnarray}
By employing (\ref{qmom}) and (\ref{Ijk}) one has
\begin{equation}
{\frac{d^3I^{ij}}{dt^3}} =4\xi mr^2\Omega^3
\left(\begin{array}{cc}
\sin\Omega t & \cos\Omega t\\
\cos\Omega t & -\sin\Omega t
\end{array}\right).
\end{equation}
Substituting into Equation (\ref{dEgw/dt}) finally gives
\begin{equation}
L_{gw}=\frac{32}{5}\xi^2m^2 r^4\Omega^6.
\label{EQN_ALGW}
\end{equation}
The quadrupole formula for a circular binary of point masses is obtained upon taking $\xi=1$, $m=\mu$ and $r=a$, here $\mu$ is the reduced mass and $a$ is the binary separation. 

In (\ref{EQN_ALGW}), the limit $\xi=1$ obtains the canonical formula of quadrupole gravitational wave emission. 
A circular binary of masses $M_i$ $(i=1,2)$ with chirp mass $\mu = (M_1M_2)^{5/3}(M_1+M_2)^{-1/5}$. For a 
matched filtering detection method, the relevant quantity is the amplitude that takes into account the square root
of the associated number of wave periods. In the frequency domain, the corresponding quantity is the
{\em characteristic strain amplitude}, given by the square root of the energy per unit logarithmic frequency interval
\citep{fla98a},
\begin{eqnarray}
h_{char}(f) = \frac{\sqrt{2}}{\pi D}\sqrt{\left|\frac{dE}{df}\right|}.
\label{EQN_hchar0}
\end{eqnarray}
Consider a circular binary with small mass-ratio $\sigma=M_2/M_1<<1$, so that $M=M_1+M_2\simeq M_2$.
At an orbital separation $a$, it emits quadrupole gravitational radiation at a frequency $\pi M f = (M/a)^\frac{2}{3}$.
The total energy $E=-\frac{1}{2}\sigma M^2/a=-\frac{1}{2}M\sigma (M\pi f)^\frac{2}{3}$ hereby shrinks, whereby (cf. \citep{mis73})
\begin{eqnarray}
\frac{dE}{df} = - \frac{\pi \sigma M^2}{3 (\pi M f)^\frac{1}{3}}
\label{EQN_hchar1}
\end{eqnarray}
and hence (cf. \citep{tho98,ju00}) 
\begin{eqnarray}
\begin{array}{l}
h_{char}(f) = 8.6 \times 10^{-22} \,\sigma^\frac{1}{2} M_1{^\frac{1}{3}} \left(\frac{D}{100\,\mbox{Mpc}}\right)^{-1} \left(\frac{f}{100\,\mbox{Hz}}\right)^{-\frac{1}{6}}
\end{array}
\label{EQN_hchar2}
\end{eqnarray}
for a central mass $M=M_1\times 10\,M_\odot$. This shows that the low frequency emission at early in spiral is particularly important
for detection.

In the presence of ellipticity $e$, the luminosity in gravitational waves is greater than (\ref{EQN_ALGW}) by additional
radiation at frequency harmonics $m>2$ \citep{pet63}. The result can be expressed by an enhancement factor $F(e)\ge1$. 
The time rate of change in orbital frequency satisfies
\begin{eqnarray}
\dot{f}_{orb} = \frac{96}{5} (2\pi)^{\frac{8}{3}} f_{orb}^{\frac{11}{3}} F(e), F(e)=\frac{1+\frac{73}{24}e^2+\frac{37}{96}e^4}{(1-e^2)^\frac{7}{2}}.
\label{EQN_Aforb1}
\end{eqnarray}
Given an initial ellipticity $e_0$, the orbital separation $a=a(t)$ hereby satisfies
\begin{eqnarray}
a(t) = a_0 \left( 1 - \frac{t}{\tau_0}\right)^\frac{1}{4}, \tau_0 = \frac{5a^4_0}{256 M_1M_2(M_1+M_2) }f(e_0),  
\label{EQN_Aforb2}
\end{eqnarray}
where $f(e_0)\le 1$ $(e_0\ge 0)$ obtains as an integral over $0\le e \le e_0$ (see, e.g., \cite{pos06} for a detailed expression).

It should be mentioned that (\ref{EQN_Tho1a}-\ref{EQN_Tho1b}) are derived for sources about the Minskowski background space-time. In an applying to the  multipole mass-moments of a strongly magnetized torus about the ISCO of a rotating black hole \citep{bro06}, the radiation is emitted in a strongly curved space-time. It requires extending (\ref{EQN_Tho1a}) by an additional grey body factor, that represents suppression of radiation at low $l$ for a given $I_{lm}$. (For a related discussion, see, e.g., \cite{van99}.) The grey body factor derives from scattering of relatively low frequency gravitational waves in the curved space-time around black holes, that results in partial absorption by the black hole. However, in a suspended accretion state which balances heating by input from the black hole and cooling in gravitational radiation \citep{van12}, $I_{lm}$ is self-regulated such that $L_{lm}$ in (\ref{EQN_Tho1a}) times such grey body factor balances with the energetic input from the black hole. Emissions from $I_{lm}$ beyond the ISCO are relatively less affected by space-time curvature. At large distances away from the black hole, the grey body factor is effectively one. A detailed derivation of the grey body factor is beyond the scope of this review. 

\section{{\bf Appendix B.} Relativistic frame dragging}

The Kerr metric in Boyer-Lindquist coordinates $(t,r,\theta,\phi)$ explicitly brings about the Killing vectors $k^b=(\partial_t)^b$ and $m^b=(\partial_\phi)^b$ of time slices of constant coordinate time $t$. It gives an exact solution of frame dragging in terms of the angular velocity $\omega$ of particles of zero-angular momentum. In Boyer-Lindquist coordinates $(t,r,\theta,\phi)$ of the Kerr metric \citep{tho86}, the world line of zero-angular momentum observers (ZAMOs, \citep{tho86} are orthogonal to slices of constant time-at-infinity. The angular velocity $\omega=d\phi/dt$ decays with the cube of the distance to the black hole at large distances. 

By frame dragging, the ISCO of corotating orbits shrinks to the event horizon from $6M$ around a non-rotating Schwarzschild black hole. This appears in X-ray spectroscopy of MCG 6-30-15 \citep{iwa96}. These results are time variable over a time scale of a year, that may reflect intermittency in the inner radius of the disk \citep{fab95} or, alternatively, in circumnuclear clouds intermittently absorbing disk emissions.

The complete gravitational field induced by the angular momentum and mass of a rotating black hole is described by the Riemann tensor. For completeness, we here include a brief summary of earlier derivations on the associated energetic interactions \citep{van12}. 

\subsection{Gravitational spin-orbit energy}

Consider the tetrad 1-forms
\begin{eqnarray}
%\begin{array}{l}
e_{(0)}= \alpha dt,~~ 
e_{(1)} =\frac{\Sigma}{\rho}(d\phi - \omega dt)\sin\theta,~~
e_{(2)}= \frac{\rho}{\sqrt{\Delta}} dr,~~
e_{(3)} = \rho d\theta,
%\end{array}
\end{eqnarray}
where $\alpha={\rho}{\Sigma}^{-1}\sqrt{\Delta}$ is the redshift factor, $\Sigma^2=(r^2+a^2)^2-a^2\Delta\sin\theta$, $\rho=r^2+a^2\cos^2\theta$, $\Delta=r^2-2Mr+a^2$ and $\omega={2aMr}{\Sigma}^{-2}$ is the angular 
velocity of frame dragging. The Riemann tensor has the following non-zero components \citep{cha83}
\begin{eqnarray}
\begin{array}{rcl}
R_{0123} &=& A,~~R_{1230} = AC,~~R_{1302} = AD\\
-R_{3002} &=& R_{1213} = -3aA\sqrt{\Delta}\Sigma^{-2}(r^2 + a^2)\sin\theta\\
-R_{1220} &=& R_{1330} = -3aB\sqrt{\Delta}\Sigma^{-2}(r^2 + a^2)\sin\theta\\
-R_{1010} &=& R_{2323} =  B = R_{0202} + R_{0303}\\
-R_{1313} &=& R_{0202} =  BD,~~-R_{1212} = R_{0303} = -BC,
\end{array}
\label{EQN_R}
\end{eqnarray}
where
\begin{eqnarray}
\begin{array}{lcl}
A = aM\rho^{-6}(3r^2 - a^2 \cos^2\theta),~~
B  =  Mr\rho^{-6}(r^2 - 3a^2 \cos^2\theta),\\
C = \Sigma^{-2}[(r^2 + a^2)^2 + 2a^2\Delta\sin^2\theta], ~~
D  =  \Sigma^{-2}[2(r^2 + a^2)2 + a^2\Delta\sin^2\theta].
\end{array}
\end{eqnarray}
About the black hole spin axis ($\theta=0$), $2A = -\partial_r\omega = {2aM}{\rho^{-6}}(3r^2 - a^2)$, $C = 1$, $D = 2,$
$J$ induced components appear in the first three of (\ref{EQN_R}). Integrating the Papapetrou force on a test particle with velocity
four-vector $u^b$ satisfying \citep{pap51}
\begin{eqnarray}
F_2 =\frac{1}{2}\epsilon_{abef}R^{cf}_{cd}J_p^au^bu^d = J_pR_{3120} = J_pAD = -\partial_2\omega J_p
\end{eqnarray}
gives 
\begin{eqnarray}
{E}=\int_r^\infty F_2 ds.
\label{EQN_E2}
\end{eqnarray}

Alternatively, consider the angular velocity $\Omega=u^{\phi}/u^t$. The normalization $-1 = u^cu_c = \left[g_{tt} + g_{\phi\phi}\Omega(\Omega-2\omega)\right](u^t)^2$ gives two roots 
\begin{eqnarray}
\Omega_{\pm} = \omega \pm \sqrt{\omega^2 - (g_{tt} + (u^t)^{-2})/g_{\phi\phi}}.
\label{EQN_ompm}
\end{eqnarray}
Two particles with the same angular momentum in absolute value,
\begin{eqnarray}
%\begin{array}{ll}
J_{p,\pm} = g_{\phi\phi}u^t(\Omega_{\pm}+\omega) = 
  g_{\phi\phi}u^t \sqrt{\omega^2-(g_{tt}+(u^t)^{-2})/g_{\phi\phi}}=\pm J_p
%  \end{array}
\end{eqnarray}
hereby have (with the same $u^t$) the total energies $E_{\pm} = (u^t)^{-1} + \Omega_{\pm}J_{\pm}$. One-half the difference satisfies
\begin{eqnarray}
{E} =\frac{1}{2}(E_+ - E_-) = \omega J_p.
\label{EQN_E3}
\end{eqnarray}

As a gravitational interaction, the curvature-spin coupling (\ref{EQN_E3}) acts universally on angular momentum,
whether mechanical or electromagnetic in origin.

The result of (\ref{EQN_E2}) combined with canonical pair-creation processes will
be a possibly force-free outflow along open magnetic field lines along the black hole spin axis, such as envisioned in \citep{bla77}. Intermittent inner engines hereby produce outgoing Alfv\'en fronts, that communicate the raw Faraday induced potential within the inner engine (roughly, on the event horizon of the black hole) out to large distance. The result may thus produce a linear accelerator ahead of the Alfv\'en front in regions of relatively low opacity, facilitating the production of UHECRs \citep{van09}.

The structure of force-free outflows is a limit of ideal MHD \citep{glo14}, which neglects inertia (and hence Reynolds stresses) 
in addition to being free of dissipation of the electromagnetic field. Originally, this limit was motivated to model
extragalactic outflows, e.g., \citep{fan74}, but increasingly this limit appears to be relevant also to 
extreme sources such as GRBs (e.g. \citep{lyu03,lyu03b}). 

Let $p_B=B^2/8\pi$ and $e_B=B^2/8/pi$ denote the magnetic pressure and energy density.
In a magnetic flux tube of radius $R$, the dissipationless limit implies adiabatic compression:
$p_B(2\pi RdR)=d(\pi e_BR^2)$, i.e., the magnetic flux $\Phi=\pi BR^2$ is {\em frozen} 
into the fluid. In contrast, a torsional perturbation mediating angular momentum outflow creates 
an {\em Alfv\'en wave} with velocity \citep{lic67}
\begin{eqnarray}
v_A=\frac{B}{\sqrt{4\pi\rho+B^2}},
\label{EQN_VA}
\end{eqnarray}
where $\rho$ denotes the fluid density as seen in the comoving frame. The Alfv\'en wave
is purely rotational, leaving density (and magnetic flux) invariant.  It should be mentioned that (\ref{EQN_VA}) is unique to 
MHD in U(1). It does not generalize to colored MHD \citep{van94}. 

Neglecting inertia, the Alfv\'en velocity reaches the velocity of light. Neglecting Reynolds stresses,
\begin{eqnarray}
F_{ab}j^b=0,
\label{EQN_FF}
\end{eqnarray}
which reduces the number of degrees of freedom in the electromagnetic field to two. 
For an electric current $j^b=\rho_e v^b$ associated with a charge density $\rho_e$ with four-velocity $v^b$, 
(\ref{EQN_FF}) implies $v^i\partial_iA_\phi=0$ and $v^i\partial_iA_0=0$ for a time-independent tube 
$A_\phi=$const. along the polar axis $\theta=0$. The electric potential hereby satisfies $A_0=A_0(A_\phi)$,
and the electric field $\partial_iA_0=A_0^\prime\partial_iA_\phi$, 
in the Boyer-Lindquist frame of reference, is normal to the flux surfaces. Force-free flux 
surfaces are equipotential surfaces (\citep{gol69,bla77,tho86}). 

Alfv\'en surfaces in force-free outflows from Intermittent inner engines can thus transmit Faraday induced potentials outwards. They can produce linear accelerators upstream at large distances from the source, providing a suitable condition for
the creation of UHCRs from ionic contaminants by, e.g., UV-irradiation from a surrounding torus in AGN.

\subsection{Alfv\'en waves in a torus magnetosphere}

Consider the electromagnetic two-tensor $F_{ab}$ \citep{lic67} 
\begin{eqnarray}
{\bf F} = {\bf u}\wedge {\bf e} + * {\bf u} \wedge {\bf h}
\end{eqnarray}
in the four-vector representation $(u^b,e^b,h^b)$ associated with a time like unit tangent $u^b$, $u^cu_c=-1$, 
of ZAMOs. Following \citep{bar72,tho86}, %tho82
we have the one-form ${\bf u} = -\alpha {\bf d}t$ 
with redshift $\alpha$. Then ${\bf u}=\alpha^{-1}({\bf k}+\omega {\bf m})$
is linear combination of the Killing vectors, satisfying $\nabla_cu^c=0$. ZAMOs measure an electric field $e^b$ and a magnetic field $h^b$, $e^b=u_cF^{ac}$ and $h^b=u_c*F^{cb},$ each with three degrees of freedom given $u^ce_c=u^ch_c=0$. The same ZAMOs
observe ${\bf e}=(0,E^i)$ and ${\bf h}=(0,B^i)$, where $i=1,2,3$ refers to the coordinates of the 
surfaces of constant $t$. The star $*$ denotes the Hodge dual, satisfying $*^2=-1$ in four dimensions. 

Faraday's equation 
\begin{eqnarray}
\nabla_a*F^{ab}=0
\end{eqnarray}
can be expanded by considering $\nabla_a (u^ah^b-u^bh^a) = {L}_u h^b + (\nabla_cu^c)h^b-(\nabla_ch^c)u^b,$
where ${L}_u h^b = (u^c\nabla_c)h^b - (h^c\nabla_c)u^b$ denotes the Lie-derivative of $h^b$ with respect to $u^b$. 
Projected onto surfaces of constant $t$ (orthogonal to $u^b$), we have
\begin{eqnarray}
\left({L}_{\bf u} {\bf h}\right)_\perp = \alpha^{-1}\left(\partial_t {\bf B} + {L}_\omega {\bf B}\right)
\end{eqnarray}
evaluated in the frame of ZAMOs, where $L_\omega$ is the Lie-derivative with respect to $\omega^i\equiv\omega m^i$ 
($m^i$ is not a unit three-vector). Next, $\nabla_a = D_a- u_a(u^c\nabla_c)$ and $(*{\bf u}\wedge {\bf h})_{abcd} = \epsilon_{abcd}u^ce^d$. With acceleration 
$(u^c\nabla_c)u_b=\alpha^{-1}\nabla_b\alpha$, consider
$\nabla^b(\epsilon_{abcd}u^ce^d)=\epsilon_{abcd}(D^bu^c)e^d-\epsilon_{abcd}u^ba^ce^d
+\epsilon_{abcd}u^c\nabla^be^d.$ Projection of the right hand side onto the space like coordinates
$i=(r,\theta,\phi)$ normal to $u^b$ satisfies
\begin{eqnarray}
%\begin{array}{ll}
\epsilon_{ibcd}(D^bu^c)e^d+\tilde{\epsilon}_{ijk}a^je^k+\tilde{\epsilon}_{ijk}\nabla^je^k= 
\epsilon_{ibcd}(D^bu^c)e^d+\alpha^{-1}\tilde{\epsilon}_{ijk}\nabla^j(\alpha e^k),
%\end{array}
\end{eqnarray}
where $\epsilon_{aijk}u^a=\tilde{\epsilon}_{ijk}=\sqrt{h}\Delta_{ijk}$ with
$\sqrt{-g}=\alpha\sqrt{h}$ over the space like volume element $\sqrt{h}$,
where $\Delta_{ijk}$, $\Delta_{123}=1$. The first term on the right hand side vanishes, since $D_bu_c$ is spacelike:
$u^b(D_bu_c)=0$ by construction and $u^cD_bu_c=0$ by $u^2=-1$. Consequently, Faraday's law includes an 
additional term (derived alternatively in \citep{tho86} and references therein)
\begin{eqnarray}
\tilde{\nabla}\times \alpha {\bf E} = -\partial_t {\bf B} + 4\pi {J}_m,
\label{EQN_FAR}
\end{eqnarray}
where $\tilde{\nabla}_i=D_i$ and
\begin{eqnarray}
{J}_m = -\frac{1}{4\pi} { L}_\omega {\bf B}.
\label{EQN_JM}
\end{eqnarray}
$J_m$ appears analogously to a current of virtual magnetic monopoles. 

Applied to a torus magnetosphere, (\ref{EQN_JM}) satisfies
\begin{eqnarray}
%\begin{array}{ll}
\omega_i { J}_m^i \simeq \frac{1}{8\pi} {\bf B}\cdot \tilde{\nabla}(\omega_i\omega^i) > 0, ~~
\omega_i\omega^i = 4\frac{z^2\sin^2\lambda}{(z^2+\sin^2\lambda)^3} ~(\theta=\frac{\pi}{2}),
%\end{array}
\label{EQN_LOM}
\end{eqnarray}
where the inequality refers to a poloidal ingoing magnetic field.

By (\ref{EQN_LOM}), frame dragging induced poloidal current loops in the inner torus magnetosphere. 
The resulting poloidal Afv\'en waves produces Maxwell stresses by which rotating black holes lose angular
momentum to surrounding matter. The black hole hereby spins down, which should have an imprint on
any light curve derived from high energy emissions derived from (\ref{EQN_E2}), while surrounding
matter is brought into a state of forced turbulence by competing torques acting on the inner and outer
faces \citep{van99}, possibly related to forced turbulence in Taylor-Couette flows \citep{ste09}. 

The Alfv\'en waves effectively transport angular momentum out an onto the torus, provided they are not canceled by Reynolds stresses from a baryon-rich torus wind back into the black hole. In this event, the inner face of the torus will be spun up, whereby it assumes a state of super-Keplerian motion. The resulting differential rotation can induce non-axisymmetric wave instabilities. The associated surface gravity tends to suppress baryon-rich outflows from the inner face of the torus. 
A detailed description of this suppression of Reynolds stresses falls outside the scope of the present discussion, however.

%\newpage\clearpage


\begin{thebibliography}{99}
\bibitem[LIGO(2016)]{LIG16} LIGO Virgo Collaboration, 2016, Phy. Rev. Lett., 116, 241102
\bibitem[Akutsu et al.(2019)]{aku19} Akutsu, T., et al., 2019, Class. Quantum Grav., 36, 165008
\bibitem[Cutler \& Thorne(2002)]{cut02}Cutler, C., \& Thorne, K.S., 2002, in Proc. GR16, Durban, South Afrika
\bibitem[Sathyaprakash \& Schutz(2009)]{sat09} Sathyaprakash, B.S., \& Schutz, B.F., 2009, Liv. Rev. Relativity, 12, 2 %(http://www.livingreviews.org/lrr-2009-2)
\bibitem[Kalogera(2017)]{kal17} Kalogera, V., 2017, Nat. Astron., 1, 0088
\bibitem[Kinugawa et al.(2014)]{kin14} Kinugawa T., Inayoshi K., Hotokezaka K., Nakauchi D., Nakamura T., 2014, MNRAS, 442, 2963
\bibitem[Inayoshi et al.(2017)]{ina17} Inayoshi, K., Hirai, R., Kinugawa, T., \& Hotokezaka, K., 2017, MNRAS, 468, 5020
\bibitem[Qin et al.(2019)]{qin19} Qin, Y., Marchant, P., Fragos, T., Meynet, G., \& Kalogera, V., 2019, ApJ, 870, L18
\bibitem[van Putten \& Della Valle(2017)]{van17a} van Putten, M.H.P.M., \& Della Valle, 2017, MNRAS, 464, 3219
\bibitem[Verbunt(1997)]{ver97} Verbunt, F., 1997, Class. Quantum Grav. 14, 1417
\bibitem[Taylor \& Weissberg(1989)]{tay89} Taylor, J.H., \& Weisberg, J.M., 1989, ApJ, 345, 434
\bibitem[Taylor(1994)]{tay94} Taylor, J.H., 1994, Rev. Mod. Phys., 66, 711
\bibitem[Weisberg et al.(2010)]{wei10} Weisberg, J.M., Nice, D.J., \& Taylor, J.H., 2010, ApJ, 722, 1030
\bibitem[Lyne et al.(2004)]{lyn04} Lyne, A.G., Burgay, M., Kramer, M., Possenti, A., et al., 2004, Science, 303, 1153
\bibitem[Smal(1967)]{sma67} Smak, J. 1967, Acta Astron., 17, 255
\bibitem[Paczy\'nski(1967)]{pac67} Paczy\'nski, B.P., 1967, Acta. Astron., 17, 287
\bibitem[Faulkner(1971)]{fau71} Faulkner, J., 1971, ApJ, 170, L99
\bibitem[Faulkner et al.(1972)]{fau72} Faulkner, J., Flannery, B.P., Warner, B., 1972, ApJ, 175, L79
\bibitem[Nelemans(2005)]{nel05} Nelemans, G., 2005, $in$ The Astrophysics of Cataclysmic Variables and Related Objects, eds. J.-M. Hameury \& J.-P. Lasota, ASP Conf. Ser., 330
\bibitem[Postnov \& Yungelson(2006)]{pos06} Postnov, K.A., \& Yungelson, L.R., 2006, Living Rev. Relativity, 9, 6 (http://www.livingreviews.org/lrr-2006-6)
\bibitem[Bidsten et al.(2006)]{bil06} Bildsten, L., Townsley, D.M., Deloye, C.J., \& Nelemans, G., 2006, ApJ, 640, 466
\bibitem[Woudt \& Warner(2003)]{wou03} Woudt, P.A., \& Warner, B., $in$ Proc. IAU JD5, ``White Dwarfs: Galactic and Cosmological Probes," eds. Ed Sion, Stephane Vennes and Harry Shipman; astro-ph/0310494v1
\bibitem[Espaillat et al.(2005)]{esp05} Espaillat, C., Patterson, J., Warner, B., \& Woudt, P., 2005, PASP, 117, 189
\bibitem[Burrows \& Lattimer(1987)]{bur87} Burrows, A., \& Lattimer, J.M., 1987, ApJ, 318, L63
\bibitem[Garrison et al.(1987)]{gar87} Garrison, R., Shelton, I., Madore, B., Cassatella, A., Wamsteker, W., Sanz, L., Gry, C., 1987, IAUC 4330, 1
\bibitem[Herald et al.(1987)]{her87} Herald, D., McNaught, R.H., Morel, M., et al., 1987, IAUC 4317, 1 
\bibitem[Kunkel et al.(1987)]{kun87} Kunkel, W., Madore, B., Shelton, I., et al., 1987, IAUC 4316, 1
\bibitem[Abbott et al.(2017a)]{abb17} Abbott, B.P., Abbott, R., Abbott, T.D., et al., 2017, Phys. Rev. Lett., 119, 161101
\bibitem[Klebesadel et al.(1973)]{kle73} Klebesadel, R., Strong I. and Olson R., 1973, ApJ, 182, L85
\bibitem[van Putten(2000)]{van00} van Putten, M.H.P.M., 2000, Phys. Rev. Lett., 84, 3752
\bibitem[Galama et al.(1998)]{gal98}Galama, T.J., Vreeswijk, P.M., van Paradijs, J., et al. 1998, Nature, 395, 670
\bibitem[Mirabel et al.(1994)]{mir94} Mirabel, I.F., \& Rodriguez, L.F., 1994, Nature, 371, 46
\bibitem[Thorsett \& Chakrabarty(1999)]{tho99} Thorsett, S.E., \& Chakrabarty, D., 1999, ApJ, 512, 288
\bibitem[Nice et al.(2004)]{nic04} Nice, D.J., Splaver, E.M., \& Stairs, I.H., 2004, $in$ Rasio, F.A., and Stairs, I.H., eds., ``Binary Radio Pulsars, Meeting at the Aspen Center for Physics," ASP Conf. Ser., 328, 371
\bibitem[Bailyn et al.(1998)]{bai98} Bailyn, C.D., Jain, R.K., Coppi, P., \& Orosz, J.A., 1998. ApJ, 499, 367
\bibitem[Frail et al.(2001)]{fra01} Frail, D.A., et al., 2001, ApJ, 567, L41
\bibitem[Ghirlanda et al.(2006)]{ghi06} Ghirlanda, G., Ghisillini, G., \& Firmani, C., 2006, New J. Phys., 8, 123
\bibitem[Ghirlanda et al.(2013)]{ghi13} Ghirlanda, G., Ghisillini, G., Salvaterra, R., et al., 2013, MNRAS, 428, 123 
\bibitem[van Putten et al.(2011b)]{van11b} van Putten, M.H.P.M., Della Valle, M., \& Levinson, A., 2011b, A\&A, 535, L6
\bibitem[BATSE(2001)]{bat01} http://www.batse.msfc.nasa.gov/batse/%grb/duration/images/4b$_-$t90.gif
\bibitem[Maurer et al.(2010)]{mau10} Maurer, J. I., Mazzali, P. A., Deng, J., et al. 2010, MNRAS, 402, 161
\bibitem[van Putten(2001b)]{van01b}van Putten, M.H.P.M., 2001, Phys. Rev. Lett., 87, 091101
\bibitem[Levinson et al.(2002)]{lev02} Levinson, A., et al., 2002, ApJ,  576, 923
\bibitem[Maeda et al.(2002)]{mae02} Maeda, K., et al. 2002, ApJ, 565, 405
\bibitem[Maeda et al.(2008)]{mae08} Maeda, K., et al., 2008, Science, 319, 1220
\bibitem[Fruchter et al.(2006)]{fru06} Fruchter, A.S., et al., 2006, Nature, 441, 463
\bibitem[Soker(2016)]{sto16} Soker, N., 2016, NewA Rev.., 75, 1
\bibitem[Soker(2017)]{sto17} Soker, N., 2017, Res. Astron. Astrophys., 17, 113
\bibitem[Piran et al.(2017)]{pir17} Piran, T., Nakar, E., Mazzalli, P., \& Pian, E., 2018, arXiv:1704.08298
\bibitem[Mazzali et al.(2008)]{maz08} Mazzali, P.A., Valenti, S., Della Valle, S., et al., 2008, Science, 321, 1185
\bibitem[Couch et al.(2011)]{cou11} Couch, S.M., Pooley, D., Wheeler, J.G., \& Milosavljevi\'c, M., 2011, ApJ, 727, 104 
\bibitem[Bromberg et al.(2012)]{bro12} Bromberg, )., et al., 2012, ApJ, 749, 110
\bibitem[Soderberg et al.(2008)]{sod08} Soderberg, A. M., Berger, E., S. Page, K.L., et al. 2008, Nature, 453, 469
\bibitem[Margutti et al.(2014)]{mar14} Margutti, R., Milisavljevic, D., Soderbert, A.M., Guidorzi, C., Morsony, B.J., et al., 2014, ApJ, 797, 107
\bibitem[Nakar \& Sari(2012)]{nak12} Nakar, E., \& Sari, R., 2012, ApJ, 747, 88
\bibitem[Nakar(2015)]{nak15} Nakar, E., 2015, ApJ, 807, 172
\bibitem[Granot et al.(2017)]{gra17} Granot, A., Nakar, E., \& Levinson, A., 2017, arXiv:1708.0518
\bibitem[Obsergaulinger \& Aloy(2017)]{obe17} Obergaulinger, M., \& Aloy, \'A, 2017, MNRAS, 469, L43
\bibitem[van Putten(2015b)]{van15a} van Putten, M.H.P.M., 2015b, MNRAS, 447, L113
\bibitem[M\'eszaros \& Waxman(2001)]{mes01} M\'esz\'aros, P.M., and Waxman, E., 2001, Phys. Rev. Lett., 87, 171102
\bibitem[EHT Collaboration(2019)]{EHT19} EHT Collaboration, 2019, ApJ, 875, L5
\bibitem[van Putten \& Levinson(2012)]{van12b} van Putten, M.H.P.M., \& Levinson, A., 2012, {\em Relativistic Astrophysics of the Transient Universe}  (Cambridge: Cambridge University Press)
\bibitem[Kerr(1963)]{ker63} Kerr, R.P., 1963, Phys. Rev. Lett., 11, 237
\bibitem[Ott(2009)]{ott09} Ott, C.D., 2009, Class. Quant. Grav., 2009, 26, 063001
\bibitem[R\"over et al.(2009)]{rov09} R\"over, C. Bizouard, M.-A., Christensen, N., Dimmelmeier, H., Heng, I.-S., \& Meyer, R., 2009, Phys. Rev. D, 80, 102004
\bibitem[Burrows et al.(2007)]{bur07} Burrows, A., Dessart, L., Livne, E., Ott., C.D., \& Murphy, J., 2007, ApJ, 664, 416 
\bibitem[Dessart et al.(2008)]{des08} Dessart, L., Burrows, A., Livne, E., \& Ott, C.D., 2008, ApJ, 673, L43
\bibitem[Heo et al.(2015)]{heo15} Heo, J.-E., Yoon, S., Lee, D.-S., et al. 2015, NewA, 42, 24
\bibitem[Ando et al.(2013)]{and13} Ando, S., Baret, B., Bartos, I., et al., 2013, Rev. Mod. Phys., 85, 2013
\bibitem[Aasi et al.(2014)]{aas14} Aasi, J., Abbott, B.P., Abbott, T., et al., 2014, Phys. Rev. D, 89, 122004
\bibitem[Haensel et al.(2009)]{hae09} Haensel, P., Zdunik, J. L., Bejger, M., et al. 2009, A\&A, 502, 605

\bibitem[van Putten \& Levinson(2003)]{van03} van Putten, M.H.P.M., \& Levinson, A., 2003, ApJ, 584, 937
\bibitem[van Putten \& Levinson(2002)]{van02b} van Putten, M.H.P.M., \& Levinson, 2002, Science, 294, 1837
\bibitem[Bekenstein(1973)]{bek73} Bekenstein, J. 1973, ApJ, 183, 657
\bibitem[Thuan \& Ostriker(1974)]{thu74} Thuan, T.X., \& Ostiker, J.P., 1974, ApJ, 191, L105
\bibitem[Novikov(1975)]{nov75} Novikov, I.D., 1975, Astron. Zh. 52, 657 (Transl. 1976, Sov. Astron. 19, 398)
\bibitem[Epstein(1976)]{eps76} Epstein, R., Ph.D. Thesis, (Stanford University, Stanford, 1976)
\bibitem[Detweiler \& Lindblom(1981)]{det81} Detweiler, S., \& Lindblom, L., 1981, ApJ, 250, 739
\bibitem[Kotake et al.(2006)]{kot06} Kotake, K., Sato, K., \& Takahashi, K., 2006, Rep. Prog. Phys., 69, 971
\bibitem[Fryer \& New(2011)]{fry11} Fryer, C.L., \& New, K.C.B., 2011, Living Rev. Relativity, 1

\bibitem[van Putten et al.(2014b)]{van14b} van Putten, M.H.P.M., Gyeong-Min, Lee, Della Valle, M., Amati, L., \& Levinson, A., 2014b, MNRASL, 444, L58

\bibitem[Ciufolini et al.(2004)]{ciu04} Ciufolini, I., \& Pavlis, E.C., 2004, Nature, 431, 958.
\bibitem[Ciufolini et al.(2007)]{ciu07} Ciufolini, I., 2007, Nature 449, 41
\bibitem[Ciofolini et al.(2009)]{ciu09} Ciufolini, I., Paolozzi, A., Pavlis, E.C., et al., 2009, Space Sci Rev., 148, 71
\bibitem[Everitt et al.(2011)]{eve11} Everitt, C.W.F., et al., 2011, Phys. Rev. Lett. 106, 221101
\bibitem[van Putten(2013)]{van13} van Putten, M.H.P.M., 2013, Act. Polytechnica, 52, 736
\bibitem[Levinson et al.(2015)]{lev15} Levinson, A., van Putten, M.H.P.M., \& Pick, G., 2015, ApJ, 812, 124

\bibitem[van Putten(2017b)]{van17b} van Putten, M.H.P.M., 2017, PTEP, 93F01
\bibitem[van Putten(2002)]{van02}van Putten, M.H.P.M., 2002, ApJ, 575, L71
\bibitem[Kobayashi \& Meszaros(2003)]{kob03} Kobayashi, S., \& Meszaros, P. 2003, ApJ, 589, 861
\bibitem[Piro \& Pfahl(2007)]{pir07} Piro, A.L., \& Pfahl, E., 2007, ApJ, 658, 1173 
\bibitem[Gamma(2001)]{gam01} Gammie, C. F., 2001, ApJ, 553, 174
\bibitem[Rice et al.(2005)]{ric05} Rice, W.K.M., Lodato, G., \& Armitage, P.J., 2005, MNRAS, 364, L56
\bibitem[Mejia et al.(2005)]{mej05} Mejia, A.C., et al., 2005, ApJ, 619, 1098
\bibitem[Lovelace et al.(2014)]{lov14} Lovelace, R.V.E., \& Romanova,  M.M., 2014, Fluid Dyn. Res., 46, 041401
\bibitem[Hadley et al.(2014)]{had14} Hadley, K.Z., \& Fernandez, P., 2014, Astrophys. Space Sci., 353, 191
\bibitem[Tagger et al.(1990)]{tag90} Tagger, M., Henriksen, R.N., Sygnet, J.F., \& Pellat, R., 1990, ApJ 353, 654
\bibitem[Tagger \& Pellat(1999)]{tag99} Tagger, M., \& Pellat, R., 1999, A\&A, 349, 1003
\bibitem[Tagger(2001)]{tag01} Tagger, M., 2001, A\&A, 380, 750
\bibitem[Bromberg et al.(2006)]{bro06} Bromberg, O., Levinson, A., \& van Putten, M.H.P.M., 2006, NewA, 619, 627
\bibitem[Tagger \& Verni'ere(2006)]{tag06} Tagger, M., \& Varni\`ere, P., 2006, ApJ, 642, 1457
\bibitem[Tagger \& Mella(2006)]{tag06b} Tagger, M., \& Melia, F., 2006, ApJ, 636, L33
\bibitem[Thorne(1980)]{tho80} Thorne, K.S., 1980, Rev. Mod. Phys., 52, 299
\bibitem[Anninos et al.(1993)]{ann93} Anninos, P., Hobill, D., Seidel, E., Smarr, L., \& Suen, W.M., 1993, Phys. Rev. Lett. 71, 2851
\bibitem[Gibbons(1972)]{gib72} Gibbons, G.W., 1972, Commun. Math. Phys. 27, 87
\bibitem[van Putten(2012b)]{van12c} van Putten, M.H.P.M., 2012b, Phys. Rev. D, 2012, 85, 064046
\bibitem[Kyutoku(2013)]{kyu13} Kyutoku, K., 2013, {\em The Black Hole-Neutron Star Binary Merger in General Relativity} (Springer-Verlag)
\bibitem[Szil\'agyi et al.(2015)]{szi15} Szil\'agyi, B., Blackman, J., Buonanno, A., et al., 2015, Phys. Rev. Lett., 115, 031102

\bibitem[Bernuzzi et al.(2015b)]{ber15} Bernuzzi, S., Radice, D., Ott, C.D., et al., 2015, arXiv:1512.06397

\bibitem[Shapiro \& Teukolsy(1983)]{sha83} Shapiro, S.L., Teukolsky, S.A., 1983, Black Holes, White Dwarfs and Neutron Stars. Wiley, New York
\bibitem['t Hooft(2002)]{tho02} 't Hooft, G., 2002, Introduction to General Relativity (Princeton, NJ: Rinton Press)
\bibitem[Peters \& Mathews(1963)]{pet63} Peters, P.C., \& Mathews, J., 1963, Phys. Rev., 131, 435
\bibitem[Thorne(1992)]{tho92} Thorne, K.S., 1992, $in$ Advances in General Relativity, eds. A Janis and J Porter (Boston: Birkhauser)
\bibitem[Ju et al.(2000)]{ju00} Ju, L., Blair, D.G., \& Zhao,  C., 2000, Rep. Prog. Phys., 63, 1317

\bibitem[Bernuzzi et al.(2015a)]{ber15a} Bernuzzi, S., Dietrich, T., \& Nagar, A., 2015, 115, 091101
\bibitem[Baiotti et al.(2008)]{bai08} Baiotti, L., Giacomazzo, B., and Rezzolla, L., 2008, Phys. Rev. D, 78, 084033
%\bibitem[Bernuzzi et al.(2015a)]{ber15a} Bernuzzi, S., Dietrich, T., \& Nagar, A., 2015, 115, 091101

\bibitem[Turtle(1987)]{tur87} Turtle, A.J., et al., 1987, Nature, 327, 38
\bibitem[Papaliosis et al.(1989)]{pap89} Papaliosis, C., Krasovska, M., Koechlin, L., Nisenson, P., \& Standley, C., 1989, Nature, 338, 565 
\bibitem[Nisenson \& Papaliosios(1999)]{nis99} Nisenson, P., \& Papaliosios, C., 1999, ApJ, 518, L29
\bibitem[Mattei et al.(1979)]{mat79} Mattei, J., Johnson, G.E., Rosino, L., Rafanelli, P., Kirshner, R., 1979. IAU Circ. 3348, 1
\bibitem[Pagnaude et al.(2011)]{pat11}Patnaude, D.J., Loeb, A., \& Jones, C., 2011, NewA, 16, 187
\bibitem[Filippenko(1997)]{fil97} Filippenko, A.V., 1997, ARA\&A, 35, 309 
\bibitem[Gilmozzi et al.(1987)]{gil87} Gilmozzi, R., Cassatella, A., Clavel, J., et al., 1987, Nature, 328, 318
\bibitem[Kirshner et al.(1987)]{kir87} Kirshner, R.P., Sonneborn, G., Grenshaw, D.M., Nassiopoulos, G.E., 1987, ApJ, 320, 602
\bibitem[Mazzali et al.(2005)]{maz05} Mazzali, P.A., et al., 2005, Science, 308, 1284
\bibitem[Tauberger et al.(2009)]{tau09} Taubenberger, S., et al., 2009, MNRAS, 397, 677
\bibitem[Modjaz et al.(2014)]{mod14} Modjaz, M., et al., 2014, AJ, 147, 99M

\bibitem[Connaughton(2017)]{con17} Connaughton, V., 2017 GCN, 21505 
\bibitem[Savchenko et al.(2017)]{sav17} Savchenko, V., Ferrigno, C. , Kuulkers, E., et al., 2017, ApJ, 848, L15
\bibitem[Goldstein et al.(2017)]{gol17} Goldstein, A., et al., 2017, ApJ, 848, L14
\bibitem[Pozanenko et al.(2018)]{poz18} Pozanenko, A.S., Barkov, M.V., Minaev, P.Y., et al., 2018, ApJ, 852, L30
\bibitem[Kasliwal et al.(2017)]{kas17b} Kasliwal, M.M., Nakar, E., Singer, L.P., 2017, Science, 358, 1559

\bibitem[van Putten(2005)]{van05b} van Putten, M.H.P.M., 2005, {\em Gravitational radiation, Luminous Black Holes and Gamma-Ray Burst Supernovae} (Cambridge: Cambridge University Press)

\bibitem[Coughlin \& Dietrich(2019)]{cou19} Coughlin, M.W., \& Dietrich, T., 2019, PRD, 100, 043001
\bibitem[Kasen et al.(2017)]{kas17} Kasen, D., Metzger, B., Barnes, J., et al., 2017, Nat., 551, 80
\bibitem[Smartt et al.(2017)]{sma17} Smartt, S.J., Chen, T.-W., Jerkstrand, A., Couthlin, M., Kankare, E., et al., 2017, Nat., 551, 75
\bibitem[Pian et al.(2017)]{pia17} Pian, E., D'Avanzo, P.,Benetti, S., Branchesi, M., Brocato, E., et al. 2017, Nat., 551, 67
\bibitem[D'Avenzo et al.(2017)]{dav17} D'Avanzo, E., Benetti, S., Branchesi, M., Brocato, E., et al, 2017, Nat., 551, 67

\bibitem[van Putten(2019a)]{van19a} van Putten, M.H.P.M., \& Della Valle, M., 2019, MNRAS, 482, L46
\bibitem[van Putten(2019b)]{van19b} van Putten, M.H.P.M., \& Della Valle, M., \& Levinson, A., 2019, ApJ, 876, L2
\bibitem[Mooley et al.(2018a)]{moo18a} Mooley, K.P., Deller, A.T., Gottlieb, O., et al., 2018, Nat. 554, 207
\bibitem[Mooley et al.(2018b)]{moo18b} Mooley, K.P., Deller, A.T., Gottlieb, O., et al., 2018, Nat. 561, 355

%\bibitem[\'O Colgain et al.(2018)]{ogo18} \'O Colg\'ain, E., van Putten, M.H.P.M., \& Yavartanoo, H., 2019, PLB, 793, 126
\bibitem[van Putten(2017)]{van17} van Putten, M.H.P.M., 2017, ApJ, 848, 28
\bibitem[Riess et al.(2019)]{rie19} Riess, A.G., et al., 2019, ApJ, 876, 85

\bibitem[Guidorzi et al.(2017)]{gui17} Guidorzi, C., Margutti, R., Brout, D., Scoling, D., Fong, W., et al., 2017, ApJ, 851, 36

\bibitem[van Putten(2003)]{van03a} van Putten, M.H.P.M., 2003, ApJ, 583, 374

\bibitem[van Putten et al.(2014a)]{van14} van Putten, M.H.P.M., Guidorzi, C.., \& Frontera, P., 2014a, ApJ, 786, 146

%\bibitem[van Putten(2014)]{van14c} van Putten, M.H.P.M., Lee, G.M., Della Valle, M., Amati, L., \& Levinson, A., 2014, MNRASL, 444, L58

\bibitem[Drout et al.(2011)]{dro11} Drout, M.R., Soderberg, A.M., Gal-Yam, A., et al., 2011, ApJ, 741, 97
\bibitem[Li et al.(2011a)]{li11a} Li, W., Leaman, J., Chornock, R., et al. 2011a, MNRAS, 412, 1441
\bibitem[Scolnic et al.(2011)]{pan11} Scolnic, D., Riess, A., Huber, M., et al., 2011, AAS, \#218, 127.09
 \bibitem[Kulkarni et al.(2014)]{kul14} Kulkarni, S.R., Zwicky Transient Factory Proposal, priv. commun.
\bibitem[Belkin(2015)]{bel15} Bellm, E.C., \& Kulkarni, S.R., 2015, AAS Meeting \#225, \#328.04
\bibitem[Colgate(1968)]{col68} Colgate, S. A. 1968, Canadian J. Phys., 46, S476
\bibitem[Colgate(1970)]{col70} Colgate, S. A. 1970, Acta Physica Academiae Scientiarum Hungaricae, 29, Suppl. 1, 353
\bibitem[Colgate(1974)]{col74} Colgate, S. A. 1974, ApJ, 187, 333
\bibitem[Woosley \& Bloom(2006)]{woo06} Woosley, S.E., \& Bloom, J.S., 2006, Ann.Rev.Astron.Astrophys., 44, 507
\bibitem[Gezari(2008)]{gez08} Gezari, S. 2008, ApJ, 683, L131
\bibitem[Campana et al.(2006)]{cam06}Campana, S., Mangano, V., Blustin, A. J., et al. 2006, Nature, 442, 1008
\bibitem[Weaver(1976)]{wea76} Weaver, T. A. 1976, ApJS, 32, 233
\bibitem[H\"offlich et al.(2009)]{hof09} H\"oflich, P., \& Schaefer, B.E., 2009, ApJ, 705, 483
\bibitem[Katz et al.(2010)]{kat10} Katz, B., Budnik, R., \& Waxman, E. 2010, ApJ, 716, 781
\bibitem[Nakar \& Sari(2010)]{nak10} Nakar, E., \& Sari, R., 2010, ApJ, 725, 904
\bibitem[Svirski et al.(2012)]{svi12} Svirski, G, Nakar, E, \& Sari, R. 2012, ApJ, 759, 108
\bibitem[Eichler et al.(1989)]{eic89} Eichler, D., Livio, M., Piran, T., \& Schramm, D. 1989, Nature, 340,126
\bibitem[Paczy\'nski(1991)]{pac91} Paczy\'nski, B.P., 1991, Acta. Astron., 41, 257
\bibitem[van Putten \& Ostriker(2001)]{van01} van Putten, M.H.P.M., \& Ostriker, E., 2001, ApJ, 552, L31

\bibitem[Bromberg et al.(2013)]{bro13} Bromberg, O., Nakar, E., Piran, T., \& Sari, R., 2013, ApJ, 764, 179
\bibitem[Kouveliotou et al.(1993)]{kov93} Kouveliotou, C., et al., 1993, ApJ, 413, L101
\bibitem[Band et al.(1993)]{ban93}Band, D., Matteson, J., Ford, L., et al., 1993, ApJ, 413, 281
\bibitem[Costa et al.(1997)]{cos97} Costa, E., et al., 1997, Nature, 387, 878
\bibitem[van Paradijs et al.(1997)]{par97} van Paradijs, J., et al., 1997, Nature, 386, 686
\bibitem[Metzger et al.(1997)]{met97}Metzger, M., Djorgovski, S.G., Kulkarni, S.R., et al., 1997 Nature, 387, 879
\bibitem[Amati et al.(1998)]{ama98} Amati, L., Piro, L., \& Antonelli, L.A., et al., 1998, Nucl. Phys. B., 69, 656
\bibitem[Gehrels et al.(2005)]{geh05} Gehrels, N., et al., 2005, Nature, 437, 851
\bibitem[Fox et al.(2005)]{fox05} Fox, D.B., et al., 2006, Nature, 437, 845
\bibitem[Hj\"orth et al.(2011)]{hjo11} Hj\'orth, J., \& Bloom, J.S., 2011, {\em in} Gamma-Ray Bursts, eds. C. Kouveliotou, R. A. M. J. Wijers, S. E. Woosley (Cambridge University Press)
\bibitem[Melandri et al.(2013)]{mel13} Melandri, A., et al., 2013, GCN Circ. 14673
\bibitem[Della Valle et al.(2003)]{del03} Della Valle, M., et al., 2003, A\&A, 406, L33
\bibitem[Kelly et al.(2008)]{kel08} Kelly, P.L., Kirshner, R.P., \& Pahre, M., 2008, ApJ, 687, 1201
\bibitem[Raskin et el.(2008)]{ras08} Raskin, C., et al., 2008, ApJ, 689, 358
\bibitem[Wanderman \& Piran(2010)]{wan10} Wanderman, D., \& Piran, T., MNRAS, 2010, 406, 1944
\bibitem[Grieco et al.(2012)]{gri12} Grieco, V., Matteucci, F., Meynet, G., Longo, F., Della Valle, M., Salvaterra, R., 2012, MNRAS, 423, 3049 
\bibitem[Amati et al.(2017)]{ama17} Amati, L. O'Brien, P.O., G\"otz, D., Bezoo, E., Tenzer, C., et al., 2017, arXiv:1710.04638v2
\bibitem[H\"offlich et al.(1999)]{hof99} H\"offlich, P., Wheeler, J.C., \& Wang, L., 1999, ApJ, 521, 179
\bibitem[Bisnovatyi-Kogan(1970)]{bis70}Bisnovatyi-Kogan, G. S., 1970, Astron. Zh., 47, 813
\bibitem[MacFadyen \& Woosley(1999)]{mac99} MacFadyen A. I., \& Woosley S. E., 1999, ApJ, 524, 262
\bibitem[Bisnovatyi-Kogan et al.(1966)]{bis66} Bisnovatyi-Kogan, G.S., \& Kazhdan, Ya.M., 1966, Astron. Zh. 43, 761 (Transl. 1967, Sov. Astron. 10, 604)
\bibitem[Barkat et al.(1967)]{bar67} Barkat, Z., Rakavy, G. \& Sack, N., 1967, Phys. Rev. Lett. 18, 379
\bibitem[Gal-Yam et al.(2009)]{gal09}Gal-Yam, A., Mazzali, P., Ofek, E. O., et al. 2009, Nature, 462, 624
\bibitem[Chardonet et al.(2010)]{cha10} Chardonet, P. et al., 2010, ApJ Supp., 325, 153
\bibitem[van Putten(2012a)]{van12} van Putten, M.H.P.M., 2012a, Prog. Theor. Phys., 127, 331
 \bibitem[Quinby et al.(2009)]{qui11} Quimby, R.M., et al., 2009, 2011, Nature, 474, 487
\bibitem[Malesani et al.(2004)]{mal04} Malesani, D., Tagliaferri, G., Chincarini, G., et al. 2004, ApJ, 609, L5
\bibitem[Masetti et al.(2006)]{mas06}Masetti, N., Palazzi, E., Pian, E., et al. 2006, GCN, 4803, 1
\bibitem[Modjaz et al.(2006)]{mod06} Modjaz, M., Stanek, K. Z., Garnavich, P. M., et al. 2006, ApJ, 645, L21
\bibitem[Sollerman et al.(2006)]{sol06} Sollerman, J., Jaunsen, A. O., Fynbo, J. P. U., et al. 2006, A\&A, 454, 503
\bibitem[Mirabal et al.(2006)]{mir06} Mirabal, N., Halpern, J. H., An, D., Thorstensen, J. R., \& Terndrup, D. M. 2006, ApJ, 643, L99
\bibitem[Pian et al.(2006)]{pia06} Pian, E., Mazzali, P. A., Masetti, N., et al. 2006, Nature, 442, 1011
\bibitem[Cobb et al.(2006b)]{cob06b} Cobb, B. E., Bailyn, C. D., van Dokkum, P. G., \& Natarajan, P. 2006b, ApJ, 645, L116
\bibitem[Chornock et al.(2010)]{cho10} Chornock, R., Berger, E., Levesque, E. M., et al. 2010, arXiv:1004.2262
\bibitem[Bufeno et al.(2011)]{buf11}Bufano, F., Benetti, S., Sollerman, J., Pian, E., \& Cupani, G. 2011, Astron. Nachr., 332, 262
\bibitem[Stanek et al.(2003)]{sta03}Stanek, K. Z., Matheson, T., Garnavich, P. M., et al. 2003, ApJ, 591, L17
\bibitem[Hj\"orth et al.(2003)]{hjo03} Hj\"orth, J., Sollerman, J., Moller, P., et al. 2003, Nature, 423, 847
\bibitem[Matheson et al.(2003)]{mat03} Matheson, T., Garnavich, P. M., Stanek, K. Z., et al. 2003, ApJ, 599, 394
\bibitem[Cenko et al.(2010)]{cen10} Cenko, S. B., Frail, D. A., Harrison, F. A., et al. 2010, ApJ, 711, 641
\bibitem[de Ugarte Postigo et al.(2011)]{deu11}de Ugarte Postigo, A., Goldoni, P., Milvang-Jensen, B., et al. 2011, GCN, 11579
\bibitem[Chandra et al.(2008)]{cha08} Chandra, P., Cenko, S. B., Frail, D. A., et al. 2008, ApJ, 683, 924
\bibitem[Woosley(2010)]{woo10} Woosley, S.E., 2010, ApJ, 719, L204
\bibitem[Kasen \& Bildsten(2010)]{kas10} Kasen, D., \& Bildsten, L., 2010, ApJ, 717, 245
\bibitem[Nicholl et al.(2013)]{nic13} Nicholl, M., et al., 2013, Nature, 502, 346
\bibitem[Dong et al.(2015)]{don15} Dong, D., Shappee, B.J., Prieto, J.L., et al., 2015, Science, 351, 6270
\bibitem[Abramovici et al.(1992)]{abr92} Abramovici, A., Althouse, W.E., Drever, R.W.P., et al., 1992, Science, 256, 325
\bibitem[Acernese et al.(2006)]{ace06} Acernese, F. et al. (Virgo Collaboration), 2006, Class. Quantum Grav., 23, S635
\bibitem[Acernese et al.(2007)]{ace07} Acernese, F. et al. (Virgo Collaboration), 2007, Class. Quantum Grav., 24, S381
\bibitem[Somiya(2012)]{som12} Somiya, K., (for the KAGRA Collaboration), 2012, Class. Quantum Grav., 29, 124007 
\bibitem[KAGRA(2014)]{kag14} KAGRA Project (NAOJ), http://gwcenter.icrr.u-tokyo.ac.jp/en/
\bibitem[van Putten(2004)]{van04} van Putten, M.H.P.M., 2004, ApJ Lett, 611, L81


\bibitem[Guetta \& Della Valle(2007)]{gue07} Guetta, D., \& Della Valle, M., 2007, ApJ, 657, L73
\bibitem[Paczy\'nski(1998)]{pac98} Paczy\'nski, B.P., 1998, ApJ, 494, L45
\bibitem[van Putten(2015c)]{van15} van Putten, M.H.P.M., 2015c, ApJ, 810, 7
\bibitem[Capellaro et al.(1999)]{cap99} Cappellaro, E., Evans, R., \& Turatto, M., 1999, A\&A, 351, 459
\bibitem[Barbon et al.(1999)]{bar99} Barbon, R., Buond\'i, V., Cappellaro, E., \& Turatto, M., 1999, A\&A Suppl. Ser., 139, 531
\bibitem[Li et al.(2011b)]{li11b} Li, W., Chornock, R., Leaman, J., et al. 2011b, MNRAS, 412, 1473
\bibitem[Taylor et al.(2014)]{tay14} Taylor, M., Cinabro, D., Dilday, B., Galbany, L., Gupta, R., et al., 2014, ApJ, 792, 135
\bibitem[Capellaro et al.(2015)]{cap15} Cappellaro, E., Botticella, M.T., Pignata, G., Grado, A., Greggio, L., et al., 2015, A\&A, 584, A62
\bibitem[Kasen et al.(2013)]{kas13} Kasen,D., Badnell, N. R. \& Barnes, J., 2013, ApJ, 774, 25
\bibitem[Barnes \& Kasen(2013)]{bar13} Barnes, J. \& Kasen, D., 2013, ApJ, 775, 18
\bibitem[Tanaka \& Hotokezaka(2013)]{tan13a} Tanaka, M., \& Hotokezaka, K., 2013, ApJ, 775, 113

\bibitem[HEASARC(2016)]{HEA} HEASARC, $http://swift.gsfc.nasa.gov/archive/grb_table/$
\bibitem[Perley et al.(2009)]{per09} Perley, D.A., et al., 2009, ApJ, 696, 1871
\bibitem[Norris et al.(2010)]{nor10} Norris, J.P., et al., 2010, ApJ, 717, 411
\bibitem[Coward et al.(2012)]{cow12} Coward, D.M., et al., 2012, MNRAS, 425, 2668
\bibitem[Gompertz et al.(2014)]{gom14} Gompertz, B.P., et al., 2014, MNRAS, 438, 240
\bibitem[Berger et al.(2007c)]{ber07c} Berger, E., et al., 2007d, GCN circ. 5995
\bibitem[Fong et al.(2010)]{fon10} Fong, W., et al., 2010, ApJ, 708, 9 %n=1 exp disk
\bibitem[Page et al.(2006)]{pag06} Page, K.L., et al., 2006, ApJ, 637, L13
\bibitem[Bloom et al.(2006)]{blo06} Bloom, J.S., et al., 2006, ApJ, 638, 354
\bibitem[Bloom et al.(2007)]{blo07} Bloom, J.S., et al., 2007, ApJ, 654, 878 %old red galaxy
\bibitem[Cucchiara et al.(2013)]{cuc13} Cucchiara, A., et al., 2013, ApJ, 777, 94
\bibitem[Frederiks(2013)]{fre13} Frederiks, D., 2013, GCN Circ. 14772
\bibitem[Kocevski et al.(2010)]{koc10} Kocevski, D., et al., 2010, MNRAS, 404, 963
\bibitem[Berger \& Soderberg(2005)]{ber05} Berger, E., \& Soderberg, A.M., 2005, GCN Circ. 4384
\bibitem[Berger(2005b)]{ber05a} Berger, E., 2005b, GCN Circ. 3801 %cluster, 1.7e50
\bibitem[Berger et al.(2005)]{ber05c} Berger, E., et al., 2005c, Nature, 438, 15
\bibitem[Berger et al.(2007)]{ber07} Berger, E., et al., 2007a, ApJ, 660, 496
\bibitem[Berger et al.(2007a)]{ber07a} Berger, E., et al., 2007b, ApJ, 664, 1000 %Eiso=7e50
\bibitem[Berger et al.(2007b)]{ber07b} Berger, E., et al., 2007c, GCN Circ. 7151
\bibitem[Berger et al.(2007d)]{ber07d} Berger, E., et al., 2007e, GCN circ. 7154
\bibitem[Golenetskii et al.(2005)]{gol05} Golenetskii, S., et al., 2005, GCN Circ. 4394
\bibitem[Perley et al.(2013)]{per13} Perley, D.A, 2013, GCN circ. 15319
\bibitem[Perley et al.(2010)]{per10} Perley, D,A., et al., 2010, GCN circ. 11464
\bibitem[Berger et al.(2006)]{ber06b} Berger, E., at al., 2006, GCN Circ. 5965 %061217
\bibitem[de Ugarte Postigo et al.(2006)]{deu06} de Ugarte Postigo, A., et al., 2006, GCN circ. 5951
\bibitem[Rau et al.(2009)]{rau09} Rau, A., et al., 2009, GCN Circ. 9353
\bibitem[Nicuesa Guelbenzu et al.(2012)]{gue12} Nicuesa Guelbenzu, A., et al. 2012, A\&A, 538, L7
\bibitem[Cenko et al.(2008)]{cen08} Cenko, S.B., et al., 2008, arxiv:0802.0874v1 
\bibitem[Cucciara et al.(2006)]{cuc06} Cucchiara, A., et al., 2006, GCN Circ. 5470
\bibitem[Ukwatta et al.(2010)]{ukw10} Ukwatta, T., et al., 2010, GCN circ. 10976
\bibitem[Prochaska et al.(2006)]{pro06} Prochaska, J.X., et al., 2006, ApJ, 642, 989 %Eiso=0.017; cluster
\bibitem[Berger(2006)]{ber06a} Berger, E., 2006, $in$ Gamma-ray bursts in the Swift era, 16$^{th}$ Maryland Astroph. Conf., AIP Conf. Proc., 836, 33; arXiv:astro-ph/0602004v1 
\bibitem[Ferraro et al.(2007)]{fer07} Ferrero, P., et al., 2007, ApJ, 134, 2118 %spiral, irregular
\bibitem[Antonelli et al.(2009)]{ant09} Antonelli, L. A., et al., 2009, A\&A, 507, L45
\bibitem[Della Valle et al.(2006)]{del06} Della Valle, M., et al., 2006, Nature, 444, 1050
\bibitem[Fynbo et al.(2006)]{fyn06a} Fynbo, J.P.U., et al., 2006, Nature, 444, 1047
\bibitem[Cobb et al.(2006a)]{cob06} Cobb, B.E., et al., 2006a, GCN Circ. 5282
\bibitem[Prochaska et al.(2005)]{pro05} Prochaska, J.X., et al., 2005, GCN circ. 3700
\bibitem[Cenko et al.(2006)]{cen06} Cenko, S.B., et al., 2006, GCN Circ. 5946
\bibitem[Graham et al.(2009)]{gra09} Graham, J.F., et al., 2009, ApJ, 698, 1620
\bibitem[Graham et al.(2007)]{gra07} Graham, J.F., et al., 2007, GCN Circ. 6836
\bibitem[Jakobsson \& Fynbo(2007)]{jak07} Jakobsson, P., Fynbo, J. P. U., arXiv:0704.1421
\bibitem[Moretti et al.(2006)]{mor06} Moretti, A., et al., 2006, GCN-Report-9.1, gcn.gsfc.nasa.gov/reports/report$_-$9$_-$1.pdf

\bibitem[Amati et al.(2002)]{ama02} Amati, L., et al., 2002, A\&A, 390, 81
\bibitem[Amati et al.(2006)]{ama06} Amati, L., et al., 2006, MNRAS, 372, 233
\bibitem[Amati et al.(2008)]{ama08} Amati, L., et al., 2008, MNRAS, 391, 577
\bibitem[Amati et al.(2009)]{ama09} Amati, L., Frontera, F., Guidorzi, C., A\&A, 2009, 508, 173 
\bibitem[Cano et al.(2014)]{can14} Cano, Z., de Ugarte Postigo, A., Pozanenko, A., et al., 2014, A\&A; arXiv:1405.3114.
\bibitem[Swift(2016)]{swi14} http://swift.gsfc.nasa.gov/analysis/threads/bat$_-$threads.html

%\bibitem[Dichiara et al.(2013a)]{dic13a} Dichiara, S., Guidorzi, C., Amati, L. A., \& Frontera, F. 2013a, MNRAS, 431, 3608
\bibitem[Gal-Yam et al.(2006)]{gal06} Gal-Yam, A., et al., 2006, Nature, 444, 1053
\bibitem[Tanvir et al.(2005)]{tan05} Tanvir, N.R., Chapman, R., Levan, A.J., \& Priddey, R.S., 2005, Nature, 438, 991
\bibitem[van Putten(2008)]{van08} van Putten, M.H.P.M., 2008, ApJ, 684, L91

\bibitem[Caito et al.(2010b)]{cai09}Caito, L., Bernardini, M.G.,  Bianco, C.L., Dainotti, M.G., Guida, R., Ruffini, R.,  et al., 2010, A\&A, 498, 501 
\bibitem[Zhang(2006)]{zha06} Zhang, B., 2006, Nature, 444, 1010
\bibitem[Zhang et al.(2007)]{zha07} Zhang, B., Zhang, B.-B., Liang, E.-W., Gehrels, N., Burrows, D.N., \& M\'esz\'aros, P., 2007, ApJ, 655, L25

\bibitem[van Putten(2009)]{van09} van Putten, M.H.P.M., 2009, MNRAS, 396, L81

\bibitem[van Putten \& Gupta(2009)]{van09a} van Putten, M.H.P.M., \& Gupta, A.C., 2009, MNRAS, 394, 2238
\bibitem[Kumar et al.(2008a)]{kum08} Kumar, P., Narayan, R., \& Johnson, J. L., 2008a, Science, 321, 376
 

\bibitem[Usov(1994)]{uso94} Usov, V., 1994, MNRAS, 267, 1035
\bibitem[Metzger et al.(2011)]{met11} Metzger, D.B., et al., 2011, Mon. Not. R. Astron. Soc. 413, 2031
\bibitem[Levinson \& Eichler(1993)]{lev93} Levinson, A., \& Eichler, D., 1993, ApJ, 418, 386
\bibitem[Eichler(2011)]{eic11} Eichler, D., 2011, ApJ, 730, 41
\bibitem[Zalamea \& Beloborodov(2011)]{zal11} Zalamea, I., \& Beloborodov, A., 2011, MNRAS, 410, 2302
\bibitem[Levinson \& Glubus(2013)]{lev13b}  Levinson, A., \& Glubus, N., 2013, ApJ, 770, 159
\bibitem[Bogovalov(1995)]{bog95} Bogovalov, S. V. 1995, Astronomy Letters, 21, 565
\bibitem[Chiueh et al.(1991)]{chi91} Chiueh, T, Li, Z-Y. \& Begelman, M. C, 1991, ApJ, 377, 462
\bibitem[Heyvaerts \& Norman(1989)]{hey89} Heyvaerts, J.  \& Norman, C. 1989, ApJ, 347, 1055
\bibitem[Lyubarsky(2009)]{lyu09} Lyubarsky, Y. 2009, ApJ, 698, 1570
\bibitem[Granot et al.(2011)]{gra11} Granot, J., Komissarov, S.S., \& Spitkovsky, A., 2011, MNRAS, 411, 1323
\bibitem[Lyutikov(2011)]{lyu11} Lyutikov M., 2011, MNRAS, 411, 422
\bibitem[Giannios \& Spruit(2007)]{gia07} Giannios, D. \& Spruit, H. 2007, A\&A, 469, 1
\bibitem[Levinson \& van Putten(1997)]{lev97} Levinson, A., \& Van Putten, M., 1997, ApJ, 488, 69
\bibitem[Lyubarsky(2010)]{lyu10} Lyubarsky, Y. 2010, ApJ, 725, L234
\bibitem[Lyutikov \& Blandford(2003)]{lyu03} Lyutikov, M., \& Blandford, R., 2003, arXiv:0312347
\bibitem[Lyutikov et al.(2003)]{lyu03b} Lyutikov, M., Pariev, V.I., \& Blandford, R., 2003, ApJ, 597, 998
\bibitem[Zhang \& Yahn(2011)]{zha11} Zhang, B., \& Yahn, H., 2011, ApJ, 726, 90
\bibitem[McKinney et al.(2012)]{mck12} McKinney, J. C.\& Uzdensky, D. A. 2012, MNRAS, 419, 573 
\bibitem[Levinson \& Begelman(2013)]{lev13} Levinson, A., \& Begelman, M. C., 2013, ApJ, 764, 148 
\bibitem[M\'eszaros \& Rees(1993)]{mes93} M\'esz\'aros, P., \& Rees, M.J., 1993, ApJ 405, 278 
\bibitem[Rees \& M\'eszaros(1992)]{ree92} Rees, M.J. \& M\'esz\'aros, P., 1992, MNRAS, 258, P41 
\bibitem[Bromberg \& Levinson(2007)]{bro07} Bromberg, O., \& Levinson, A., 2007, ApJ, 671, 678
\bibitem[Lazzati et al.(2009)]{laz09} Lazzati, D. Morsony, B. \& Begelman, M. 2009, ApJ, 700, L47 
\bibitem[Piran(1999)]{pir99} Piran, T., 1999, 314, 575
\bibitem[Piran(2004)]{pir04} Piran, T., 2004, RvMP, 76, 1143
\bibitem[Beloborodov(2013)]{bel13} Beloborodov, A. 2013, ApJ, 764, 157
\bibitem[Crider et al.(1997)]{cri97}  Crider, A. Liang, E. P., Smith, I. A. et al.  1997, ApJL, 479, L39
\bibitem[Eichler \& Levinson(2000)]{eic00} Eichler, D. \& Levinson,A. 2000, ApJ, 529, 146 
\bibitem[Preece et al.(1998)]{pre98} Preece, R. D.  et al. 1998, ApJL, 506, L23
\bibitem[Ryde(2004)]{ryd04} Ryde, F., 2004, ApJ, 614, 827
\bibitem[Ryde(2005)]{ryd05} Ryde, F., 2005, ApJ, 625, L95
\bibitem[Frontera et al.(2001)]{fro01} Frontera, F., et al., 2001, ApJ, 550, 47
\bibitem[Giannios(2012)]{gia12} Giannios, D. 2012, MNRAS, 422, 3092
\bibitem[Peer et al.(2006)]{pee06} Peer, A. Meszaros, P. \& Rees, M. 2006, ApJ, 642, 995 
\bibitem[Ryde \& Peer(2009)]{ryd09} Ryde, F. \& Peer, A. 2009,  702, 1211 
\bibitem[Vurm et al.(2013)]{vur13} Vurm, I. Lyubarsky, Y. \& Piran, T. 2013, ApJ, 764, 143

\bibitem[Eichler(1994)]{eic94} Eichler, D.  1994, ApJS, 90, 877
\bibitem[Bromberg et al.(2011b)]{bro11b} Bromberg, )., et al. 2011b, ApJ, 733, 85
\bibitem[Morsony et al.(2010)]{mor10} Morsony, B. Lazzati, D. \& Begelman, M. C. 2010, ApJ, 723, 267 
\bibitem[Budnik et al.(2010)]{bud10} Budnik, R. et al. 2010, ApJ, 725, 63
\bibitem[Levinson \& Brombert(2008)]{lev08} Levinson, A. \& Bromberg,O. 2008, Phys. Rev. Lett., 100, 131101 
\bibitem[Levinson(2012)]{lev12}  Levinson, A. 2012, ApJ, 756, 174
\bibitem[Keren \& Levinson(2014)]{ker14} Keren, S., \& Levinson, A., 2014, ApJ, 789, 128
\bibitem[Ito et al.(2017)]{ito17} Ito, H., Levinson, A., Stern, B.E., \& Nagataki, S., 2017, arXiv:1709.08955

\bibitem[Frontera et al.(2013)]{fro13} Frontera, F., Amati, L., Farinelli, R., et al., 2013, ApJ, 779, 175
\bibitem[Chincarini et al.(2010)]{chi10} Chincarini, G., Mao, J., Margutti, R., et al. 2010, MNRAS, 406, 2113
\bibitem[Bernardini et al.(2011)]{ber11} Bernardini, M.G., et al., 2011, A\&A, 526, A27
\bibitem[Margutti et al.(2011)]{mar11} Margutti, R., Chincarini, G., Granot, J., Guidorzi, C., Berger, E., et al., 2011, MNRAS, 417, 2144
\bibitem[Sari \& Piran(1997)]{sar97} Sari, R., \& Piran, T., 1997, ApJ, 485, 270
\bibitem[Piran \& Sari(1997)]{pir97} Piran, T., \& Sari, R., 1997, arXiv:9702093
\bibitem[Kobayashi et al.(1997)]{kob97} Kobayashi, S., Piran, T., \& Sari, R., 1997, ApJ, 490, 92
\bibitem[Nakar \& Piran(2002)]{nak02} Nakar, E., \& Piran, T., 2002, MNRAS, 330, 920
\bibitem[Thompson(1994)]{tho94} Thompson, C., 1994, Mon. Not. R. Astron. Soc. 270, 480
\bibitem[Sari et al.(1998)]{sar98} Sari, R., Piran, T., \& Narayan, R., 1998, ApJ, 347, L17
\bibitem[Thompson et al.(2007)]{tho07} Thompson, C, M\'esz\'aros, P., \& Rees, M.J., 2007, ApJ, 666, 1012
\bibitem[Beloborodov et al.(1998)]{bel98} Beloborodov, A. M., Stern, B. E., \& Svensson, R. 1998, ApJ, 508, L25
\bibitem[Beloborodov et al.(2000)]{bel00} Beloborodov, A. M., Stern, B. E., \& Svensson, R. 2000, ApJ, 535, 158
\bibitem[Guidorzi et al.(2012)]{gui12} Guidorzi, C., Margutti, M., Amati, L. A., et al. 2012, MNRAS, 422, 1785

\bibitem[Dichiara et al.(2013a)]{dic13a} Dichiara, S., Guidorzi, C., Amati, L. A., \& Frontera, F. 2013a, MNRAS, 431, 3608
\bibitem[Dichiara et al.(2013b)]{dic13b} Dichiara, S., Guidorzi, C., Frontera, F., \& Amati, L. A. 2013b, ApJ, 777, 132
\bibitem[Zrake \& MacFadyen(2013)]{zra13} Zrake, J, \& MacFadyen A.I., 2013, ApJ, 763, L12
\bibitem[Calafut \& Wiita(2014)]{cal14} Calafut, V., \& Wiita, P., 2015, JApA, 36, 225
\bibitem[Pastorello et al.(2007)]{pas07} Pastorello, A., et al., 2007, Nature, 449, 1

\bibitem[Villasenor et al.(2005)]{vil05} Villasenor, J. S., Lamb, D. Q., Ricker, G. R., et al. 2005, Natur, 437, 855
\bibitem[Hj\"orth et al.(2005)]{hjo05} Hj\"orth, J., Watson, D., Fynbo, J. P. U., et al. 2005, Natur, 437, 859
\bibitem[Nathanail et al.(2015)]{nat15} Nathanail, A., Strantzalis, A., \& Contopoulos, I. 2015, MNRAS, 455, 4479
\bibitem[van Putten(2008)]{van08b} van Putten, M.H.P.M., 2008, ApJ, 685, L63
\bibitem[Shahmoradi \& Nemiroff(2015)]{sha15} Shahmoradi, A., \& Nemiroff, R. J. 2015, MNRAS, 451, 126

\bibitem[van Putten(1999)]{van99} van Putten, M.H.P.M., 1999, Science, 284, 115
\bibitem[Bondi(1952)]{bon52} Bondi, H., 1952, MNRAS, 112, 195
\bibitem[Bardeen(1970)]{bar70} Bardeen, J. M. 1970, Nature, 226, 64

\bibitem[McKinney(2005)]{mck05} McKinney, J.C., 2005, ApJ, 630, L5
\bibitem[Kumar et al.(2008b)]{kum08b} Kumar, P., Narayan, R., \& Johnson, J. L., 2008b, MNRAS, 388, 1729

\bibitem[King \& Pringle(2006)]{kin06} King, A.R., \& Pringle, J.E., 2006, MNRAS, 373, L90
\bibitem[Globus \& Levinson(2014)]{glo14} Globus, N., \& Levinson, A., 2014, ApJ, 796, 26

\bibitem[Wald(1974)]{wal74} Wald, R.M., 1974. Phys. Rev. D. 10, 1680
\bibitem[Dokuchaev(1986)]{dok86} Dokuchaev, V.I., 1986. Sov. Phys. JETP 65, 1079
\bibitem[van Putten(2001a)]{van01p} van Putten, M.H.P.M., 2001, Phys. Rep., 354, 1 

\bibitem[Macchetto et al.(1997)]{mac97} Macchetto, F., Marconi, A., Axon, D.J., Capetti, A., Sparks, W., \& Crane, P., 1997, ApJ, 489, 579 
\bibitem[Walsh et al.(2013)]{wal13} Walsh, J.L., Barth, A.J., Ho, L.C., \& Sarzi, M., 2013, ApJ, 770, 86
\bibitem[Greiner et al.(2001)]{gre01} Greiner, J., Cuby, J.G., \& McCaughrean, M.J.,  2001, Nature, 414, 522
\bibitem[Ruffini \& Wilson(1975)]{ruf75} Ruffini, R., \& Wilson, J.R., 1975, Phys. Rev. D,12, 2959
\bibitem[Bisnovatyi-Kogan et al.(1976)]{bis76} Bisnovatyi-Kogan, G.S., \& Ruzmaikin, A.A., 1976, Ap. Space. Sci. 42, 401
\bibitem[Blandford \& Znajek(1977)]{bla77} Blandford, R.D., \& Znajek, R.L., 1977, Mon. Not. R. Astron. Soc. 179, 433
\bibitem[Balbus \& Hawley(1991)]{bal91} Balbus, S.A., \& Hawley, J.F., 1991, ApJ, 376, 214
\bibitem[Lubow et al.(1994)]{lub94} Lubow, S.H., Papaloizou, J.C.B., \& Pringle, J.E., 1994, MNRAS, 267, 235

\bibitem[Carter(1968)]{car68}Carter, B., 1968, Phys. Rev., 174, 1559
\bibitem[Cohen et al.(1973)]{coh73} Cohen, J.M., Tiomno, J., \& Wald, R.M., 1973, Phys. Rev. D. 7, 998
\bibitem[Papapetrou(1951)]{pap51} Papapetrou, A., 1951, Proc. Roy. Soc., 209, 248
\bibitem[Pirani(1956)]{pir56}Pirani, F.A.E., 1956, Act. Phys. Pol., XV, 389

\bibitem[Gupta \& van Putten(2012)]{gup12} Gupta, A.C., \& van Putten, M.H.P.M., 2012, {\em in} Astron. Soc. India Conf. Ser., 5, p123

\bibitem[Eichler et al.(2009)]{eic09} Eichler, D., Guetta, D., \& Manis, H., 2009, ApJ, 690, L61
\bibitem[Lee \& Kluzniak(1998)]{lee98} Lee, W.H., \& Kl\'uzniak, 1998, ApJ, 494, L53
\bibitem[Lee \& Kluznian(1999)]{lee99} Lee, W.H., \& Kl\'uzniak, 1999, ApJ, 526, 178
\bibitem[Rosswog(2007)]{ros07} Rosswog, S., 2007, MNRAS, 376, 48
\bibitem[van Putten et al.(2004)]{van04b} van Putten, M.H.P.M., Lee, H.K., Lee, C.H., \& Kim, H., 2004, Phys. Rev. D, 2004, 69, 104026 
\bibitem[Mangano et al.(2007)]{man07} Mangano V., et al., 2007, A\&A, 470, 105
\bibitem[Gehrels et al.(2009)]{geh09} Gehrels, N., Ramirez-Ruiz, E., \& Fox, D.B., 2009, ARAA. 47, 567 
\bibitem[Lei et al.(2008)]{lei08} Lei, W.H., Wang, D.X., Zou, Y.C., \& Zhang, L., 2008, ChJAA, 8, 405

\bibitem[Pringle(1981)]{pri81} Pringle, J.E., 1981, ARAA, 19, 137
\bibitem[Toomre(1964)]{too64} Toomre, A., 1964, ApJ, 139, 1217
\bibitem[Goldreich \& Lynden-Bell(1965)]{gol65} Goldreich, P., \& Lynden-Bell, D., 1965, MNRAS, 130, 125
\bibitem[Griv(2011)]{gri11} Griv, E., 2011, ApJ, 733, 43
\bibitem[Popham et al.(1999)]{pop99} Popham, R., Woosley, S.E., \& Fryer, C., 1999, ApJ, 518, 356
\bibitem[Chen \& Beloborodov(2007)]{che07} Chen, W-X, \& Beloborodov, A., 2007, ApJ, 657, 383
\bibitem[Fishbone(1972)]{fis72} Fishbone, L.G., 1972, ApJ., 175, L155
\bibitem[Lattimer \& Schramm(1974)]{lat74} Lattimer, J.M., \& Schramm, D.N., 1974, ApJ, 192, L145
\bibitem[Lattimer \& Schramm(1976)]{lat76} Lattimer, J.M., \& Schramm, D.N., 1976, ApJ, 210, 549
\bibitem[Rees et al.(1974)]{ree74} Rees, M. J., Ruffini, R., \& Wheeler, J. A. 1974, {\em Black Holes, Gravitational Waves and Cosmology: An Introduction to Current Research} (New York: Gordon \& Breach), Ch.7
\bibitem[Duez et al.(2004)]{due04} Duez, M. D., Shapiro, S. L., \& Yo, H.-J. 2004, Phys. Rev. D, 69, 104016
\bibitem[Lipunov(1983)]{lip83} Lipunov, V.M., 1983, Ap\&SS, 97, 121
\bibitem[Leung et al.(1997)]{leu97} Leung, P.T., Liu, Y.T., Suen, W.-M., Tam, C.Y., \& Young, K., 1997, Phys. Rev. Lett., 78, 2894
\bibitem[Font \& Papadopoulos(2001)]{fon01} Font, J.A., \& Papadoupoulos, P., 2001, $in$ Proc. Spanish Rel. Meeting, eds. J.F. Pascual-S\'anchez, L. Flor\'ia,
\bibitem[Nagar et al.(2007)]{nag07} Nagar, A., Zanotti, O., Font, J.A., \& Rezolla, L., 2007, Phys. Rev. D, 2007, 75, 044016
\bibitem[Lipunova et al.(2009)]{lip09} Lipunova, G.V., Gorbovskoy, E.S., Bogomazov, A.I., \& Liponov, V.M., 2009, MNRAS, 397, 1695
\bibitem[Fryer et al.(2002)]{fry02} Fryer, C. L., Holz, D.E., \&  Hughes, S. A., 2002, ApJ, 565, 430
\bibitem[Thorne et al.(1986)]{tho86} Thorne, K.S., Price, R.H., \& McDonald, D.H., 1986, Black Holes: The Membrane Paradigm (New Haven, CT: Yale University Press)
\bibitem[Kokkotas \& Schmidt(1999)]{kok99} Kokkotas, K.D., \& Schmidt, B.G., 1999, Living Rev. Relativity, 2, http://www.livingreviews.org/lrr-1999-2 
\bibitem[Echeverria(1998)]{ech98} Echeverria F., 1988, Phys. Rev. D, 40, 3194

\bibitem[Shakura \& Sunyaev(1973)]{sha73} Shakura, N.I., \& Sunyaev, R.A., 1973, Astron. Astophys., 24, 337

\bibitem[Papaloizou \& Pringle(1984)]{pap84} Papaloizou, J.C.B., \& Pringle, J.E., 1984, MNRAS, 208, 721
\bibitem[Kiuchi et al.(2011)]{kiu11} Kiuchi, K., et al., 2011, Phys. Rev. Lett., 106, 251102
\bibitem[Toscani et al.(2019)]{tos19} Toscani, M., Lodato, G., \& Nealon, R., 2019, MNRAS, 489, 699
\bibitem[van Putten et al.(2011a)]{van11} van Putten, M.H.P.M., Kanda, N., Tagoshi, H., Tatsumi, D., Masa-Katsu, F., \& Della Valle, M., 2011a, Phys. Rev. D 83, 044046
\bibitem[van Putten(2016)]{van16} van Putten, M.H.P.M., 2016, ApJ,  819, 169

\bibitem[Flanagan \& Hughes(1998a)]{fla98a} Flanagan, E.E., \& Hughes, S.A., 1998a, Phys. Rev. D, 57, 4535
\bibitem[Flanagan \& Hughes(1998b)]{fla98b} Flanagan, E.E., \& Hughes, S.A., 1998b, Phys. Rev. D, 57, 4566

\bibitem[Porciani \& Madau(2001)]{por01} Porciani, C., \& Madau, P., 2001, ApJ, 548, 522
\bibitem[Phinney(2001)]{phi01} Phinney, E.S., 2001, astro-ph/0108028
\bibitem[Sutton et al.(2010)]{sut10} Sutton, P.J., Jones, G., Chatterji, S., et al., 2010, N. J. Phys.,  12,053034
\bibitem[Prestegard \& Thrane(2012)]{pre12} Prestegard, T., \& Thrane, E., 2009, https://dcc.ligo.org/public/0093/L1200204/001/burstegard.pdf  
\bibitem[Thrane \& Coughlin(2013)]{thr13} Thrane, E., \& Coughlin, M., 2013, Phys. Rev. D 88, 083010
\bibitem[Thrane \& Coughlin(2014)]{thr14} Thrane, E., \& Coughlin, M., 2014, Phys. Rev. D 89, 063012
\bibitem[Coughlin et al.(2015)]{cou15} Coughlin, M., Meyers, P., Kandhasamy, S., et al., 2015, Phys. Rev. D, 92, 43007
\bibitem[Abbott et al.(2015)]{abb15} Abbott, B.P., Abbott, R., Abbott, T.D., et al., 2015, Phys. Rev. D, 93, 042005
\bibitem[Gossan et al.(2015)]{gos15} Gossan, S.E., Putton, P., Stuver, A., et al., 2015, arXiv:1511.02836v1
\bibitem[Mohapatra et al.(2011)]{moh12} Mohapatra, S., Nemtzow, Z., Chassande-Mottin, E., \& Codanati, L., 2012, J. Phys. Conf. Ser., 363, 012031
\bibitem[Chassande-Mottin et al.(2007)]{cha07} Chassande-Mottin, E., Pai, A., Rabaste, O., 2007, Proc. SPIE Int.Soc. Opt. Eng. 6701 (2007) 670112
\bibitem[Chassande-Mottin et al.(2017)]{cha17} Chassande-Mottin, E., Lebigot, E., Magaldi H., Chase, E., Pai, A., V., G., \& Vedovato, G., 2017, arXiv:1710.09256

\bibitem[Accadia et al.(2010)]{acc10} Accadia, T., et al., 2010, Class. Quantum Grav., 27 194011
\bibitem[Biscanns et al.(2018)]{bis18} Biscans, S., et al., 2018, Class. Quantum Grav., 35, 055004
\bibitem[van Putten(2019)]{van19c} van Putten, M.H.P.M., 2019, KAGRA f2f meeting, Toyama; GWDOC JGW-G1910639-v1

\bibitem[Arnowitt et al.(1962)]{arn62} Arnowitt, R., Deser, R., \& Misner, C.W., 1962, in Gravitation: An Introduction to
\bibitem[van Putten(1996)]{van96} van Putten, M.H.P.M., \& Eardley, D.M., 1996, Phys. Rev. D, 53, 3056
\bibitem[van Putten(2010)]{van10} van Putten, M.H.P.M., 2010, Class. Quant. Grav., 27, 075011
\bibitem[Wald(1984)]{wal84} Wald, R.M., 1984, General Relativity (Chicago: University of Chicago Press)
\bibitem[Riess et al.(1998)]{rie98} Riess, A., et al., 1998, ApJ,  116, 1009
\bibitem[Perlmutter et al.(1999)]{per99} Perlmutter, S., et al., 1999, ApJ, 517, 565 
\bibitem[Misner et al.(1973)]{mis73} Misner, C.W., Thorne, K.S., \& Wheeler, J.A., 1973, Gravitation (San Francisco:
\bibitem[Fierz et al.(1939)]{fie39} Fierz, M., \& Pauli, W., 1939, Proc. R. Soc. Lond., A173, 211
\bibitem[Thorne(1992)]{tho98} Thorne, K.S., 1992, $in$ Recent Advances in General Relativity, eds. A Janis and J Porter (Boston: Birkhauser)
\bibitem[Iwasawa(1996)]{iwa96} Iwasawa, K., Fabian, A.C., \& Reynolds, C.S., et al., 1996, MNRAS, 282, 672
\bibitem[Fabian et al.(1995)]{fab95} Fabian, A.C., et al., 1995, MNRAS, 277, L11
\bibitem[Chandrasekhar(1983)]{cha83} Chandrasekhar, S., 1983, {\em The Mathematical Theory of Black Holes} (New York: Oxford University Press)
\bibitem[Fanaroff \& Riley(1974)]{fan74} Fanaroff, B.L., \& Riley, J.M., 1974, MNRAS, 167, 31P
\bibitem[Lichnerowich(1967)]{lic67} Lichnerowicz A., 1967, Relativistic hydrodynamics and magneto-hydrodynamics (W.A. Benjamin Inc., New York)
\bibitem[van Putten(1994)]{van94} van Putten, M.H.P.M., 1994, Phys. Rev. D., 50, 6640
\bibitem[Goldreich \& Julien(1969)]{gol69} Goldreich, P., \& Julian, W.H., 1969, ApJ, 157, 869
\bibitem[Bardeen et al.(1972)]{bar72} Bardeen, J.M., Press, W.H., \& Teukolsky, S.A., 1972, Phys. Rev. D, 178, 347 
\bibitem[Stefani et al.(2009)]{ste09} Stefani, F., Gerbeth, G., Gundrum, T., et al., Phys. Rev. E, 2009, 80, 066303 
%\bibitem[van Putten(2005)]{van05} van Putten, M.H.P.M., 2005, Nuov. Cim. C, 28, 597; ibid., 2008, ApJ, 685, L63

\end{thebibliography}
\end{document}